\documentclass[11pt]{report}
\usepackage{amsmath}
\usepackage{sw20mitt}
\usepackage{makeidx}
\usepackage{graphicx}

\input{tcilatex}
\makeindex
\graphicspath{{C:/swp55/Common/thesisnew/}}

\begin{document}

\title{The Generalized Quantization Schemes for Games and its Application to
Quantum Information}
\author{Ahmad Nawaz}
\prevdegrees{M.Sc. Quaid-i-Azam University Islamabad,\\
\ \ \ \ \ \ \ \ \ \ \ \ \ \ \ \ Pakistan (1988)\ \ \ \ \ \ \ \ \ \ \ \ \ \ \
\ \ \ \ \ \ \ \ \ \ \ \ \ \ }
\department{Department of Electronics}
\thisdegree{Doctor of Philosophy }
\university{Quaid-i-Azam University Islamabad, Pakistan}
\copyrightnotice{Quaid-i-Azam University Islamabad, Pakistan (2004)}
\degreemonth{January}
\degreeyear{2007}
\date{12 June 2007}
\chairmanname{Professor Dr. S. Azhar Abbas Rizvi}
\chairmantitle{Head of Department}
\super{Abdul Hameed Toor}
\supertitle{Associate Professor}
\maketitle
\tableofcontents
\listoffigures

\begin{abstract}
Theory of quantum games is relatively new to the literature and its
applications to various areas of research are being explored. It is a novel
interpretation of strategies and decisions in quantum domain. In the earlier
work on quantum games considerable attention was given to the resolution of
dilemmas present in corresponding classical games. Two separate quantum
schemes were presented by Eisert \textit{et al.} \cite{eisert} and Marinatto
and Weber \cite{marinatto} to resolve dilemmas in Prisoners' Dilemma and
Battle of Sexes games respectively. However for the latter scheme it was
argued \cite{benjamin} that dilemma was not resolved. We have modified the
quantization scheme of Marinatto and Weber to resolve the dilemma. We have
developed \textrm{a} generalized quantization scheme for two person non-zero
sum games which reduces to the existing schemes under certain conditions.
Applications of this generalized quantization scheme to quantum information
theory are studied. Measurement being ubiquitous in quantum mechanics can
not be ignored in quantum games. With the help of generalized quantization
scheme we have analyzed the effects of measurement on quantum games. Qubits
are the important elements for playing quantum games and\ are generally
prone to decoherence due to their interactions with environment. An analysis
of quantum games in presence of quantum correlated noise\textrm{\ }is
performed in the context of generalized quantization scheme. Quantum key
distribution is one of the key issues of quantum information theory for the
purpose of secure communication. Using mathematical framework of generalized
quantization scheme we have proposed a protocol for quantum key
distribution. This protocol is capable of transmitting four symbols for key
distribution using a two dimensional quantum system. Quantum state
tomography has a substantial place in quantum information theory. Much like
its classical counterpart, its aim is to reconstruct a three dimensional
image through a series of different measurements. Making use of the
mathematical framework of generalized quantization scheme we have presented
a technique for quantum state tomography.
\end{abstract}

\begin{center}
\bigskip

\bigskip

\bigskip

\bigskip\ \bigskip \bigskip \vspace{2.2in}\bigskip \\[0pt]
\begin{figure}[th]
\centering
\par
\includegraphics[scale=1.2]{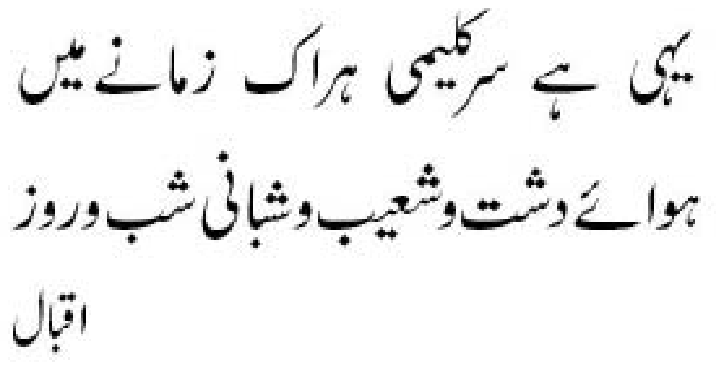} \label{fig:image1}
\end{figure}

\ \ \ \ \ \ \ \ \pagebreak \pagebreak\ \ \ \ 

\bigskip

\pagebreak

\bigskip

\bigskip

\bigskip

\bigskip\ \bigskip \bigskip \vspace{2.2in}\bigskip

\bigskip

\bigskip

\bigskip

{\Huge Dedications\bigskip }

\textrm{To the memories of my father who never compromised on
studies,\medskip }

\textrm{To the prayers of my mother without them this task was plainly
impossible\medskip }

\textrm{and\medskip }

\textrm{To the honour and dignity of my teachers seen and unseen.}

\bigskip

\allowbreak \pagebreak

{\huge Acknowledgement}
\end{center}

All glory be to Allah who helped me to manage the unmanageable.

I am extremely grateful to my supervisor, Dr. Abdul Hameed Toor for his
invaluable assistance, expert guidance, amiable mood and the provision of
friendly and affable environment during the entire research work on and off
the campus. A huge note of thanks must go to my teachers Dr. Azhar Abbas
Rizvi, Dr. Farhan Saif, Dr. Qaiser Naqvi and all other faculty members of
the Department of Electronics, Quaid-i-Azam University Islamabad for their
help and guidance during my course work. I am extremely thankful to Dr.
Azhar Iqbal for introducing me to the fascinating field of quantum game
theory and for motivating discussions on the subject. I am very grateful to
Prof. J\'{a}nos A. Bergou\ for his helpful suggestions and invaluable
discussions on quantum information theory during his visit to Pakistan. An
honourable note of thanks must go to Prof. Dr. Moiz Hussain who taught me to
be determined, concentrated and feisty in complicated, impenetrable and
challenging situations of life.

My respectful thanks are also due to my senior colleagues Dr. Altaf Hussain,
Mr. Ejaz Ahmad Mukhtar, Mr. Fahim Ahmad and Mr. Shahid Noman-ul-Haq\ for
providing me the opportunity and facilities for studies. To my friends and
colleagues, Mr. Khalil Ahmad, Mr. Muhammad Azam Ghori, Mr. Muhammad Salim,
Mr. Muhammad Qadir Asad, Mr. Athar Rasool, Mr. Muhammad Israr Khan, Mr.
Muhammad Iqbal and Z. Z. Bhatty- thank you for your constant encouragement,
empowering remarks, regular inspiration and stimulating discussions in these
challenging hours of my life. I am very thankful to my best friends and
caring fellows Mr. Mehboob Hussain, Mr. Mustansar Nadeem and Miss. Nigum
Arshed whose company imparted me indelible memories and converted my stay at
university into a memorable time of my life.

How can I forget my mother at this occasion who always prayed for my success
and never complained about my long absence during the research period. Thank
you, mother! thanks a lot. I am very grateful to my brothers Mr. Ayaz Khan,
Mr. Haroon Khan, Mr. Sheraz Khan, Mr. Aamer Shehzad and sisters for their
support and care during these busy hours. At last and not at least I am
thankful to my wife, Saima Rasti and children, Minahil Yaqub, Laiba Yaqub
and Muhammad Obaidullah Yaqub who energized me with their innocent remarks
and kept me reminding that there is also a world outside the
academia.\pagebreak

\begin{center}
This thesis is based on the following publications:\bigskip
\end{center}

\begin{itemize}
\item Ahmad Nawaz and A. H. Toor, Dilemma and quantum battle of sexes, J.
Phys. A: Math. Gen. \textbf{37}, 4437\ (2004).

\item Ahmad Nawaz and A. H. Toor, Generalized quantization scheme for two
person non-zero sum game, J. Phys. A: Math. Gen. \textbf{37}, 11457 (2004).

\item Ahmad Nawaz and A. H. Toor, Role of measurement in quantum games, J.
Phys. A: Math. Gen. \textbf{39}, 2791\ (2006).

\item Ahmad Nawaz and A. H. Toor, Quantum games with correlated noise, J.
Phys. A: Math. Gen. \textbf{39},\ 9321 (2006).

\item Ahmad Nawaz, A. H. Toor and J. Bergou, Efficient quantum key
distribution: submitted.

\item Ahmad Nawaz and A. H. Toor, Quantum games and quantum state
tomography: in preparation.

\pagebreak
\end{itemize}

\chapter{Introduction}

Game theory%
\index{Game theory} deals with the situations where two (or more) players or
the decision makers compete to maximize their respective gains. The player's
gain known as payoff%
\index{Payoff} can be in the form of money or some sort of spiritual
happiness which one feels on one's success. The players%
\index{Player} are rational in nature therefore, while taking any action to
achieve their objectives they keep an eye on the expectations and objectives
of the other players and they also know well the strategies to achieve these
objectives \cite{dixit}. Furthermore these interactions are strategic in
nature as the payoff of one player depends on his own and as well as on the
strategies adopted by other player/players \cite{neumann}. The strategy%
\index{Strategy} of players is a complete plan of actions depending on the
sensitivity and nature of a particular situation (game). The rational
reasoning of the players for selection of those strategies that maximizes
their payoffs decides the outcome of a game. A set of strategies from which
unilateral deviation of a player reduces his/her payoff is called Nash
equilibrium%
\index{Nash equilibrium (NE)} (NE) of the game which is a key concept in the
solution of a game \cite{nash}.

Game theory was developed by von Neumann and Morgenstern \cite{neumann} and
John Nash \cite{nash} as a tool to understand economic behaviors. Since then
it has been widely used in various fields including warfare%
\index{Warfare}, anthropology,%
\index{Anthropology} social psychology,%
\index{Social psychology} economics,%
\index{Economics} politics,%
\index{Politics} business,%
\index{Business} international relations,%
\index{International relations} philosophy%
\index{Philosophy} and biology%
\index{Biology}. It is also used by computer scientists in artificial
intelligence \cite{Emilia,Schulte}%
\index{Artificial intelligence} and cybernetics%
\index{Cybernetics} \cite{Gubko,Kuntsevich}. There\ is an increasing
interest in applying the game theoretic concepts to physics%
\index{Physics} \cite{abbott}. Some algorithms and protocols of quantum
information%
\index{Quantum information} theory has also been formulated in the language
of game theory \cite{wiesner,goldenberg,vaidman,gisin,ekert-1,cloning}.

The problems of classical game theory can be implemented into an
experimental (physical) set-up by using classical bits%
\index{Classical bit}. A classical bit can be represented by any two level
system such as a coin i.e. it can be encoded on any system that can take one
of the two distinct possible values. For example a bit on a compact disk
means whether a laser beam is reflected or not reflected from its surface. A
bit is represented by the Boolean states $0$ and $1$. To play\ two players
classical games experimentally we need an arbiter having two similar coins
in same state. He hands over a coin to each of the player. The strategies of
the players are to flip or not to flip the coin. The players return their
coins to arbiter after playing the respective strategies. Checking the state
of coins the arbiter announces the payoffs for players using the payoff
matrix known to both the players.

Quantum games, on the other hand, are played using quantum bits (qubits) and
the qubits are much different than their classical counterpart. A qubit%
\index{Qubit} is a microscopic system such as an electron or nuclear spin or
a polarized photon. In this case the Boolean states $0$ and $1\ $are
represented by a pair of reliably distinguished states of the qubit \cite%
{bennett00}. Spin up and spin down of an electron or the horizontal and
vertical polarizations of photon are very remarkable examples in this
regard. Qubits can exist in form of superposition in two dimensional Hilbert
space spanned by the unit vectors. Furthermore qubits can also exist in a
state totally different than classical states called entangled state.
Computers that work on the basis of these quantum resources are known as
quantum computers 
\index{Quantum computer} \cite{feynmann,deutch,shor1,shor2,grover1,grover2}.
Extensive\ study of quantum computation motivated the study of quantum
information theory%
\index{Quantum information}. This relatively new research field taught to
think physically about computation and provided with the exciting
capabilities for the information storage, processing and communication \cite%
{chuang}. Processing of information in quantum domain started an interesting
debate among scientists for faster than light communication, a task that is
impossible according the Einstein's theory of relativity \cite{herbert}. It
was directly linked to a question whether it was possible to clone an
unknown quantum state. However, no cloning theorem \cite{wootters} proved
that the task that was easy to accomplish with classical information is
impossible for quantum information. Quantum information%
\index{Quantum information} theory gave a new brand of cryptography%
\index{Cryptography} where security does not depend upon the computational
complexity but depends upon fundamental physics and introduced quantum
computers that can provide the mathematical solutions to certain problems
very fast. It is stated that information theory based on quantum principles
extends and completes the classical information theory just as the complex
numbers extend and complete the real numbers \cite{bennett00,bennett1}.
These fascinating ideas led to translate the problems of game theory into
physical set-up that uses qubits instead of classical bits \cite%
{meyer,eisert,marinatto}.

Quantum game theory%
\index{Quantum game!story of spaceship} started with an interesting story of
success of a hypothetical quantum player over a classical player in quantum%
\emph{\ }penny flip game \cite{meyer}. David Meyer described this game%
\index{Quantum penny flip} by the story of a spaceship which faces a
catastrophe during its journey.\ Suddenly a quantum being, Q, appears to
help save the spaceship if Picard%
\index{Captain Picard}, the captain of the spaceship, beats him in a penny
flipping game. According to the game, Picard is to place the penny with head
up in a box. Q has an option to either flip the penny or leave it unchanged
without looking at it. Then Picard has the same options without having a
look at the penny. Finally Q takes the turn with the same options without
looking at the penny. If in the end penny is head up then Q wins otherwise
Picard wins. Captain Picard being expert of game theory knows that this game
has no deterministic solution and deterministic Nash equilibrium \cite%
{neumann,nash}. In other words, there exist no such pair of pure strategies
from which unilateral withdrawal of any player can enhance his/her payoff.
Therefore, he agrees to play with Q. But to Picard's surprise, Q always
wins. Since the quantum being Q is capable of playing quantum strategies%
\index{Quantum strategy} which is the superposition of head and tail in the
two dimensional Hilbert space%
\index{Hilbert space}, thus he is always the winner.

In non-zero sum classical games Nash equilibrium (NE)%
\index{Nash equilibrium (NE)!shortcommings of} is central to analysis,
however, this concept has some shortcomings as well. First, it is not
necessarily true that each game has a unique Nash equilibrium.\ There are
examples of the games with multiple Nash equilibria where the players cannot
choose the Nash equilibrium e.g. Battle of Sexes%
\index{Battle of sexes} and Chicken%
\index{Chicken game} games. Second, in some cases Nash equilibrium could
result outcomes being very far from the benefit of players. Prisoners'
Dilemma%
\index{Prisoners' dilemma} is an interesting example depicting such a
situation where the players trying to maximize their respective payoffs\
fall in a dilemma and end up with worst outcomes. Quantum game theory%
\index{Quantum game} helps resolve such dilemmas \cite{eisert,marinatto} and
shows that quantum strategies can be advantageous\ over classical strategies 
\cite{meyer,eisert,flitney}. To deal with such situations one of the
foremost and elegant quantization schemes 
\index{Quantization scheme!of Eisert et al.|textit} is introduced by Eisert 
\textit{et al}. \cite{eisert} taking Prisoners' Dilemma as an example. In
this quantization scheme the strategy space of the players is a two
parameter set of $2\times 2$ unitary matrices. Starting with maximally
entangled initial quantum state the authors showed that for a suitable
quantum strategy the dilemma disappears. They also pointed out a quantum
strategy which always wins over all the classical strategies. Marinatto and
Weber%
\index{Quantization scheme!of Marinatto and Weber} \cite{marinatto}
introduced another interesting and simple scheme for the analysis of
non-zero sum classical games in quantum domain. They gave Hilbert structure
to the strategic spaces of the players. They used maximally entangled
initial state and allowed the players to play their tactics by applying
probabilistic choice of unitary operators. They applied their scheme to an
interesting game of Battle of Sexes%
\index{Battle of sexes} and found out the strategy for which both the
players can achieve equal payoffs. Both Eisert's and Marinatto and Weber's%
\emph{\ }schemes give interesting results for various quantum analogue of
classical games \cite{flitney,azhar,azhar1,azhar2,jiang,rosero}.

Meyer \cite{meyer,meyer-1}, in his pioneering work pointed out a connection
between quantum games and quantum information processing. Lee and Johnson 
\cite{lee-2} presented a game theoretic model for quantum state estimation%
\index{Quantum state estimation} and quantum cloning%
\index{Quantum cloning}. They also developed a connection between quantum
games and quantum algorithms \cite{lee}. In this thesis we introduced a
generalized quantization scheme for two person non zero sum games and by
using the mathematical framework of this generalized quantization scheme
(chap. \ref{general}) we have proposed an efficient protocol for quantum key
distribution. This protocol can be used to transmit four symbols for key
distribution between sender and receiver\ using a two dimensional system,
whereas in other quantum key distribution schemes higher dimensional systems
are used for this purpose \cite{ekert}. Using the framework of generalized
quantization scheme a protocol for quantum state tomography%
\index{Quantum state tomography} is also presented. It can safely be stated
that this work is a step forward for strengthening the established link
between quantum games and quantum information theory.

\begin{center}
{\large Thesis Layout and Statement of Original Contribution}
\end{center}

Chapter \ref{CGT} is a brief introduction to classical game theory\textrm{\ }%
while chapter \ref{QM} contains some basic concepts of quantum mechanics
required to understand quantum games. Chapter \ref{QGT} and chapter \ref{QIT}
give reviews of quantum game theory and quantum information theory
respectively.

In section (\ref{dilemma}) we show that the worst case payoffs scenario in
quantum Battle of Sexes%
\index{Battle of sexes}, as pointed out by Benjamin \cite{benjamin}, is not
due to the quantization scheme itself but it is due to the restriction on
the initial state parameters of the game. If the game is allowed to start
from a more general initial entangled state then a condition on the initial
state parameters can be set such that the payoffs for the mismatched or the
worst case situation are different for different players which results in a
unique solution of the game.

Chapter \ref{general} deals with the generalized quantization scheme%
\index{Quantization scheme!generalized} for two person non-zero sum games
which gives a relationship between Eisert \textit{et al.} \cite{eisert} and
Marinatto and Weber \cite{marinatto} quantization schemes.\textrm{\ }%
Separate set of parameters are identified for which this scheme reduces to
that of Marinatto and Weber and Eisert \textit{et al.} schemes. Furthermore
there have\ been identified some other interesting situations which are not
apparent within the exiting quantizations schemes.\textrm{\ }In section (\ref%
{measurement}) the effects of measurement on quantum games%
\index{Quantum game!role of measurement} are analyzed under the generalized
quantization scheme. It is observed that as in the case of quantum channel
capacities \cite{king} , one can have four types of payoffs in quantum games
for different combinations of input states and measurement basis.
Furthermore a relation among these payoffs is also established.

In chapter \ref{correlated noise} we analyze quantum games%
\index{Quantum game!with correlated noise} in presence of quantum correlated
dephasing channel in the context of our generalized quantization scheme for
non-zero sum games. It is shown that in the limit of maximum correlation the
effect of decoherence vanishes and the quantum game behaves as a noiseless
game.

In chapter \ref{crypto} and \ref{QST},\ using the mathematical framework of
generalized quantization scheme, we present protocols for quantum key
distribution and quantum state tomography respectively.

\chapter{\label{CGT}Game Theory}

Game theory provides us with mathematical tools to help understand the
phenomena that we observe when two or more players with conflicting
interests interact. The physical situations arising in daily life are
represented by abstract models and the contestants are supposed to be
rational in nature who reason strategically \cite{dixit,martin,anatol}.
Players play their strategies while keeping an eye on the objectives and
expectations of other players and\ hence the resulted payoffs are functions
of the strategies adopted by all the players involved in the contest.

In the following some basic definitions and terminology required to help
understand the mathematical models of game theory are given following with
some interesting examples from classical game theory. For these definitions
and examples we consulted the Refs. \cite{dixit,martin,anatol}.

\section{Basic Definitions}

\textbf{Game}%
\index{Game!definition of}:- A game consists of a set of players, a set of
rules that dictates what actions the players can perform and a payoff
function that tells about the reward of a player against given set of
strategies. Mathematically it is a triple $\left( N,\Omega ,\$\right) $
where $N$ is the number of players, $\Omega =_{\times k}\Omega $ with $1\leq
k\leq N$ such that each $\ \Omega _{k}$ is the set of strategies for the $%
kth $ player and $\$:\Omega \longrightarrow R^{N}$ where $\$$ is the payoff
of the $kth$ player.

\textbf{Player}%
\index{Player!definition of}:- In all game theoretic models the basic entity
of a game is a player. It is an agent taking part in a game. Player can be
an individual or a set of individuals.

\textbf{Payoff}%
\index{Payoff!definition of}:-These are the real numbers associated with
each possible outcome of a game.

\textbf{Move}%
\index{Move!definition of}:- These are the actions or choices available to a
player in a game.

\textbf{Strategy}%
\index{Strategy!definition of}:- It is the complete plan of actions of
players for all possible circumstances during the course of the play.

\textbf{Pure strategy}%
\index{Strategy!pure}:- Pure strategy is a nonrandom course of action for
players. These are the moves that are specified without any uncertainty.
Unless otherwise stated a strategy refers to a pure strategy.

\textbf{Mixed strategy}%
\index{Strategy!mixed}:- This is a rule that tells the player to use each or
some of their pure strategies with specific probabilities.

\textbf{Dominant strategy}%
\index{Strategy!dominant}:- A pure strategy is referred to as dominant
strategy if it results higher payoff than any alternate strategy for all
possible choices of the opposing players. Mathematically a strategy $i$ is
dominant strategy of player $i$ if%
\begin{equation*}
\$_{i}(s_{1},%
\text{......}s_{i-1},s_{i},\text{......}s_{n})\geq \$_{i}(s_{1},\text{.....}%
s_{i-1},\acute{s}_{i},.\text{......}s_{n}).
\end{equation*}

\textbf{Rationality}%
\index{Rationality}:- Reasoning strategically while keeping an eye on the
objects and expectations of other players.

\textbf{Zero sum game}%
\index{Game!zero sum}:- A game is zero-sum if the sum of the players'
payoffs is always zero. A two players zero sum game is also called a duel.

\textbf{Non zero sum game}%
\index{Game!non zero sum}:- A game in which the sum of the players' payoffs
is not zero.

\textbf{Information}%
\index{Information}:- What each player knows at each point of a game.
Information may be perfect or imperfect, symmetric or asymmetric, complete
or incomplete and certain or uncertain.

\textbf{Symmetric game}%
\index{Game!symmetric}:- \textrm{A game }$G=(I,S,\$)$\textrm{\ is a
symmetric two player game if }$I=\{1,2\}$, $S_{1}=S_{2}$\textrm{\ and }$%
\$_{2}\left( s_{1},s_{2}\right) =\$_{2}\left( s_{2},s_{1}\right) $\textrm{\
for all }$\left( s_{1},s_{2}\right) \in S.$\textrm{\ In symmetric games} all
the players face exactly the same choices and exactly the same outcomes
associated with their choices. Otherwise the game is asymmetric.

\textbf{Nash equilibrium}%
\index{Nash equilibrium (NE)!definition of} \textbf{(NE)}:- It is set of
strategies from which unilateral deviation of a player reduces his/her
payoff.

\textbf{Maximin}:-%
\index{Maximin} The largest minimum payoff in a zero sum game.

\textbf{Minimax}:-%
\index{Minimax} The smallest maximum payoff in a zero sum game.

\textbf{Pareto optimal} (PO)%
\index{Pareto optimal}:- A solution set is Pareto Optimal means that there
are no other solutions in which all the players simultaneously do better.

\textbf{Evolutionary stable strategy (ESS)}%
\index{ESS}:- The concept of ESS is refinement to Nash equilibrium. An ESS
is a strategy if adopted by a population then no mutants can invade it by
playing any other strategy.

\textbf{Sequential games}%
\index{Game!sequential}:- These are the game where the players act on strict
turns.

\textbf{Simultaneous games}%
\index{Game!simultaneous}:- These are the games where the players act at the
same time.

\section{Representation of Games}

There are different ways to represent a game, however, the following two
ways are most commonly used \cite{rasmusen}.

\subsection{Normal Form}

In the normal form%
\index{Game!normal form} the game is represented by a payoff matrix which
shows the players, strategies and payoffs. This representation is also
called strategic form representation. The normal form representation for the
Prisoners' Dilemma%
\index{Prisoners' dilemma} game, for example, is given by the following
payoff matrix 
\begin{equation}
\text{{\large Alice}}%
\begin{array}{c}
C \\ 
D%
\end{array}%
\overset{%
\begin{array}{c}
\\ 
\text{{\large Bob}}%
\end{array}%
}{\overset{%
\begin{array}{cc}
C\text{ \ \ \ \ } & D%
\end{array}%
}{\left[ 
\begin{array}{cc}
\left( 3,3\right) & \left( 0,5\right) \\ 
\left( 5,0\right) & \left( 1,1\right)%
\end{array}%
\right] }},
\end{equation}%
In this case, there are two players; one chooses the row and the other
chooses the column. Each player has two strategies $C$ and $D$. The payoffs
are provided in the interior as the elements of the bi-matrix. The first
number is the payoff received by the row player, Alice and the second is the
payoff for the column player, Bob. Suppose that Alice plays $C$ and that Bob
plays $D$, then Alice gets 0, and Bob gets 5. When a game is presented in
normal form, it is presumed that each player acts simultaneously or, at
least, without knowing the actions of the other player.

\subsection{Extensive Form}

In the extensive form%
\index{Game!extensive form} games are presented by trees. The points of
choice for a player are at each vertex or node of the tree. The number
listed at vertex is the identification for the players and the lines going
out of the vertex specifies the moves of the players. The payoffs are
written at the end of branches of the tree. For example, we can represent
Prisoners' Dilemma game in extensive form as shown in figure \ref{extensive
form} 
\begin{figure}[th]
\centering
\includegraphics[scale=.8]{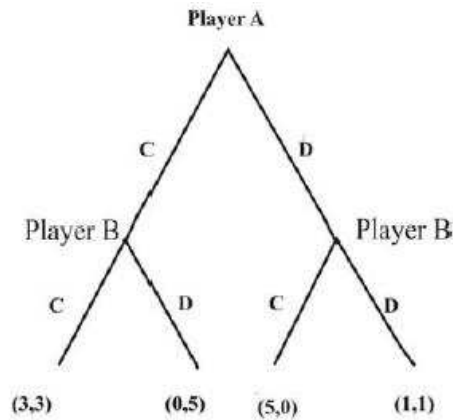}
\caption{Prisoners' Dilemma in extensive form}
\label{extensive form}
\end{figure}

Both the sequential move game and simultaneous move game can be represented
by extensive form. In the case of simultaneous move games either a dotted
line or circle is drawn around two different vertices to show that they are
the part of the same information set which means that the players do not
know at which point they are.

\section{\label{examples}Examples}

In the following we give some examples of classical games that very often
appear in the literature on game theory.

\subsection{Matching Pennies}

Matching pennies%
\index{Matching pennies} is a simple example from a class of zero sum games.
In this game two players Alice and Bob show heads or tails from a coin. If
both are heads or both are tails then Alice wins, otherwise Bob wins. The
payoff matrix for this game is 
\begin{equation}
\overset{%
\text{ \ \ \ \ \ \ \ \ \ \ \ \ \ \ \ \ \ \ \ }%
\begin{array}{c}
\text{{\large Bob}} \\ 
\begin{array}{cc}
H\text{ \ \ \ \ } & T%
\end{array}%
\end{array}%
}{\text{{\large Alice}}%
\begin{array}{c}
H \\ 
T%
\end{array}%
\left[ 
\begin{array}{cc}
\left( 1,-1\right) & \left( -1,1\right) \\ 
\left( -1,1\right) & \left( 1,-1\right)%
\end{array}%
\right] }
\end{equation}

\subsection{Prisoners' Dilemma\textbf{\ }}

This game%
\index{Prisoners' dilemma!introduction of} starts with a story of two
suspects, say Alice and Bob, who have committed a crime together. Now they
are being interrogated in a separate cell. The two possible moves for each
player are to cooperate ($C$) or to defect ($D$) without any communication
between them according to the following payoff matrix%
\begin{equation}
\text{{\large Alice }}%
\begin{array}{c}
C \\ 
D%
\end{array}%
\overset{}{\overset{%
\begin{array}{c}
\text{{\large Bob}} \\ 
\begin{array}{cc}
C\text{ \ \ \ \ } & D%
\end{array}%
\end{array}%
}{\left[ 
\begin{array}{cc}
\left( 3,3\right) & \left( 0,5\right) \\ 
\left( 5,0\right) & \left( 1,1\right)%
\end{array}%
\right] }}.  \label{matrix-prisoner}
\end{equation}%
\emph{\ }It is clear from the above payoff matrix that $D$\ is the dominant
strategy for both players. Therefore, rational reasoning forces each player
to play $D$. Thus ($D,D$) results as the Nash equilibrium%
\index{Nash equilibrium (NE)} of this game with payoffs $(1,1),$ which is
not Pareto Optimal%
\index{Pareto optimal}. However, it was possible for the players to get
higher payoffs if they would have played $C$\ instead of $D$. This is the
origin of dilemma in this game \cite{flood}. A generalized payoff matrix for
Prisoners Dilemma is given as 
\begin{equation}
\text{{\large Alice }}%
\begin{array}{c}
C \\ 
D%
\end{array}%
\overset{}{\overset{%
\begin{array}{c}
\text{{\large Bob}} \\ 
\begin{array}{cc}
C\text{ \ \ \ \ } & D%
\end{array}%
\end{array}%
}{\left[ 
\begin{array}{cc}
\left( r,r\right) & \left( s,t\right) \\ 
\left( t,s\right) & \left( u,u\right)%
\end{array}%
\right] }},  \label{matrix-prisoner-general}
\end{equation}%
where $t>r>u>s.$

The games like Prisoners' Dilemma%
\index{Prisoners' dilemma} are important for the study of game theory for
two reasons. First the payoff structure of this game is applicable to many
different strategic situations arising in economics, social, political and
biological competitions. Second the nature of equilibrium outcome is very
strange. The players rational reasoning to maximize the payoffs gives them
the payoff which is lower than they could have achieved if they used their
dominated strategies. This particular feature of the game received much
attention that\textrm{\ how the players} can achieve better payoffs \cite%
{dixit}.

\subsection{Chicken Game%
\index{Chicken game}}

The payoff matrix for this game is%
\begin{equation}
\text{{\large Alice }}%
\begin{array}{c}
C \\ 
D%
\end{array}%
\overset{}{\overset{%
\begin{array}{c}
\text{{\large Bob}} \\ 
\begin{array}{cc}
C\text{ \ \ \ \ } & D%
\end{array}%
\end{array}%
}{\left[ 
\begin{array}{cc}
\left( 3,3\right) & \left( 1,4\right) \\ 
\left( 4,1\right) & \left( 0,0\right)%
\end{array}%
\right] }}.  \label{matrix chicken}
\end{equation}%
In this game two players drove their cars straight towards each other. The
first to swerve to avoid a collision is the loser (chicken) and the one who
keeps on driving straight is the winner. There is no dominant strategy in
this game. There are two Nash equilibria%
\index{Nash equilibrium (NE)} $CD$ and $DC,$ the former is preferred by Bob
and the latter is preferred by Alice. The dilemma of this game is that the
Pareto Optimal%
\index{Pareto optimal} strategy $CC$ is not NE.

\subsection{\label{bos}Battle of Sexes}

In the usual exposition of this game%
\index{Battle of sexes} two players Alice and Bob are trying to decide a
place to spend Saturday evening. Alice wants to attend Opera while Bob is
interested in watching TV at home and both would prefer to spend the evening
together. The game is represented by the following payoff matrix:%
\begin{equation}
\text{{\large Alice}}\overset{%
\begin{array}{c}
\end{array}%
}{%
\begin{array}{c}
O \\ 
T%
\end{array}%
}\overset{%
\begin{array}{c}
\text{{\large Bob}} \\ 
\begin{array}{cc}
\text{ }O\text{ \ \ \ \ \ \ } & T%
\end{array}%
\end{array}%
}{\left[ 
\begin{array}{cc}
(\alpha ,\beta ) & (\sigma ,\sigma ) \\ 
(\sigma ,\sigma ) & (\beta ,\alpha )%
\end{array}%
\right] ,}  \label{matrix-BoS}
\end{equation}%
where $O$ and $T$ represent Opera and TV, respectively, and $\alpha $, $%
\beta $, $\sigma $ are the payoffs for players for different choices of
strategies with $\alpha >\beta >\sigma $. There are two Nash equilibria%
\index{Nash equilibrium (NE)} $(O,O)$ and $(T,T)$ existing in the classical
form of the game. In absence of any communication between Alice and Bob,
there exists a dilemma as Nash equilibria $\left( O,O\right) $ suits Alice
whereas Bob prefers $(T,T).$ As a result both players could end up with
worst payoff in case they play mismatched strategies.

\subsection{Rock-Scissors-Paper}

In this game%
\index{Rock-Scissors-Paper} Alice and Bob make one of the symbols with their
hand simultaneously, a rock, paper, scissors. In this game a player wins,
loses or ties. The simple rule of the game is that paper covers rock so a
player who makes the symbol of paper wins over the player who makes the
symbol of rock. Scissors cuts paper so a player making the symbol of
scissors win over the player making the symbol of paper. The rock breaks
scissors therefore, the player who makes the symbol of rock wins over the
player who makes scissors. If both make the same symbol then the game ties.
The payoff matrix for this game is 
\begin{equation}
\text{{\large Alice}}\overset{%
\begin{array}{c}
\end{array}%
}{%
\begin{array}{c}
R \\ 
S \\ 
P%
\end{array}%
}\overset{\overset{}{%
\begin{array}{c}
\text{{\large Bob}} \\ 
\begin{array}{ccc}
R & S & P%
\end{array}%
\end{array}%
}}{\left[ 
\begin{array}{ccc}
0 & 1 & -1 \\ 
-1 & 0 & 1 \\ 
1 & -1 & 0%
\end{array}%
\right] }.
\end{equation}

\section{Applications of Game Theory}

Game theory%
\index{Game theory!uses of} models the real life situations in an abstract
manner. Due to their abstraction these models can be applied to study a wide
range of phenomena \cite{dixit,martin,anatol,martin-1}. The best examples
are the application of the theory of Nash equilibrium concept to study
oligopolistic and political competitions, explanation of the distribution of
tongue length in bees and tube length in flowers with the help of the theory
of mixed strategy equilibrium, the use of the theory of repeated games in
social phenomena like threats and promises \cite{martin}. Furthermore the
models of game theory are successfully being used in fields including
warfare, anthropology, social psychology, economics, politics, business,
international relations, philosophy and biology. It is said that the
importance of game theory for social sciences is the same as the importance
of mathematics is for natural sciences \cite{shaun}. Now there is an
increasing interest of applying it to physics \cite{abbott}.

A. Dixit and S. Skeath \cite{dixit} explained the role of games%
\index{Game!role in real life} in real life as: Our life is full of events
that resemble games. Many events and their outcomes around us force us to
ask why did it happen like this? If we can find the decision makers involved
in these situations who have different aims and interests then game theory
provides us the answer. One of the interesting examples is the cutthroat
competition in business where the rivals are trapped in Prisoners' Dilemma
like situation. Similarly in situations where multiple decision makers
interact strategically, game theory can help to foresee the actions of
rivals and the outcome of their actions. On the other hand we can provide
services to a participant involved in any game like situation to advise him
what strategies are good and which one leads to disaster.

\chapter{\label{QM}Review of Quantum Mechanics}

Quantum mechanics%
\index{Quantum mechanics!basic concepts of} is the mathematical theory for
the description of nature. Its concepts are very different than those of
classical physics. It was developed in response to the failure of classical
physics to explain the atomic structure and some properties of
electromagnetic radiations. Consequently there developed a theory that not
only can explain the structure and the properties of the atoms and how they
interact in molecules and solids but also the properties of subatomic
particles such as protons and neutrons. In this chapter we explain some
concepts of quantum mechanics. In preparation of this chapter we used the
Refs. \cite{chuang,preskill}.

\section{Basic Concepts}

A state is the complete description of the quantum system. For a physical
state of a system it is a ray in Hilbert space.

\subsection{Hilbert Space}

The Hilbert space%
\index{Hilbert space} is specified by the following properties :

\begin{enumerate}
\item It is a vector space over the complex numbers $C$. In Dirac's ket-bra
notation the vectors are denoted by \textit{ket vectors} $\left| \psi
\right\rangle $.

\item It has an inner product $\left\langle \phi \right. \left| \psi
\right\rangle $\ that maps an ordered pair of vectors to $C$ defined by the
following\ properties.

\begin{enumerate}
\item Positivity: $\left\langle \psi \right. \left| \psi \right\rangle >0$
for $\left| \psi \right\rangle \neq 0$, where $\left\langle \psi \right| $\
is called \textit{bra vector}.

\item Linearity: For any two vectors $\left| \psi _{1}\right\rangle ,$ $%
\left| \psi _{2}\right\rangle $\ and $\left| \phi \right\rangle $ we have 
\begin{equation*}
\left\langle \phi \right| \left( a\left| \psi _{1}\right\rangle +b\left|
\psi _{2}\right\rangle \right) =a\left\langle \phi \right. \left| \psi
_{1}\right\rangle +b\left\langle \phi \right. \left| \psi _{2}\right\rangle .
\end{equation*}

\item Skew symmetry: $\left\langle \phi \right. \left| \psi \right\rangle
=\left\langle \psi \right. \left| \phi \right\rangle ^{\ast }$ where $\ast $
denotes the complex conjugate.
\end{enumerate}

\item It is complete in the norm $\left| \left| \psi \right| \right|
=\left\langle \psi \right. \left| \psi \right\rangle ^{%
\frac{1}{2}}$.
\end{enumerate}

\subsection{Observable}

It is the physical property of quantum system that can be measured e.g.
position, spin, and energy of a system. The observables%
\index{Observable} are represented by Hermitian operators in the Hilbert
space. Every observable $%
\hat{A}$ has a spectral decomposition of the form 
\begin{equation}
\hat{A}=\underset{m}{\sum }\lambda _{m}\hat{P}_{m},
\end{equation}%
where $\hat{P}_{m}$ is the projector onto the eigen space of $\hat{A}$\ with
eigenvalue $\lambda _{m}.$

\subsection{Pure State}

A pure quantum state is the state that can be described by a ket vector%
\index{Pure state}. Mathematically it is written as 
\begin{equation}
\left| \psi \right\rangle =\underset{i}{\sum }a_{i}\left| \psi
_{i}\right\rangle ,
\end{equation}%
where $a_{i}$ are complex numbers.

\subsection{Mixed State}

Mixed state%
\index{Mixed state} is a statistical mixture of two or more pure states. For
example 
\begin{equation}
\rho =%
\frac{1}{2}\left| \psi \right\rangle \left\langle \psi \right| +\frac{1}{2}%
\left| \phi \right\rangle \left\langle \phi \right| ,
\end{equation}%
is a mixed state where $\left| \psi \right\rangle $ and $\left| \phi
\right\rangle $ are two pure states.

\subsection{Density Matrix}

A density matrix or density operator describes the statistical state of a
quantum system. Its analogous concept in classical statistical mechanics is
phase-space density which gives the probability distribution of position and
momentum. The need for a statistical description via density matrices arises
when it is not possible to describe a quantum mechanical system by states
represented by ket vectors.

For any pure state density matrix%
\index{Density matrix} is given by the projection operator of the state and
for a mixed state it is the sum of projectors i.e. 
\begin{equation}
\rho =\underset{i}{\sum }p_{i}\left| \psi _{i}\right\rangle \left\langle
\psi _{i}\right| ,
\end{equation}%
where $p_{i}$\ is the probability of the system being in a
quantum-mechanical state $\left| \psi _{i}\right\rangle .$\ The expectation
value of of any operator $%
\hat{M}$\ can be found by density operator using the formula 
\begin{equation}
M=\text{Tr}\left[ \rho \hat{M}\right] =\underset{i}{\sum }p_{i}\left\langle
\psi _{i}\right| \hat{M}\left| \psi _{i}\right\rangle ,
\end{equation}%
where Tr\ represents the trace of a matrix. The probabilities $p_{i}$\ are
nonnegative numbers and normalized i.e. the sum of all the probabilities
equals one. For the case of density matrix it is stated that $\rho $\ is a
positive semidefinite Hermitian operator and its trace is one i.e. its
eigenvalues are nonnegative and sum to one.

\subsection{Qubit}

The unit of classical information is bit. A bit is indivisible and has only
two possible values $0$ or $1$. The corresponding unit of quantum
information is qubit%
\index{Qubit} or quantum bit. The simplest possible Hilbert state is two
dimensional Hilbert space with orthonormal basis $\left| 0\right\rangle $
and $\left| 1\right\rangle .$ These basis correspond to classical bits $0$
and $1$. The difference between bits and qubits is that a qubit can also
exist in a state other than $\left| 0\right\rangle $ or $\left|
1\right\rangle $ in the form of linear combination called superposition.
Mathematically it is written as 
\begin{equation}
\left| \psi \right\rangle =a\left| 0\right\rangle +b\left| 1\right\rangle ,
\end{equation}%
when $a$ and $b$ are complex numbers with $\left| a\right| ^{2}+\left|
b\right| ^{2}=1.$ If a measurement which distinguishes $\left|
0\right\rangle $\ from $\left| 1\right\rangle $\ is performed on this qubit
then the outcome is $\left| 0\right\rangle $ with probability $\left|
a\right| ^{2}$ and $\left| 1\right\rangle $ with probability $\left|
b\right| ^{2}$. Furthermore except for the special cases $a=0$ or $b=0$ the
measurement disturbs the state of a qubit. If a qubit is unknown then there
is no way to determine $a$ and $b$ with single measurement. However with
this measurement the qubit is prepared in known state $\left| 0\right\rangle 
$ or $\left| 1\right\rangle $ which is different from its initial form. The
difference between the qubits and bits in this respect is that a classical
bit can be measured without disturbing it and all the information that was
encoded can be deciphered where as measurement disturbs the qubit. The
physical quantities corresponding to the qubits $\left| 0\right\rangle $ and 
$\left| 1\right\rangle $ can be spin up and spin down state of an electron
or the horizontal and vertical polarization of a photon respectively.

\begin{figure}[th]
\centering
\includegraphics[scale=.6]{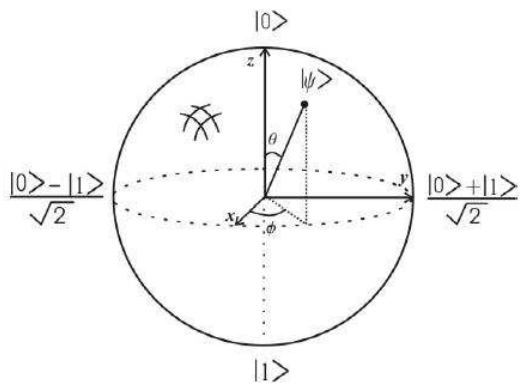}
\caption{Bloch sphere representation of a qubit}
\label{bloch}
\end{figure}

A geometrical representation which provides a useful means of visualizing
the state of a single qubit is known as Bloch sphere representation%
\index{Bloch sphere!representation of} as shown in figure \ref{bloch}.An
arbitrary single qubit state can be written as 
\begin{equation}
\left\vert \psi \right\rangle =e^{i\gamma }\left( \cos 
\frac{\theta }{2}\left\vert 0\right\rangle +e^{i\phi }\sin \frac{\theta }{2}%
\left\vert 1\right\rangle \right)
\end{equation}%
where $\theta ,\phi $ and $\gamma $ are real numbers. The factor $e^{i\gamma
}\ $has no observable effects, therefore, it can be ignored. Furthermore $%
0\leq \theta \leq \pi $ and $0\leq \phi \leq 2\pi $ define a point on a unit
three-dimensional sphere. In this representation the pure states lie on the
surface of the sphere and the mixed states lie inside the sphere. 

\section{Postulates of Quantum Mechanics}

The formulation of quantum mechanics%
\index{Quantum mechanics!postulates of} is based on the following postulates.

\subsection{Postulate 1: State Space}

The state of the system is completely described by a state vector which is a
ray in Hilbert space%
\index{State space}. In Dirac ket-bra notation the states of the system are
denoted by ket vectors $\left| \psi \right\rangle $. In this space the
states $\left| \psi \right\rangle $ and $e^{i\alpha }\left| \psi
\right\rangle $ describe the same physical state. For two given states $%
\left| \psi \right\rangle $ and $\left| \phi \right\rangle $ we can form
another state by superposition as $a\left| \psi \right\rangle +b\left| \phi
\right\rangle .$ The relative phase in this superposition state is
physically significant, this means that $a\left| \psi \right\rangle +b\left|
\phi \right\rangle $ is identical to $e^{i\alpha }\left( a\left| \psi
\right\rangle +b\left| \phi \right\rangle \right) $ but different from $%
a\left| \psi \right\rangle +e^{i\alpha }b\left| \phi \right\rangle .$

\subsection{Postulate 2: Evolution}

The evolution%
\index{Evolution} of the state of a closed system is described by
Schrodinger equation%
\index{Schrodinger equation} 
\begin{equation*}
i\hbar 
\frac{d\left| \psi \right\rangle }{dt}=\hat{H}\left| \psi \right\rangle ,
\end{equation*}%
where $\hbar $ is constant known as Planck's constant and its value is
determined experimentally. $\hat{H}$ is a Hermitian operator called the
Hamiltonian of the system and it gives the energy of the system.

\subsection{\label{POVM}Postulate 3: Measurement}

Quantum measurements are described by a collection $\left\{ \hat{M}%
_{m}\right\} $ of measurement operators. These operators act on the state
space of the system being measured. The index $m$ corresponds to one of the
possible measurement outcomes. If the state of the quantum system is $\left|
\psi \right\rangle $ immediately before the measurement%
\index{Measurement} then the probability that an outcome $m$ will occur is 
\begin{equation}
p(m)=\left\langle \psi \right| 
\hat{M}_{m}^{\dagger }\hat{M}_{m}\left| \psi \right\rangle ,
\end{equation}%
and the state of the system just after the measurement is 
\begin{equation}
\left| \acute{\psi}\right\rangle =\frac{\hat{M}_{m}\left| \psi \right\rangle 
}{\sqrt{\left\langle \psi \right| \hat{M}_{m}^{\dagger }\hat{M}_{m}\left|
\psi \right\rangle }}.
\end{equation}%
The measurement operators satisfy the completeness relation 
\begin{equation}
\underset{m}{\sum }\hat{M}_{m}^{\dagger }\hat{M}_{m}=1,  \label{completeness}
\end{equation}%
which ensures the fact that probabilities sum to $1$.

There are two important special cases for the measurement process. One is
the Projective measurement and the other is POVM (Positive Operator Value
Measure).

\subsubsection{Projective Measurement}

In this case the measurement operators%
\index{Measurement!projective measurement} $%
\hat{M}$ in addition to completeness relation (\ref{completeness}) also
satisfy the condition that $\hat{M}_{m}$ are orthogonal projectors.
Mathematically it can be written as 
\begin{equation}
\hat{M}_{m^{^{\prime }}}\hat{M}_{m}=\delta _{m,m^{^{\prime }}}\hat{M}_{m}.
\end{equation}%
A projective measurement is described by a Hermitian operator $\hat{M}$ on
the state space of the system. This Hermitian operator is termed as
observable. The spectral decomposition of this observable is 
\begin{equation}
\hat{M}=\underset{m}{\sum }mP_{m},
\end{equation}%
where $P_{m}$ is the projector onto the eigen space of $\hat{M}$ with
eigenvalues $m$. On measuring the state $\left| \psi \right\rangle $ the
probability of getting result $m$ is 
\begin{equation}
p(m)=\left\langle \psi \right| P_{m}\left| \psi \right\rangle ,
\end{equation}%
and the state of the system just after the measurement is 
\begin{equation}
\left| \acute{\psi}\right\rangle =\frac{P_{m}\left| \psi \right\rangle }{%
\sqrt{\left\langle \psi \right| P_{m}\left| \psi \right\rangle }}.
\end{equation}%
If the system is subjected to same measurement immediately after the
projective measurement the same outcome occurs with certainty.

\subsubsection{POVM}

In certain experiments the post measurement state of the system is of little
interest whereas the main item of interest is the probabilities of the
respective measurements. One of the examples of such experiment is the Stern
Gerlach experiment. The mathematical tool for measurement in such a case is
POVM%
\index{Measurement!POVM}. A POVM on quantum system is a collection, $\{%
\hat{E}_{m}\}$\ of positive operators satisfying 
\begin{equation}
\underset{m}{\sum }\hat{E}_{m}=I,
\end{equation}%
where $I$ is the identity operator. When a state $\left| \psi \right\rangle $
is subjected to POVM the probability of the outcome $m$ is 
\begin{equation}
p(m)=\left\langle \psi \right| \hat{E}_{m}\left| \psi \right\rangle .
\end{equation}%
The state after measurement is not specified and therefore the measurement
cannot be repeated.

\subsection{\label{postulate 4}Postulate 4 : Composite System}

The state space of the composite physical system%
\index{Composite system} is the tensor product of the component systems. If
we have a quantum mechanical system composed of $n$ quantum systems such
that for each system $i$ the state is $\left| \psi _{i}\right\rangle .$\
Then the joint state for the whole system is given as 
\begin{equation}
\left| \Psi \right\rangle =\left| \psi _{1}\right\rangle \otimes \left| \psi
_{2}\right\rangle 
\text{........}\otimes \left| \psi _{n}\right\rangle .
\end{equation}%
One of the interesting properties of the composite system which is unique to
quantum system is entanglement.

\subsubsection{\label{entanglement}Entanglement}

The state of a composite quantum system can be written as a tensor product
of its component system states. For example, the state of a system composed
of two qubits is specified by a vector in a tensor product space spanned by
the basis $\left| 00\right\rangle ,\left| 01\right\rangle ,\left|
10\right\rangle ,\left| 11\right\rangle $. The quantum mechanical system can
also exist as a linear combination or superposition of the states. Out of
these states there exist some states in which there is a strong correlation
between the components as compared to classical systems. These states are
non-separable i.e. cannot be written as a product of the component systems.
The state of a composite system that cannot be written as product of the
states of its component systems is called entangled state. The well known
examples of maximally entangled states are 
\begin{subequations}
\label{Bell states}
\begin{eqnarray}
\left| \psi ^{+}\right\rangle &=&\frac{1}{\sqrt{2}}\left( \left|
0_{A}0_{B}\right\rangle +\left| 1_{A}1_{B}\right\rangle \right)
\label{Bell states-a} \\
\left| \psi ^{-}\right\rangle &=&\frac{1}{\sqrt{2}}\left( \left|
0_{A}0_{B}\right\rangle -\left| 1_{A}1_{B}\right\rangle \right)
\label{Bell states-b} \\
\left| \phi ^{+}\right\rangle &=&\frac{1}{\sqrt{2}}\left( \left|
0_{A}1_{B}\right\rangle +\left| 1_{A}0_{B}\right\rangle \right)
\label{Bell states-c} \\
\left| \phi ^{-}\right\rangle &=&\frac{1}{\sqrt{2}}\left( \left|
0_{A}1_{B}\right\rangle -\left| 1_{A}0_{B}\right\rangle \right)
\label{Bell states-d}
\end{eqnarray}%
where the first element in the ket refers to system A (first system) and the
second to system B (second system). The states given by Eqs. (\ref{Bell
states}) are known as Bell states%
\index{Bell states}. Note that none of these states can be written as the
product of two states describing the state of the particles. Whenever
measurement is performed on any member of the set then entanglement is
destroyed and the particles obtain the definite state. In an entangled
system the observables are strongly correlated hence required to be
specified with reference to other objects even if they are far apart. For
example for the Bell state 
\end{subequations}
\begin{equation}
\left| \phi ^{-}\right\rangle =%
\frac{1}{\sqrt{2}}\left( \left| 0_{A}1_{B}\right\rangle -\left|
1_{A}0_{B}\right\rangle \right)
\end{equation}%
it is impossible to attribute a definite state to either system for the two
observers Alice and Bob observing the first and second system respectively.
Alice performs measurement on first system in computational basis $\left|
0\right\rangle $, $\left| 1\right\rangle $. There are two outcomes which are
equally likely\emph{\ }(a) if Alice gets $\left| 0\right\rangle $\ then the
system collapses to the state $\left| 01\right\rangle $\ and (b) if Alice
gets $\left| 1\right\rangle $\ then the system collapses to $\left|
10\right\rangle .$\ For the first result of Alice any subsequent measurement
by Bob always returns $\left| 1\right\rangle $ and for the second result of
Alice the subsequent measurement by Bob returns $\left| 0\right\rangle $.%
\emph{\ }It means that the measurement performed by Alice has changed the
second system even if both the systems are spatially separated.

\chapter{\label{QGT}Quantum Game Theory}

In 1970's Maynard Smith gave a new solution concept to game theory
introducing the notion of evolutionary stable strategies%
\index{ESS} (ESS) \cite{price}. He assumed that a perfectly rational being
is not a necessary element to recognize the best strategies in a game but
each player participating in the game is hardwared or programmed in with a
particular strategy by nature. When the game begins, the players contest
with the players programmed with the same or some different strategies. The
payoffs are rewarded to players against their strategies. The strategy that
fares better, multiply faster and the worst strategy declines \cite{dixit}.
As a result only the strategies with best payoff sustain while the others
are swept out. These techniques have successfully been used by biologists to
model the behavior of animals and bacteria. Exploiting these techniques
computer scientists developed some efficient algorithms for optimization
problems known as genetic algorithms%
\index{Algorithms!genetic} \cite{holland}. These algorithms are aimed to
improve the understanding of natural adaptation process, and to design
artificial systems having properties similar to natural systems \cite%
{goldberg}. On the other hand it has recently been shown that games are also
being played at microscopic level by RNA\ virus 
\index{RNA virus}\cite{turner}. Therefore, it will be very interesting to
find whether the microscopic particles such as electrons or atoms are
engaged in any type of quantum contest. One of the reasons behind these
believes is that in some situations atoms and electrons have to choose
between equally advantageous states that is a dilemma formally known as
frustration \cite{frustration}%
\index{Frustration}.\emph{\ }It is expected that quantum games might help
these frustrated atoms in resolving such dilemmas \cite{new scientist}. It
is also believed that frustration%
\index{Frustration} is involved in the phenomenon like high temperature
superconductivity%
\index{High temprature superconductivity}. If it ever becomes possible to
find the particle at play then quantum games might help to understand the
phenomenon of high temperature superconductivity \cite{new scientist}.
Quantum cloning%
\index{Quantum cloning} and quantum state estimation%
\index{Quantum state estimation} has already been proved as games \cite%
{cloning} and quantum cryptography%
\index{Quantum cryptography} is also a game played between the sender, the
receiver and the spy \cite{ekert-1}. These techniques of quantum
cryptography might help constructing a quantum stock market where the
traders would have the opportunity to encode their decisions in qubits. In
such a market entanglement could be used as a helpful resource for traders
to cooperate so that they could avoid crashes that is equivalent to the loss
of everybody in game theory \cite{new scientist}. It is also expected that
quantum games will help to introduce new business models for selling digital
contents on internet%
\index{Internet} that will discourage illegal downloading \cite{patel}. One
of the interesting phenomena that has recently been discovered is Parrondo
effect%
\index{Parrando games} in which two losing games when combined have a
tendency to win \cite{parrando}. Classical Parrondo games and their relation
to Brownian ratchet%
\index{Brownian rachet} has also gained much interest \cite%
{harner,harner1,harner2,van}. Parrondo games have been extended to quantum
domain \cite{meyer1,flitney01}. A connection between Parrondo effects and
the design of quantum algorithms has also been established \cite{lee,meyer2}
and it is further expected that quantum Parrondo games can be helpful to
control qubit decoherence \cite{lee1}. A connection between quantum games
and quantum algorithm for an oracle problem has been established as well 
\cite{meyer3}. Some search algorithms such as simulated annealing \cite%
{anealing,anealing1}%
\index{Algorithms!simulated anealing} and adiabatic algorithms\cite%
{adiabatic,adiabatic1}%
\index{Algorithms!adiabatic} are also expected to be reformulated in the
language of quantum games that might result in a strong connection between
evolutionary games and games derived from the dynamics of physical systems 
\cite{lee1}. Furthermore it is more efficient to play quantum games \cite%
{lee2}. When we entangle two qubits shared between the players then the
players have the greater number of strategies to choose from as compared to
classical games. Therefore, less information needs to be exchanged in order
to play the quantized versions of the classical games.

Quantum computation%
\index{Quantum computation}, quantum cryptography%
\index{Quantum cryptography} and quantum communication%
\index{Quantum communication} protocols are some prominent practical
manifestations of quantum mechanics where the quantum description of the
system has provided clear advantage over the classical counterparts. Simon's
quantum algorithm%
\index{Algorithms!of Simon} to identify the period of a function chosen by
oracle \cite{simon}, Shor's polynomial time quantum algorithm%
\index{Algorithms!of Shor} \cite{shor} and the key distribution protocol%
\index{Quantum key distribution} given by\emph{\ }Bennett and Brassard \cite%
{bennett-0} and by Ekert \cite{ekert}\ are some well known examples. Another
amazing manifestation of quantum mechanical effects is superdense coding%
\index{Quantum superdense coding}. Where using entanglement%
\index{Entanglement} as a resource a sender can transmit two bits of
classical information to a receiver by sending single qubit that is in her
possession \cite{chuang}. The clear superiority of the use of quantum
mechanical resources in the above well established disciplines makes it
natural to think about quantum strategies and quantum games%
\index{Quantum game} that is, if the classical strategies of the players can
be pure or mixed then why these cannot be entangled? Whether these entangled
strategies can be helpful in resolving the dilemmas in classical games such
as that in Prisoners' Dilemma%
\index{Prisoners' dilemma} and the Battle of Sexes%
\index{Battle of sexes} and whether there is any advantage in playing
quantum strategies against classical strategies? Whether this new born field
can be of any help in reformulating the protocols of quantum information
theory and is capable of introducing new protocols and new algorithms? These
are the questions mostly addressed in quantum game theory. In the following
we explain the first quantum game that was originally introduced to
demonstrate the advantage that quantum strategies can achieve over the
classical ones.

\section{Quantum Penny Flip Game}

Quantum penny flip game%
\index{Quantum penny flip} \cite{meyer} is the simplest example to
demonstrate the advantage that a quantum player, Bob can have over a
classical player, Alice. The framework of this game is as follows. Alice
places a coin with head up state in a box. Bob is given the options either
to flip the coin or to leave it unchanged. Then Alice takes her turn with
the same options without having look at the coin. Finally, Bob takes his
turn with the same options without looking at the coin. If at the end the
coin is head up then Bob wins otherwise Alice wins.

This is an example of a zero sum%
\index{Game!zero sum} game where the profit of one player means the loss of
other player. The payoff matrix for this game is%
\begin{equation}
\text{{\large Alice} }%
\begin{array}{c}
N \\ 
F%
\end{array}%
\overset{}{\overset{\overset{}{%
\begin{array}{c}
\text{{\large Bob}} \\ 
\begin{array}{cccc}
NN & NF & FN & FF%
\end{array}%
\end{array}%
}}{\left[ 
\begin{array}{cccc}
-1 & 1 & 1 & -1 \\ 
1 & -1 & -1 & 1%
\end{array}%
\right] ,}}
\end{equation}%
where $F$ stands for flipping and $N$ for not flipping the coin. According
to classical game theory this game has no deterministic solution and
deterministic Nash equilibrium%
\index{Nash equilibrium (NE)!in penny flip game} \cite{neumann,nash}. In
other words, there exist no such pair of pure strategies from which
unilateral withdrawal of a player can enhance his/her payoff.\ However,
there exists a mixed strategies Nash equilibrium which is a pair of mixed
strategies consisting of Alice flipping the coin with probability $%
\frac{1}{2}$ and Bob playing his strategies with probabilities $\frac{1}{4}$%
. When the game starts then to Alice surprise Bob, the quantum player,
always wins. Quantum mechanics tells the entre nous\ that has blindsided
Alice.

Quantum games are played using quantum objects. Therefore to see that how
Bob can win we replace the classical coin with a quantum coin%
\index{Quantum coin}. The main difference between classical coin and quantum
coin is that a classical coin can have one of two possible states i.e.
either head or tail whereas a quantum coin can also exist in a state that is
superposition%
\index{Superposition} of head and tail. In this way unlike a classical coin
a quantum coin has infinite number of states. One of the very suitable
examples of a quantum coin can be an electron defining head by the spin in
+z-axis and tail by spin pointing along -z-axis. This coin is capable of
having a linear combination of the head and tail states known as
superposition%
\index{Superposition} in quantum mechanics. On the other hand Bob is also
capable of playing quantum strategies%
\index{Quantum strategy} that Alice has never heard before. These strategies
are adept in placing the quantum coin in the superposition of head and tail
states in the two dimensional Hilbert space%
\index{Hilbert space}. Let the head of the quantum coin%
\index{Quantum coin} be represented by $\left| 0\right\rangle $ and tail by $%
\left| 1\right\rangle $ in a 2-dimensional Hilbert space%
\index{Hilbert space}. The strategies of the players can be represented by $%
2\times 2$ matrices then the move $F$, to flip and the move $N$, not to flip
the coin are of the form

\begin{equation}
F=\left[ 
\begin{array}{cc}
0 & 1 \\ 
1 & 0%
\end{array}%
\right] ,%
\text{ \ }N=\left[ 
\begin{array}{cc}
1 & 0 \\ 
0 & 1%
\end{array}%
\right] .
\end{equation}%
When the game starts Alice places the coin\ in the head up state i.e. the
initial state of coin is $\left| 0\right\rangle $. Then Bob takes his turn
and proceeds the game by applying the Hadamard gate%
\index{Hadamard gate}%
\begin{equation}
H=%
\frac{1}{\sqrt{2}}\left[ 
\begin{array}{cc}
1 & 1 \\ 
1 & -1%
\end{array}%
\right] ,  \label{hadamard-gate}
\end{equation}%
that transforms the system to $\frac{1}{\sqrt{2}}\left( \left|
0\right\rangle +\left| 1\right\rangle \right) $ that is an equal mixture of
the head and tail states.\ Now on her turn, Alice can either leave the coin
as it is (apply $N)$ or flip the coin (apply $F$). If the coherence of the
system is not effected by actions of Alice then clearly the state of the
quantum system remains unaltered.\ Bob exploiting this fact again applies
Hadamard gate%
\index{Hadamard gate} while taking his turn and the final state of the
system becomes $\left| 0\right\rangle $ resulting a certain win for Bob.

The interesting episode in competition of Alice and Bob led many scientists
to think about the quantization of non-zero sum games. In such games
although the win of a player is not a loss of the other player yet the
rational reasoning to enhance the payoffs can produce undesired outcomes.
Taking an interesting example of such a game known as Prisoners' Dilemma
(see section \ref{examples})%
\index{Prisoners' dilemma}, Eisert \textit{et al}. \cite{eisert} showed that
the dilemma which exist in the classical version of the game does not exist
in quantum version of this game. Further they succeeded in finding a quantum
strategy%
\index{Quantum strategy} which always wins over any classical strategy.
Inspired by their work, Marinatto and Weber \cite{marinatto} proposed
another interesting scheme to quantize the game of Battle of Sexes (see
section \ref{examples})%
\index{Battle of sexes}. They introduced Hilbert structure to the strategic
space of the game and argued that if the players are allowed to play quantum
strategies involving unitary operators for maximally entangled initial state
the game has a unique solution, and dilemma could be resolved.

In the following paragraphs we give a brief introduction to both these
quantization schemes one by one.

\section{\label{eisert-scheme}Eisert, Wilkens and Lewenstein Quantization
Scheme}

Eisert \textit{et al}. \cite{eisert} introduced an elegant quantization
scheme to help\ resolve the dilemma in an interesting game of Prisoners'
Dilemma%
\index{Quantization scheme!of Eisert et al.|textit} with the payoff matrix
of the form (\ref{matrix-prisoner}). This quantization scheme is a physical
model which consists of the following elements known to both the players.

\begin{enumerate}
\item A source of producing two bits, one bit for each player.

\item Physical instruments that enables the player to manipulate their own
bits in a\ strategic manner.

\item A physical measurement device which determines the players payoff from
the strategically manipulated final\ state of two bits.
\end{enumerate}

The classical strategies $C$ (Cooperate) and $D$ (Defect) are assigned two
basis vectors $\left\vert C\right\rangle $ and $\left\vert D\right\rangle $
respectively, in a Hilbert space%
\index{Hilbert space} of a two level system. The state of the game at any
instant is a vector in the tensor product space spanned by the basis vectors 
$\left\vert CC\right\rangle ,\left\vert CD\right\rangle ,\left\vert
DC\right\rangle ,\left\vert DD\right\rangle $ where the first entry in the
ket refers to the Alice's bit and the second entry is for Bob. The
experimental setup for this quantization scheme is shown in figure \ref%
{eisert-diagram}. 
\begin{figure}[th]
\centering
\includegraphics[scale=.6]{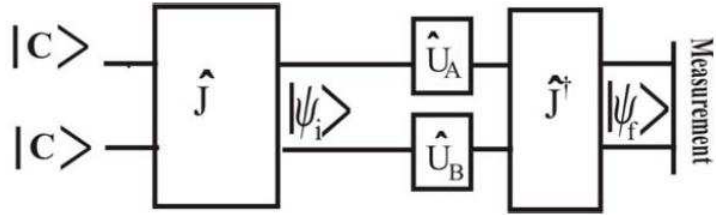}
\caption{Eisert et al. quantization scheme.}
\label{eisert-diagram}
\end{figure}

The game starts with an initial entangled state $\left| \psi
_{i}\right\rangle =%
\hat{J}\left| CC\right\rangle $ where $\hat{J}$ is a unitary operator that
entangles the players qubits and it is known to both the players. The
operator $\hat{J}$ is symmetric for fair games. The strategies of the
players are the unitary operators 
\begin{equation}
\hat{U}(\theta _{i},\phi _{i})=\left[ 
\begin{array}{cc}
e^{i\phi _{i}}\cos \frac{\theta _{i}}{2} & \sin \frac{\theta _{i}}{2} \\ 
-\sin \frac{\theta _{i}}{2} & e^{-i\phi _{i}}\cos \frac{\theta _{i}}{2}%
\end{array}%
\right] ,
\end{equation}%
with $0\leq \theta \leq \pi ,0\leq \phi \leq \frac{\pi }{2}.$ The classical
strategies to cooperate $\hat{C}=\hat{U}(0,0)$ and to defect $\hat{D}=\hat{U}%
(\pi ,0)$. To ensure that the classical Prisoners' Dilemma%
\index{Prisoners' dilemma!.} is the subset of its quantum version the
following set of subsidiary conditions were imposed by Eisert \textit{et al.}
\begin{equation}
\left[ 
\hat{J},\hat{D}\otimes \hat{D}\right] =0,\left[ \hat{J},\hat{C}\otimes \hat{D%
}\right] =0,\left[ \hat{J},\hat{D}\otimes \hat{C}\right] =0,
\label{subsidary-conditions}
\end{equation}%
From conditions (\ref{subsidary-conditions}) it comes out that 
\begin{equation}
\hat{J}=\exp \left\{ i\frac{\gamma }{2}\hat{D}\otimes \hat{D}\right\} ,
\end{equation}%
where $\gamma \in \left[ 0,\frac{\pi }{2}\right] $.

The strategic moves of Alice and Bob are the unitary operators $\hat{U}%
_{A}(\theta _{A},\phi _{A})$\ and $\hat{U}_{B}(\theta _{B},\phi _{B})$
respectively. After the application of these strategies by players the state
of the game evolves to 
\begin{equation}
\left| \psi _{f_{0}}\right\rangle =\left( \hat{U}_{A}\otimes \hat{U}%
_{B}\right) \hat{J}\left| CC\right\rangle .  \label{final-state-p}
\end{equation}%
Prior to measurement for finding the payoffs of the players a reversible
two-bit gate $\hat{J}^{\dagger }$ is applied and the state of the game
becomes 
\begin{equation}
\left| \psi _{f}\right\rangle =\hat{J}^{\dagger }\left( \hat{U}_{A}\otimes 
\hat{U}_{B}\right) \hat{J}\left| CC\right\rangle .
\end{equation}%
This follows a pair of Stern-Gerlach type detectors for measurement and the
expected payoff of Alice comes out to be 
\begin{equation}
\$_{A}=\left[ \$_{CC}\right] _{A,B}\left| \left\langle CC\right. \left| \psi
_{f}\right\rangle \right| ^{2}+\left[ \$_{DD}\right] _{A,B}\left|
\left\langle DD\right. \left| \psi _{f}\right\rangle \right| ^{2}+\left[
\$_{DC}\right] _{A,B}\left| \left\langle DC\right. \left| \psi
_{f}\right\rangle \right| ^{2}+\left[ \$_{CD}\right] _{A,B}\left|
\left\langle CD\right. \left| \psi _{f}\right\rangle \right| ^{2},
\label{Alice-payoff}
\end{equation}%
Here it is important to note that Alice's payoffs $\$_{A}$ depends on the
strategy $\hat{U}_{A}(\theta _{A},\phi _{A})$ of Alice as well as on the
Bob's strategy $\hat{U}_{B}(\theta _{B},\phi _{B}).$

In terms of density matrices the initial state $\rho _{i}=\left| \psi
_{i}\right\rangle \left\langle \psi _{i}\right| $ after the actions of the
players transform to 
\begin{equation}
\rho _{f}=\left( \hat{U}_{A}\otimes \hat{U}_{B}\right) \rho _{i}\left( \hat{U%
}_{A}\otimes \hat{U}_{B}\right) ^{\dagger }.  \label{density-matrix-eisert}
\end{equation}%
To perform measurement arbiter uses the following payoff operators 
\begin{eqnarray}
\pi _{CC} &=&\left| \psi _{CC}\right\rangle \left\langle \psi _{CC}\right| ,%
\text{ \ \ }\left| \psi _{CC}\right\rangle =\frac{\left| CC\right\rangle
+i\left| DD\right\rangle }{\sqrt{2}},  \notag \\
\pi _{CD} &=&\left| \psi _{CD}\right\rangle \left\langle \psi _{CD}\right| ,%
\text{ \ \ }\left| \psi _{CD}\right\rangle =\frac{\left| CD\right\rangle
-i\left| CD\right\rangle }{\sqrt{2}},  \notag \\
\pi _{DC} &=&\left| \psi _{DC}\right\rangle \left\langle \psi _{DC}\right| ,%
\text{ \ \ }\left| \psi _{DC}\right\rangle =\frac{\left| DC\right\rangle
-i\left| CD\right\rangle }{\sqrt{2}}  \notag \\
\pi _{DD} &=&\left| \psi _{DD}\right\rangle \left\langle \psi _{DD}\right| ,%
\text{ \ \ }\left| \psi _{DD}\right\rangle =\frac{\left| DD\right\rangle
+i\left| CC\right\rangle }{\sqrt{2}},  \label{Kraus-operators-eisert}
\end{eqnarray}%
and the expected payoffs for Alice and Bob are computed as 
\begin{equation}
\$_{A,B}=\left[ \$_{CC}\right] _{A,B}Tr\left[ \pi _{CC}\rho _{f}\right] +%
\left[ \$_{CD}\right] _{A,B}Tr\left[ \pi _{CD}\rho _{f}\right] +\left[
\$_{DC}\right] _{A,B}Tr\left[ \pi _{DC}\rho _{f}\right] +\left[ \$_{DD}%
\right] _{A,B}Tr\left[ \pi _{DD}\rho _{f}\right] ,  \label{payoffs-eisert}
\end{equation}%
where $\left[ \$_{ij}\right] _{A,B}$ are the elements of the payoff matrix
for Alice and Bob. Eisert \textit{et al.} \cite{eisert} analyzed Prisoners'
Dilemma%
\index{Prisoners' dilemma!.} game under one\ and two parameters%
\index{One parameter set of strategies}%
\index{Two parameters set of strategies!Eisert et al.} set of strategies 
\cite{eisert1} using the payoff matrix (\ref{matrix-prisoner}) as follows

\subsection{One Parameter Set of Strategies\textbf{.}}

In the one parameter set of strategies%
\index{One parameter set of strategies!Eisert et al.} the players are
restricted to apply the local operators of the form 
\begin{equation}
\hat{U}(\theta _{i})=\left[ 
\begin{array}{cc}
\cos \frac{\theta _{i}}{2} & \sin \frac{\theta _{i}}{2} \\ 
-\sin \frac{\theta _{i}}{2} & \cos \frac{\theta _{i}}{2}%
\end{array}%
\right] ,  \label{one-parameter-set}
\end{equation}%
here $0\leq \theta \leq \pi $ and $i=1,2$. For maximally entangled initial
state 
\begin{equation}
\left| \psi _{CC}\right\rangle =\hat{J}\left| CC\right\rangle =\frac{\left|
CC\right\rangle +i\left| DD\right\rangle }{\sqrt{2}},
\label{eisert-initial-state}
\end{equation}%
by the use of Eq. (\ref{density-matrix-eisert}), (\ref%
{Kraus-operators-eisert}) and (\ref{payoffs-eisert}) the payoffs of the
players become 
\begin{subequations}
\label{payoff-one-parameter-set-eisert}
\begin{eqnarray}
\$_{A}(\theta _{1},\theta _{2}) &=&3\left| \cos \frac{\theta _{1}}{2}\cos 
\frac{\theta _{2}}{2}\right| ^{2}+5\left| \sin \frac{\theta _{1}}{2}\cos 
\frac{\theta _{2}}{2}\right| ^{2}+\left| \sin \frac{\theta _{1}}{2}\sin 
\frac{\theta _{2}}{2}\right| ^{2},  \label{payoffA-one-parameter-eisert} \\
\$_{B}(\theta _{1},\theta _{2}) &=&3\left| \cos \frac{\theta _{1}}{2}\cos 
\frac{\theta _{2}}{2}\right| ^{2}+5\left| \cos \frac{\theta _{1}}{2}\sin 
\frac{\theta _{2}}{2}\right| ^{2}+\left| \sin \frac{\theta _{1}}{2}\sin 
\frac{\theta _{2}}{2}\right| ^{2}.  \label{payoffB-one-parameter-eisert}
\end{eqnarray}%
These payoffs are just like the payoffs of ordinary Prisoners' Dilemma%
\index{Prisoners' dilemma!.} when the players are playing the classical
strategies of cooperation with probabilities $\cos ^{2}%
\frac{\theta _{1}}{2}$ and $\cos ^{2}\frac{\theta _{2}}{2}.$ The
inequalities 
\end{subequations}
\begin{eqnarray}
\$_{A}(\pi ,\theta _{2}) &\geq &\$_{A}(\theta _{1},\theta _{2}),  \notag \\
\$_{B}(\theta _{1},\pi ) &\geq &\$_{B}(\theta _{1},\theta _{2}),
\end{eqnarray}%
hold for all values of $\theta _{1}$ and $\theta _{2},$ giving $(D,D)$ as
the Nash equilibrium of the game. However this Nash equilibrium is not
Pareto Optimal%
\index{Pareto optimal} as it is far from being efficient since $\$_{A}(\pi
,\pi )=\$_{B}(\pi ,\pi )=1,$ just like the classical \ version of the game.
Therefore, the one parameter set of strategies do not resolve the dilemma.

\subsection{Two Parameter Set of Strategies}

When the players are allowed to apply their local operators with two
variable $\left( \theta ,\phi \right) $ two parameters set of strategies%
\index{Two parameters set of strategies!Eisert et al.} results and their
mathematical form is 
\begin{equation}
\hat{U}(\theta _{i},\phi _{i})=\left[ 
\begin{array}{cc}
e^{i\phi _{i}}\cos \frac{\theta _{i}}{2} & \sin \frac{\theta _{i}}{2} \\ 
-\sin \frac{\theta _{i}}{2} & e^{-i\phi _{i}}\cos \frac{\theta _{i}}{2}%
\end{array}%
\right] ,  \label{two-parameter-set}
\end{equation}%
where $i=1,2.$ Using the Eq. (\ref{density-matrix-eisert}), (\ref%
{Kraus-operators-eisert}) and (\ref{payoffs-eisert}) the payoffs come out to
be 
\begin{align}
\$_{A}\left( \theta _{1},\phi _{1},\theta _{2},\phi _{2}\right) & =3\left|
\cos \frac{\theta _{1}}{2}\cos \frac{\theta _{2}}{2}\cos \left( \phi
_{1}+\phi _{2}\right) \right| ^{2}  \notag \\
& +\left| \cos \frac{\theta _{1}}{2}\cos \frac{\theta _{2}}{2}\sin \left(
\phi _{1}+\phi _{2}\right) +\sin \frac{\theta _{1}}{2}\sin \frac{\theta _{2}%
}{2}\right| ^{2}  \notag \\
& +5\left| \sin \frac{\theta _{1}}{2}\cos \frac{\theta _{2}}{2}\cos \phi
_{2}-\cos \frac{\theta _{1}}{2}\sin \frac{\theta _{2}}{2}\sin \phi
_{1}\right| ^{2},  \label{eisert-payoff-general-a}
\end{align}%
\begin{eqnarray}
\$_{B}\left( \theta _{1},\phi _{1},\theta _{2},\phi _{2}\right) &=&3\left|
\cos \frac{1}{2}\theta _{1}\cos \frac{1}{2}\theta _{2}\cos \left( \phi
_{1}+\phi _{2}\right) \right| ^{2}  \notag \\
&&+5\left| \cos \frac{1}{2}\theta _{1}\sin \frac{1}{2}\theta _{2}\cos \phi
_{1}-\sin \frac{1}{2}\theta _{1}\cos \frac{1}{2}\theta _{2}\sin \phi
_{2}\right| ^{2}  \notag \\
&&+\left| \cos \frac{1}{2}\theta _{1}\cos \frac{1}{2}\theta _{2}\sin \left(
\phi _{1}+\phi _{2}\right) +\sin \frac{1}{2}\theta _{1}\sin \frac{1}{2}%
\theta _{2}\right| ^{2}.  \label{eisert-payoff-general-b}
\end{eqnarray}%
\ In this case the Nash equilibrium $\left( \hat{D}\otimes \hat{D}\right) $
no more remains the Nash equilibrium of the game. However, there appears a
new Nash Equilibrium $\left( \hat{Q}\otimes \hat{Q}\right) $ where 
\begin{equation}
\hat{Q}=U(0,\frac{\pi }{2})=\left[ 
\begin{array}{cc}
i & 0 \\ 
0 & -i%
\end{array}%
\right] .
\end{equation}%
Eisert \textit{et al.} \cite{eisert} argued that this unique Nash
Equilibrium is Pareto Optimal%
\index{Pareto optimal} with $\$_{A}(%
\hat{Q},\hat{Q})=\$_{B}(\hat{Q},\hat{Q})=3.$ They further pointed out that
the dilemma in the classical version of game is no more present in the
quantum form of the game.

\subsection{The Miracle Move}

Imagine a situation where one of the players say Alice has the access to
whole of the strategic space where as Bob is restricted to apply classical
strategies only i.e. $\phi _{B}=0$. In this case Eisert \textit{et al.} \cite%
{eisert} pointed out that for Prisoners' Dilemma%
\index{Prisoners' dilemma!.} the quantum player Alice is always equipped
with a strategy $%
\hat{M}(\theta ,\phi )$\ that gives her a sure success against the classical
player, Bob. This quantum move $\hat{M}(\theta ,\phi )$\ is also known as
Eisert miracle move%
\index{Eisert miracle move} and is given by

\begin{equation}
\hat{M}(\frac{\pi }{2},\frac{\pi }{2})=\frac{1}{\sqrt{2}}\left[ 
\begin{array}{cc}
i & 1 \\ 
-1 & -i%
\end{array}%
\right] .
\end{equation}%
The payoffs for Alice and Bob, when Alice is playing $\hat{M}(\frac{\pi }{2},%
\frac{\pi }{2})$\ and Bob is playing any classical strategy $\hat{U}(\theta
),$\ are%
\begin{eqnarray}
\$_{A} &=&3+2\sin \theta ,  \notag \\
\$_{B} &=&\frac{\left( 1-\sin \theta \right) }{2}.  \label{miracle}
\end{eqnarray}%
It is clear from Eqs. (\ref{miracle}) that a quantum player can outperform a
classical player for all values of $\theta $. Furthermore it has also been
shown that in this unfair game the payoff for quantum player is
monotonically increasing function of $\gamma ,$\ the entanglement \textrm{%
measure} of the initial state $\left| \psi _{i}\right\rangle $ \cite%
{jiang,du-1}$.$\ For $\ \gamma =0,$\ $D$ is the dominant strategy and the
payoff\ of minimum value $1$\ is achieved however at $\gamma =\frac{\pi }{2}$
the quantum player achieves the maximum advantage of $3$.\ Furthermore there
exists a threshold value $\gamma _{th}=0.464$\ below which Alice could not
deviate form strategy $D.$\ However beyond this threshold value she will
discontinuously have to deviate from $D$\ to $Q.$\ At critical value of
entanglement parameter there is a phase like transition between the
classical and quantum domains of the game \cite{jiang,du-1}.\emph{\ }

\subsection{Extension to Three Parameters Set of Strategies}

In the Eisert \textit{et al.} \cite{eisert} scheme there seems no apparent
reason for imposing a restriction on players to apply only to two parameters
set of strategies%
\index{Two parameters set of strategies!Eisert et al.|textit}. Although this
set of strategies is not closed under composition yet it did not prevent
many authors to investigate about the quantum games using this quantization
scheme \cite{du-1,azhar-1,ozdemir,shimamura}.

Its extension to three parameters set of strategies%
\index{Three parameters set of strategies!Eisert et al.} can be accomplished
using the operators of the form

\begin{equation}
\hat{U}(\theta ,\phi ,\psi )=\left[ 
\begin{array}{cc}
e^{i\phi }\cos \frac{\theta }{2} & ie^{i\psi }\sin \frac{\theta }{2} \\ 
ie^{-i\psi }\sin \frac{\theta }{2} & e^{-i\phi }\cos \frac{\theta }{2}%
\end{array}%
\right] ,  \label{three-strategy1}
\end{equation}%
where $0\leq \theta \leq \pi ,-\pi \leq \phi ,\psi \leq \pi .$ In the case
when the players have access to full strategy space as given in (\ref%
{three-strategy1}) then for every strategy of first player Alice the second
player Bob also has a counter strategy as a result there is no pure
strategies Nash equilibrium \cite{benjamin1}. However, there can be mixed
strategies (non-unique) Nash equilibrium \cite{eisert1}.

\subsection{Applications}

An experimental demonstration of Eisert \textit{et al}. \cite{eisert}
quantization scheme%
\index{Quantization scheme!experimental demonstration} for Prisoners' Dilemma%
\index{Prisoners' dilemma!.} game has been achieved on a two qubit nuclear
magnetic resonance (NMR) computer%
\index{Nuclear magnetic resonance computer@NMR} with full range of
entanglement parameter $\gamma $ ranging from $0$ to $%
\frac{\pi }{2}$ \cite{du-2}. It is interesting to note that these results
are in good agreement with theory. Such a type of demonstration has also
been proposed on the optical computer%
\index{Optical computer} \cite{zhou}. Some other interesting issues that
have been analyzed using this quantization scheme are, the proof of quantum
Nash equilibrium theorem%
\index{Nash equilibrium (NE)!quantum theorem} \cite{lee}, evolutionarily
stable strategies (ESS)%
\index{ESS} \cite{azhar}, quantum verses classical player \cite%
{poit,flitney-1,cheon}, the difference between classical and quantum
correlations \cite{ozdemir,shimamura} and the\ model of decoherence in the
quantum games%
\index{Quantum game!model of decoherence} \cite{chen,flitney-2}. In this
model an increase in the amount of decoherence%
\index{Decoherence} degrades the advantage of a quantum player over a
classical player. However this advantage does not entirely disappear until
the decoherence is maximum. Eisert \textit{et al}. scheme can easily be
implemented to all kinds of $2\times 2$ games. A possible classification of $%
2\times 2$ games has also been given by Huertas-Rosero \cite{rosero}.

\subsection{Comments of Enk and Pike}

Enk and Pike \cite{enk} argued that the solutions of Prisoners' Dilemma%
\index{Prisoners' dilemma!.} as found by Eisert et. al \cite{eisert} are
neither quantum mechanical nor they solve classical game. But it can be
generated by extending the classical payoff matrix of the game in such a way
that it includes a pure strategy corresponding to $%
\hat{Q}$. They added that as if the quantum situation pointed out by Eisert 
\textit{et al}. can be found classically then the only defence for quantum
solution is its efficiency and it does not play any role in Prisoners'
Dilemma game%
\index{Prisoners' dilemma!.}. They also gave the suggestion to investigate
the quantum games by exploiting the non-classical correlations in entangled
states.

\section{\label{marinatto-scheme}Marinatto and Weber Quantization Scheme}

Marinatto and Weber \cite{marinatto} gave another interesting scheme for the
quantization of non-zero sum games by taking an example of\ a famous game
known as Battle of Sexes%
\index{Battle of sexes} with the payoff matrix as in (\ref{matrix-BoS}). To
analyze this game in quantum domain Marinatto and Weber \cite{marinatto}
gave Hilbert structure to the strategic space of the game by allowing the
linear combinations of classical strategies. At the beginning of the game
arbiter prepares two qubits quantum state and sends one qubit to each
player. The players apply their tactics i.e. their local operators on the
respective qubits and send them back to arbiter. The players' tactics in
this scheme are combinations of the identity operator $%
\hat{I}$ and the flip operator $\hat{C}$, with classical probabilities $p$
and $\left( 1-p\right) $, respectively for Alice and $q$ and $\left(
1-q\right) $ for Bob. This quantization scheme is depicted in fig. (\ref%
{marinatto-1})

\begin{figure}[th]
\centering
\includegraphics[scale=.9]{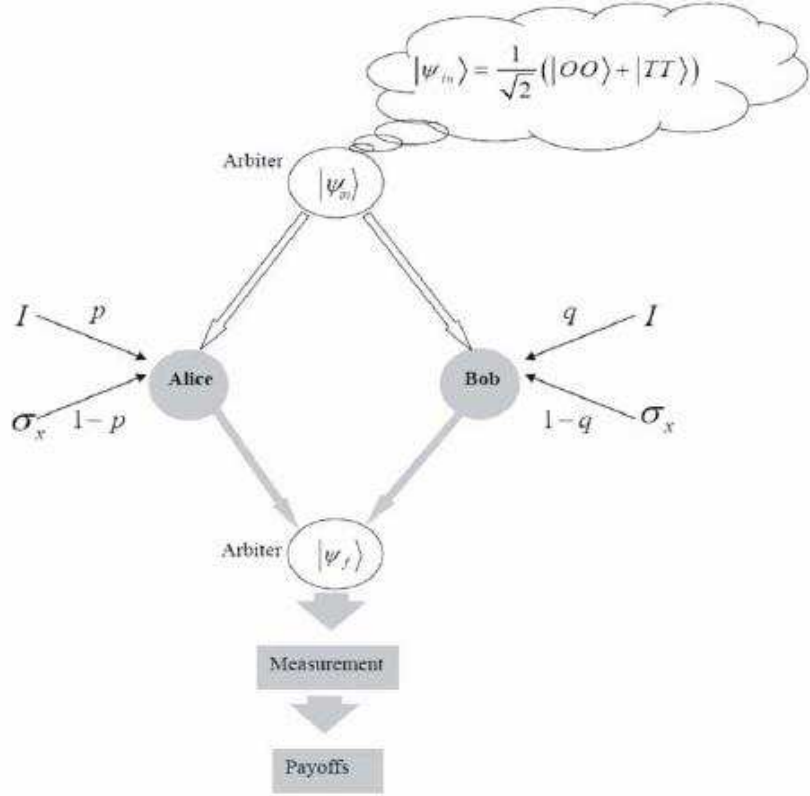}
\caption{Marinatto and Weber quantization scheme. At the beginning of the
game arbiter prepares two qubits entangled state $\left\vert \protect\psi %
\right\rangle _{in}$ and sends one qubit to each player. The players apply
their tactics i.e their local operators on their qubits and send back to
arbiter. The players' tactics in this scheme are combinations of the
identity operator $I$ and the flip operator $C$, with classical
probabilities $p$ and $\left( 1-p\right) $, respectively for Alice and $q$
and $\left( 1-q\right) $ for Bob. }
\label{marinatto-1}
\end{figure}

Marinatto and Weber \cite{marinatto} supposed that the game starts from the
initial state of the form 
\begin{gather}
\left\vert \psi \right\rangle _{in}=a\left\vert OO\right\rangle +b\left\vert
TT\right\rangle ,\text{ \ \ }  \notag \\
\text{\ }\left\vert \text{\ }a\right\vert ^{2}+\text{\ }\left\vert
b\right\vert ^{2}=1.  \label{initial-state}
\end{gather}%
Here the first entry in ket-bra $\left\vert {}\right\rangle $ is for Alice
and the second for Bob's strategy and $O$\ represents opera and $T$\
represents TV (see \ref{bos}). The density matrix for the\ quantum state \ref%
{initial-state} is defined by $\rho _{in}=\left\vert \psi _{in}\right\rangle
\left\langle \psi _{in}\right\vert $ and takes the form 
\begin{equation}
\rho _{in}=\left\vert a\right\vert ^{2}\left\vert OO\right\rangle
\left\langle OO\right\vert +ab^{\ast }\left\vert OO\right\rangle
\left\langle TT\right\vert +a^{\ast }b\left\vert TT\right\rangle
\left\langle OO\right\vert +\left\vert b\right\vert ^{2}\left\vert
TT\right\rangle \left\langle TT\right\vert .  \label{initial-density}
\end{equation}%
The unitary operators $\hat{I}$ and $\hat{C}$ \ transform the strategy
vectors\ $\left\vert O\right\rangle $ and $\left\vert T\right\rangle $\ as
follows 
\begin{equation}
\hat{C}\left\vert O\right\rangle =\left\vert T\right\rangle ,\text{ \ \ }%
\hat{C}\left\vert T\right\rangle =\left\vert O\right\rangle ,\text{ \ \ }%
\hat{C}=\hat{C}^{\dagger }=\hat{C}^{-1}.  \label{operatorss}
\end{equation}%
After the application of the tactics $\hat{I}$ and $\hat{C}$ with
probability $p$ and $1-p$ respectively by Alice and with probabilities $q$
and $1-q$ by Bob respectively, the Eq. (\ref{initial-density}) becomes 
\begin{gather}
\rho _{f}=pq\hat{I}_{A}\otimes \hat{I}_{B}\rho _{\hat{I}n}\hat{I}%
_{A}^{\dagger }\otimes \hat{I}_{B}^{\dagger }+p(1-q)\hat{I}_{A}\otimes \hat{C%
}_{B}\rho _{\hat{I}n}\hat{I}_{A}^{\dagger }\otimes \hat{C}_{B}^{\dagger } 
\notag \\
+q(1-p)\hat{C}_{A}\otimes \hat{I}_{B}\rho _{\hat{I}n}\hat{C}_{A}^{\dagger
}\otimes \hat{I}_{B}^{\dagger }+(1-p)(1-q)\hat{C}_{A}\otimes \hat{C}_{B}\rho
_{\hat{I}n}\hat{C}_{A}^{\dagger }\otimes \hat{C}_{B}^{\dagger }.
\label{final-density}
\end{gather}%
Marinatto and Weber \cite{marinatto} defined the payoff operators for Alice
and Bob as

\begin{eqnarray}
P_{A} &=&\alpha \left| OO\right\rangle \left\langle OO\right| +\beta \left|
TT\right\rangle \left\langle TT\right| +\sigma (\left| OT\right\rangle
\left\langle OT\right| +\left| TO\right\rangle \left\langle TO\right| ), 
\notag \\
P_{B} &=&\beta \left| OO\right\rangle \left\langle OO\right| +\alpha \left|
TT\right\rangle \left\langle TT\right| +\sigma (\left| OT\right\rangle
\left\langle OT\right| +\left| TO\right\rangle \left\langle TO\right| ),
\label{payoff-operator-marinatto}
\end{eqnarray}%
and payoff functions are obtained as the mean values of these operators,
i.e.,

\begin{equation}
\$_{A}(p,q)=\text{Tr}(P_{A}\rho _{f}),\text{ \ \ \ \ and\ \ \ \ \ }%
\$_{B}(p,q)=\text{Tr}(P_{B}\rho _{f}),  \label{payoff-formula}
\end{equation}%
where Tr represents the trace. With the help of Eqs. (\ref{final-density}), (%
\ref{payoff-operator-marinatto}) and (\ref{payoff-formula}) the payoffs
obtained for the players are 
\begin{eqnarray}
\$_{A}(p,q) &=&p\left[ q\left( \alpha +\beta -2\sigma \right) -\alpha \left|
b\right| ^{2}-\beta \left| a\right| ^{2}+\sigma \right] +  \notag \\
&&q\left[ -\alpha \left| b\right| ^{2}-\beta \left| a\right| ^{2}+\sigma %
\right] +\alpha \left| b\right| ^{2}+\beta \left| a\right| ^{2},
\label{marinatto-payoff-a-(BoS)}
\end{eqnarray}%
\begin{eqnarray}
\$_{B}(p,q) &=&q\left[ p\left( \alpha +\beta -2\sigma \right) -\beta \left|
b\right| ^{2}-\alpha \left| a\right| ^{2}+\sigma \right] +  \notag \\
&&p\left[ -\beta \left| b\right| ^{2}-\alpha \left| a\right| ^{2}+\sigma %
\right] +\beta \left| b\right| ^{2}+\alpha \left| a\right| ^{2}.
\label{marinatto-payoff-b-(BoS)}
\end{eqnarray}%
The payoffs of both players also depend on the tactics/ strategy played by
the other player.\ This is the explicit nature of the game. In the next we
explore the Nash equilibria as found by Marinatto and Weber \cite{marinatto}.

Let $\left( p^{\ast },q^{\ast }\right) $ be the Nash equilibrium%
\index{Nash equilibrium (NE)} (NE) of this game, then from the definition of
the Nash equilibrium it is clear that%
\begin{eqnarray}
\$_{A}(p^{\ast },q^{\ast })-\$_{A}(p,q^{\ast }) &=&\left( p^{\ast }-p\right) 
\left[ q^{\ast }\left( \alpha +\beta -2\sigma \right) -\alpha \left|
b\right| ^{2}-\beta \left| a\right| ^{2}+\sigma \right] \geq 0,  \notag \\
\$_{B}(p^{\ast },q^{\ast })-\$_{B}(p^{\ast },q) &=&\left( q^{\ast }-q\right) 
\left[ p^{\ast }\left( \alpha +\beta -2\sigma \right) -\beta \left| b\right|
^{2}-\alpha \left| a\right| ^{2}+\sigma \right] \geq 0.  \notag \\
&&  \label{NE-BoS}
\end{eqnarray}%
For the inequalities (\ref{NE-BoS}) to hold it is necessary for both the
expression in the parenthesis to be of the same sign. This gives rise to the
following cases of interest.

\textbf{Case (1) \ }When\textbf{\ }$p^{\ast }=q^{\ast }=1$ then the
inequalities (\ref{NE-BoS}) hold if%
\begin{eqnarray}
\alpha \left| a\right| ^{2}+\beta \left| b\right| ^{2}-\sigma &>&0,  \notag
\\
\beta \left| a\right| ^{2}+\alpha \left| b\right| ^{2}-\sigma &>&0.
\label{condition-11}
\end{eqnarray}%
The above conditions are satisfied for all values of $\left| a\right| ^{2}$\
and $\left| b\right| ^{2}$\ therefore, from Eqs. (\ref%
{marinatto-payoff-a-(BoS)}) and (\ref{marinatto-payoff-b-(BoS)}) the payoffs
of the players become 
\begin{eqnarray}
\$_{A}(1,1) &=&\alpha \left| a\right| ^{2}+\beta \left| b\right| ^{2}, 
\notag \\
\$_{B}(1,1) &=&\beta \left| a\right| ^{2}+\alpha \left| b\right| ^{2}.
\label{payoff-11}
\end{eqnarray}%
\textbf{Case (2) \ }When\textbf{\ }$p^{\ast }=q^{\ast }=0$ then the
inequalities (\ref{NE-BoS}) hold if

\begin{eqnarray}
\alpha \left| b\right| ^{2}+\beta \left| a\right| ^{2}-\sigma &>&0,  \notag
\\
\beta \left| b\right| ^{2}+\alpha \left| a\right| ^{2}-\sigma &>&0.
\label{condition-00}
\end{eqnarray}%
The above conditions are also satisfied for all values of $\left| a\right|
^{2}$\ and $\left| b\right| ^{2}$\ therefore, from Eqs. (\ref%
{marinatto-payoff-a-(BoS)}) and (\ref{marinatto-payoff-b-(BoS)}) the payoffs
for the players are 
\begin{eqnarray}
\$_{A}(0,0) &=&\alpha \left| b\right| ^{2}+\beta \left| a\right| ^{2}, 
\notag \\
\$_{B}(0,0) &=&\beta \left| b\right| ^{2}+\alpha \left| a\right| ^{2}.
\label{payoff-00}
\end{eqnarray}%
\textbf{Case (3) \ }When\textbf{\ }$%
\acute{p}^{\ast }=\frac{\left( \beta -\sigma \right) \left| b\right|
^{2}+\left( \alpha -\sigma \right) \left| a\right| ^{2}}{\alpha +\beta
-2\sigma },\acute{q}^{\ast }=\frac{\left( \alpha -\sigma \right) \left|
b\right| ^{2}+\left( \beta -\sigma \right) \left| a\right| ^{2}}{\alpha
+\beta -2\sigma }$ then due to the condition $\alpha >\beta >\sigma $ we see
that $0<\acute{p}^{\ast }<1$ and $0<\acute{q}^{\ast }<1$ and hence from Eqs.
(\ref{marinatto-payoff-a-(BoS)}) and (\ref{marinatto-payoff-b-(BoS)}) the
payoffs for players become 
\begin{equation}
\$_{A}\left( \acute{p}^{\ast },\acute{q}^{\ast }\right) =\$_{B}\left( \acute{%
p}^{\ast },\acute{q}^{\ast }\right) =\frac{\alpha \beta +\left( \alpha
-\beta \right) ^{2}\left| a\right| ^{2}\left| b\right| ^{2}-\sigma ^{2}}{%
\alpha +\beta -2\sigma }.  \label{payoff-mix}
\end{equation}%
\ It is clear from Eqs. (\ref{payoff-11}), (\ref{payoff-00}) and (\ref%
{payoff-mix}) that both the players will prefer to play strategies $p^{\ast
}=q^{\ast }=1$ or $p^{\ast }=q^{\ast }=0$ rather than $\left( \acute{p}%
^{\ast },\acute{q}^{\ast }\right) $. But again they are unable to decide
which of the two Nash equilibria they choose to play. It looks as if the
dilemma is still there. However this dilemma can be resolved by comparing
the payoffs of the players at these Nash equilibria. By the use of Eqs. (\ref%
{payoff-11}) and (\ref{payoff-00}) one gets\ 
\begin{eqnarray}
\$_{A}(1,1)-\$_{A}(0,0) &=&\left( \alpha -\beta \right) \left( \left|
a\right| ^{2}-\left| b\right| ^{2}\right) ,  \notag \\
\$_{B}(1,1)-\$_{B}(0,0) &=&\left( \alpha -\beta \right) \left( \left|
b\right| ^{2}-\left| a\right| ^{2}\right) .  \label{comparison}
\end{eqnarray}%
It is evident from Eq. (\ref{comparison}) that for $\left| a\right|
^{2}>\left| b\right| ^{2}$ Alice would prefer the Nash equilibrium $\left( \
p^{\ast }=q^{\ast }=1\right) $\ whereas Bob will prefer $\left( p^{\ast
}=q^{\ast }=0\right) $, but for $\left| a\right| ^{2}<\left| b\right| ^{2}$
the choices of the players are interchanged. This gives a clue for the
resolution of the dilemma. If the initial quantum state parameters are
chosen as$\ \left| a\right| ^{2}=\left| b\right| ^{2}$ $=\frac{1}{2}$ then
Eq. (\ref{initial-state}) gives 
\begin{equation}
\left| \psi \right\rangle _{in}=\frac{\left| OO\right\rangle +\left|
TT\right\rangle }{\sqrt{2}},  \label{entangled}
\end{equation}%
and by the use of Eq. (\ref{comparison}) the payoffs become 
\begin{equation}
\$_{A}=\$_{B}=\frac{\alpha +\beta }{2}.  \label{dilemma-resolved}
\end{equation}%
These payoffs for both the players are same irrespective of the choice of $%
p^{\ast }=q^{\ast }=0$ or $p^{\ast }=q^{\ast }=1.$

On the other hand for mixed strategies ( $\acute{p}^{\ast }=\acute{q}^{\ast
}=\frac{1}{2}$) the payoffs of the players for maximally entangled initial
quantum state (\ref{entangled}) with the help of Eq.\ (\ref{payoff-mix})
come out to be 
\begin{equation}
\$_{A}=\$_{B}=\frac{\alpha +\beta +2\sigma }{4}.  \label{payoff-mixed}
\end{equation}%
Comparing Eqs. (\ref{dilemma-resolved}) and (\ref{payoff-mixed}) it is clear
that initial quantum state given by Eq. (\ref{entangled}) which is maximally 
\textrm{entangled state} satisfies the Nash equilibrium conditions i.e. it
is a best rational choice which is stable against unilateral deviation and
it also gives higher reward then mixed strategy Nash Equilibrium at $\acute{p%
}^{\ast }=\acute{q}^{\ast }=\frac{1}{2}$. Marinatto and Weber \cite%
{marinatto} argued that this proves that maximally entangled strategy Eq. (%
\ref{entangled}) used as initial quantum state resolves the dilemma present
in the classical version of the Battle of Sexes%
\index{Battle of sexes!-}.

\subsection{Applications}

This quantization scheme has widely been used in various context for the
quantization of games. It gave very interesting results while investigating
evolutionarily stable strategies (ESS)%
\index{ESS} \cite{azhar} and in the analysis of repeated games \cite{azhar2}
etc. This quantization scheme has also been cast in a different manner where
the players manipulate their strategies by the application of linear
combination of the operators $%
\hat{I}$ and $\hat{C}$ as 
\begin{equation}
\hat{O}=\sqrt{p}\hat{I}+\sqrt{1-p}\hat{C}.
\end{equation}%
The operator $\hat{O}$ is termed as quantum superposed operator%
\index{Quantum superposed operator} (QSO) \cite{ma}. The explanation for
this approach is based on the argument that each player is given a handle
that can be moved continuously between $0$ and $1$. When the handle is set
to $1$ it performs the $I$ operation, when set to $0$ it performs $\sigma
_{x}$ operation and at position $1-p$ it performs the operation $%
\hat{O}=$ $\sqrt{p}\hat{I}+\sqrt{1-p}\hat{C}$.

\subsection{\label{benjamin comments}Benjamin's Comments}

In an interesting comment Benjamin \cite{benjamin1} pointed out that the
dilemma is still there as the same payoff for the two Nash equilibria make
them equally acceptable to the players and there is no way for the players
to prefer ``$1$''\ over ``$0$''. In the absence of any communication between
them they could end up with a situation $\left( 1,0\right) $ or $\left(
0,1\right) $ which corresponds to the worst payoff%
\index{Worst case payoff} for both players. Benjamin argued that this is
somewhat similar dilemma faced by players in classical version of the game.

\subsection{Marinatto and Weber's Reply}

In their response to Benjamin's comment, Marinatto and Weber \cite%
{marinatto1} insisted that since both the NE $\left( 0,0\right) $ and $%
\left( 1,1\right) $ render the initial quantum state unchanged and
corresponds to equal and maximum payoff for both the players, therefore,
both of them would prefer $\left( 1,1\right) ,$ as by choosing $p$ or $q$
equal to zero there is a danger for both the payers to get in to a situation 
$\left( 1,0\right) $ or $\left( 0,1\right) $ which corresponds to the lowest
payoff.

In the next section we show that the worst case payoff%
\index{Worst case payoff} scenario as pointed out by Benjamin is not due to
the quantization scheme itself but it is due to the restriction imposed on
initial quantum state parameters. If the game is allowed to start from more
general quantum state then the conditions on the initial quantum state
parameters can be set so that the payoffs for mismatched or worst case
situations are different for different players which results into a unique
solution of the game.

\section{\label{dilemma}Resolution of Dilemma in Quantum Battle of Sexes.}

In this section we analyze the game of quantum Battle of Sexes using the
approach developed by Marinatto and Weber \cite{marinatto}. Instead of
restricting to maximally entangled initial quantum state we consider a
general initial quantum state. Exploiting the additional parameters in the
initial state we present a condition for which unique solution of the game
can be obtained. In particular we address the issues pointed out by Benjamin 
\cite{benjamin} in Marinatto and Weber \cite{marinatto} quantum version of
the Battle of sexes game. In our approach, difference in the payoffs for the
two players corresponding to so called worst-case situation%
\index{Worst case payoff} leads to a unique solution of the game. The
results reduce to that of Marinatto and Weber under appropriate conditions.
It is further shown that initial state parameters can be controlled to make
any possible \textrm{pure} strategy pair in the game to be Nash Equilibria
and a unique solution of the game as well. However then it would not be
interesting to draw a comparison with the classical version of the game.

Since for choosing strategy on the basis of Marinatto and Weber's argument
it requires complete information on the initial quantum state and in quantum
games players\ are not supposed to measure the initial quantum state as
initial quantum state is only used to communicate their choice of local
operators to the arbiter \cite{azhar,lee,witte}.\emph{\ }The choice of these
operators depend on the payoff matrix known to them. If, however a general
initial quantum state is considered then a condition on the parameters of
the initial quantum state can be obtained for which classical dilemma can be
resolved and a unique solution of the quantum Battle of Sexes%
\index{Battle of sexes!-} is achieved. In comparison to Marinatto and Weber 
\cite{marinatto} approach a condition can also be imposed for which payoffs
corresponding to ``mismatched or worst case situation%
\index{Worst case payoff}''\ are different for two players which leads to a
unique solution of the game. Since in quantum version of the game both
players, Alice and Bob, apply their respective strategies to the initial
quantum state given to them on the basis of payoff matrix given to them. In
this approach the payoff matrix depends on the initial state and can be
controlled by its parameters. Therefore the choice of general initial
quantum state provides with additional parameters to control in comparison
with Marinatto and Weber's \cite{marinatto}.

Let Alice and Bob have the following initial entangled state at their
disposal

\begin{equation}
\left| \psi _{in}\right\rangle =a\left| OO\right\rangle +b\left|
OT\right\rangle +c\left| TO\right\rangle +d\left| TT\right\rangle ,
\label{fun}
\end{equation}%
where \ $\left| a\right| ^{2}+\left| b\right| ^{2}+\left| c\right|
^{2}+\left| d\right| ^{2}=1.$ Here the first entry in ket $\left|
{}\right\rangle $ is for Alice and the second for Bob's strategy. For $b$
and $c$ equal to zero Eq. (\ref{fun}) reduces to the initial maximally
entangled quantum state used by Marinatto and Weber \cite{marinatto}. The
unitary operators on the disposal of the players are defined as

\begin{equation}
\hat{C}\left| O\right\rangle =\left| T\right\rangle ,\text{ \ \ }\hat{C}%
\left| T\right\rangle =\left| O\right\rangle ,\text{ \ \ }\hat{C}=\hat{C}%
^{\dagger }=\hat{C}^{-1}.  \label{oper-dilemma}
\end{equation}%
Following the Marinatto and Weber, take $p\hat{I}+(1-p)\hat{C}$ and $q\hat{I}%
+(1-q)\hat{C}$ as the strategies for the two players, respectively, with $p$
and $q$ being the classical probabilities for using the identity operator $%
\hat{I}$. The final density matrix takes the form 
\begin{gather}
\rho _{f}=pq\hat{I}_{A}\otimes \hat{I}_{B}\rho _{in}\hat{I}_{A}^{\dagger
}\otimes \hat{I}_{B}^{\dagger }+p(1-q)\hat{I}_{A}\otimes \hat{C}_{B}\rho
_{in}\hat{I}_{A}^{\dagger }\otimes \hat{C}_{B}^{\dagger }  \notag \\
+q(1-p)\hat{C}_{A}\otimes \hat{I}_{B}\rho _{in}\hat{C}_{A}^{\dagger }\otimes 
\hat{I}_{B}^{\dagger }+(1-p)(1-q)\hat{C}_{A}\otimes \hat{C}_{B}\rho _{in}%
\hat{C}_{A}^{\dagger }\otimes \hat{C}_{B}^{\dagger }.  \label{def}
\end{gather}%
Here $\rho _{in}=\left| \psi _{in}\right\rangle \left\langle \psi
_{in}\right| $ which can be achieved from Eq. (\ref{fun}). The corresponding
payoff operators for Alice and Bob are

\begin{eqnarray}
P_{A} &=&\alpha \left| OO\right\rangle \left\langle OO\right| +\beta \left|
TT\right\rangle \left\langle TT\right| +\sigma (\left| OT\right\rangle
\left\langle OT\right| +\left| TO\right\rangle \left\langle TO\right| ),
\label{popa} \\
P_{B} &=&\beta \left| OO\right\rangle \left\langle OO\right| +\alpha \left|
TT\right\rangle \left\langle TT\right| +\sigma (\left| OT\right\rangle
\left\langle OT\right| +\left| TO\right\rangle \left\langle TO\right| ),
\label{popb}
\end{eqnarray}%
and payoff functions i.e. the mean values of these operators are obtained by

\begin{equation}
\$_{A}(p,q)=\text{Tr}(P_{A}\rho _{f}),\text{ \ \ \ \ and\ \ \ \ \ }%
\$_{B}(p,q)=\text{Tr}(P_{B}\rho _{f}),  \label{mean}
\end{equation}%
where Tr represents the trace.\textsl{\ }With the help of Eqs. (\ref%
{oper-dilemma}), (\ref{def}), (\ref{popa}), (\ref{popb}) and (\ref{mean})
the payoff functions for players are 
\begin{gather}
\$_{A}(p,q)=p\left[ q\Omega +\Phi \left( \left| b\right| ^{2}-\left|
d\right| ^{2}\right) +\Lambda \left( \left| c\right| ^{2}-\left| a\right|
^{2}\right) \right]  \notag \\
+q\left[ \Lambda \left( \left| b\right| ^{2}-\left| a\right| ^{2}\right)
+\Phi \left( \left| c\right| ^{2}-\left| d\right| ^{2}\right) \right]
+\Theta ,  \label{payoff1}
\end{gather}%
\begin{gather}
\$_{B}(p,q)=q\left[ p\Omega +\Phi \left( \left| b\right| ^{2}-\left|
a\right| ^{2}\right) +\Lambda \left( \left| c\right| ^{2}-\left| d\right|
^{2}\right) \right]  \notag \\
+p\left[ \Lambda \left( \left| b\right| ^{2}-\left| d\right| ^{2}\right)
+\Phi \left( \left| c\right| ^{2}-\left| a\right| ^{2}\right) \right]
+\Theta .  \label{payoff2}
\end{gather}%
In writing the above equations it is supposed that 
\begin{equation*}
\Omega =(\alpha +\beta -2\sigma )(\left| a\right| ^{2}-\left| b\right|
^{2}-\left| c\right| ^{2}+\left| d\right| ^{2}),
\end{equation*}%
\begin{equation*}
\QTR{sl}{\ }\Phi =(\alpha -\sigma )\QTR{sl}{,\ }\Lambda =(\beta -\sigma ),
\end{equation*}%
\textsl{\ }%
\begin{equation*}
\Theta =\alpha \left| d\right| ^{2}+\sigma \left| c\right| ^{2}+\sigma
\left| b\right| ^{2}+\beta \left| a\right| ^{2}.
\end{equation*}%
The Nash equilibria of the game are found by solving the following two
inequalities: 
\begin{eqnarray*}
\$_{A}(p^{\ast },q^{\ast })-\$_{A}(p,q^{\ast }) &\geq &0, \\
\$_{B}(p^{\ast },q^{\ast })-\$_{B}(p,q^{\ast }) &\geq &0,
\end{eqnarray*}%
that lead to following two conditions, respectively: 
\begin{gather}
(p^{\ast }-p)[q^{\ast }(\alpha +\beta -2\sigma )(\left| a\right| ^{2}-\left|
b\right| ^{2}-\left| c\right| ^{2}+\left| d\right| ^{2})+  \notag \\
(\sigma -\beta )\left| a\right| ^{2}+(\alpha -\sigma )\left| b\right|
^{2}+(\beta -\sigma )\left| c\right| ^{2}+(\sigma -\alpha )\left| d\right|
^{2}]\geq 0,  \label{nash1}
\end{gather}%
and

\begin{gather}
(q^{\ast }-q)[p^{\ast }(\alpha +\beta -2\sigma )(\left| a\right| ^{2}-\left|
b\right| ^{2}-\left| c\right| ^{2}+\left| d\right| ^{2})+  \notag \\
(\sigma -\alpha )\left| a\right| ^{2}+(\alpha -\sigma )\left| b\right|
^{2}+(\beta -\sigma )\left| c\right| ^{2}+(\sigma -\beta )\left| d\right|
^{2}]\geq 0.  \label{nash2}
\end{gather}%
The above two inequalities are satisfied if both the factors have same
signs. Here we are interested in solving the dilemma arising due to pure
strategies i.e. $\left( 1,\emph{\ }1\right) $ and $\left( 0,0\right) $,
therefore, we restrict ourselves to the following possible pure strategy
pairs:

\textbf{Case (a) }When $p^{\ast }=0,q^{\ast }=0$ then from the inequalities (%
\ref{nash1}) and (\ref{nash2}), reduce to

\begin{eqnarray}
(\sigma -\beta )\left| a\right| ^{2}+(\alpha -\sigma )\left| b\right|
^{2}+(\beta -\sigma )\left| c\right| ^{2}+(\sigma -\alpha )\left| d\right|
^{2} &<&0,  \notag \\
(\sigma -\alpha )\left| a\right| ^{2}+(\alpha -\sigma )\left| b\right|
^{2}+(\beta -\sigma )\left| c\right| ^{2}+(\sigma -\beta )\left| d\right|
^{2}] &<&0.  \label{cond1}
\end{eqnarray}%
All those values of the initial quantum state parameters for which the above
inequalities are satisfied, strategy pair $(0,0)$ is a Nash equilibrium.
Here we consider a particular set of values for the initial state parameter
for which unique solution of the game can be found and hence the dilemma
would be resolved, however, this choice is not unique. Let us take

\begin{equation}
\left| a\right| ^{2}=\left| d\right| ^{2}=\left| b\right| ^{2}=\frac{5}{16}%
,\left| c\right| ^{2}=\frac{1}{16}.  \label{para}
\end{equation}%
The corresponding payoffs from Eqs. (\ref{payoff1}) and (\ref{payoff2}) are

\begin{eqnarray}
\$_{A}(0,0) &=&\frac{5\alpha +5\beta +6\sigma }{16},  \notag \\
\$_{B}(0,0) &=&\frac{5\alpha +5\beta +6\sigma }{16}.  \label{zero}
\end{eqnarray}%
Physically it means that for the Nash equilibrium $(0,0)$, the two players
get equal payoff corresponding to the choice of initial state parameters
give by Eq. (\ref{para}).

\textbf{Case (b): }When $p^{\ast }=q^{\ast }=1,$ then the inequalities (\ref%
{nash1}) and (\ref{nash2}) become

\begin{eqnarray}
(\alpha -\sigma )\left| a\right| ^{2}+(\sigma -\beta )\left| b\right|
^{2}+(\sigma -\alpha )\left| c\right| ^{2}+(\beta -\sigma )\left| d\right|
^{2} &\geq &0,  \notag \\
(\beta -\sigma )\left| a\right| ^{2}+(\sigma -\beta )\left| b\right|
^{2}+(\sigma -\alpha )\left| c\right| ^{2}+(\alpha -\sigma )\left| d\right|
^{2} &\geq &0.  \label{cond2}
\end{eqnarray}%
These inequalities are again satisfied for the choice of the parameters
given by equation (\ref{para})\ for the initial quantum state and the
strategy pair $(1,1)$ is also a Nash. The corresponding payoffs for the two
players in this case are 
\begin{eqnarray}
\$_{A}(1,1) &=&\frac{5\alpha +5\beta +6\sigma }{16},  \notag \\
\$_{B}(1,1) &=&\frac{5\alpha +5\beta +6\sigma }{16}.  \label{first}
\end{eqnarray}%
For the mismatched strategies, i.e., $(p^{\ast }=0,q^{\ast }=1)$ and $%
(p^{\ast }=1,q^{\ast }=0)$ inequalities (\ref{nash1}) and (\ref{nash2}) are
not satisfied for the choice of the initial state parameters given by
equation (\ref{para}), hence these strategy pairs are not Nash. However, it
is interesting to note the corresponding payoffs for the two players i.e.

\begin{eqnarray}
\$_{A}(0,1) &=&\frac{\alpha +5\beta +10\sigma }{16},\text{\ \ \ }\$_{B}(0,1)=%
\frac{5\alpha +\beta +10\sigma }{16},  \notag \\
\$_{A}(1,0) &=&\frac{5\alpha +\beta +10\sigma }{16},\text{ \ \ }\$_{B}(1,0)=%
\frac{\alpha +5\beta +10\sigma }{16}.  \label{worst-1}
\end{eqnarray}%
Now keeping in view all the payoffs given by Eqs. (\ref{zero}), (\ref{first}%
) and (\ref{worst-1}), under the choice of Eq. (\ref{para}), the quantum
game can be represented the following payoff matrix:

\begin{equation}
\text{{\large Alice}}%
\begin{array}{c}
p=1 \\ 
p=0%
\end{array}%
\overset{}{\overset{%
\begin{array}{c}
\text{{\large Bob}} \\ 
\begin{array}{cc}
q=1\text{ \ \ \ \ } & q=0%
\end{array}%
\end{array}%
}{\left[ 
\begin{array}{cc}
\left( \acute{\alpha},\acute{\alpha}\right) & \left( \acute{\beta},\acute{%
\sigma}\right) \\ 
\left( \acute{\sigma},\acute{\beta}\right) & \left( \acute{\alpha},\acute{%
\alpha}\right)%
\end{array}%
\right] }},  \label{matrix1}
\end{equation}%
where 
\begin{eqnarray}
\acute{\alpha} &=&\frac{5\alpha +5\beta +6\sigma }{16},  \notag \\
\acute{\beta} &=&\frac{5\alpha +\beta +10\sigma }{16},  \notag \\
\acute{\sigma} &=&\frac{\alpha +5\beta +10\sigma }{16}.  \label{ours}
\end{eqnarray}%
Here $\acute{\alpha}>\acute{\beta}>\acute{\sigma}.$ On the other hand,
quantized version of Marinatto and Weber can be represented by the following
payoff matrix:

\begin{equation}
\text{{\large Alice}}\ \ \ 
\begin{array}{c}
p=1 \\ 
p=0%
\end{array}%
\overset{}{\overset{%
\begin{array}{c}
\text{{\large Bob}} \\ 
\begin{array}{cc}
q=1\text{ \ \ \ \ } & q=0%
\end{array}%
\end{array}%
}{\left[ 
\begin{array}{cc}
\left( \frac{\alpha +\beta }{2},\frac{\alpha +\beta }{2}\right) & \left(
\sigma ,\sigma \right) \\ 
\left( \sigma ,\sigma \right) & (\frac{\alpha +\beta }{2},\frac{\alpha
+\beta }{2})%
\end{array}%
\right] .}}  \label{matrix2}
\end{equation}%
In comparison with the classical version payoff matrix i.e. Eq. (\ref%
{matrix-BoS}), both Marinatto and Weber's payoff matrix (\ref{matrix2}) and
our payoff matrix (\ref{matrix1}) shows a clear advantage over the classical
version as the payoffs for the players are the same for the two pure Nash
equilibria in the quantum version of the game. Hence there is no incentive
for the players to prefer one Nash equilibrium over the other. However, as
pointed out by Benjamin \cite{benjamin}, in Marinatto's quantum version, in
absence of any communication between the players could inadvertently end up
with a mismatched strategies, i.e., $\left( 1,0\right) $ or $\left(
0,1\right) $ which corresponds to minimum possible payoff $\sigma $ for both
the players. It is important to note that in our version of the quantum
Battle of Sexes the payoffs corresponding to worst-case situation are
different for the two players. This particular feature leads to a unique
solution for the game by providing a straightforward reason for rational
players to go for one of the Nash equilibrium, i.e., $\left( 1,1\right) $
for the parameters of initial quantum state given by Eq. (\ref{para}).

It can be seen from the payoff matrix (\ref{matrix1}), that the payoff for
the two players is maximum for the two Nash equilibria, $\left( 0,0\right) $
and $\left( 1,1\right) $, but for Alice rational choice is $p^{\ast }=1$
since her payoff is maximum, i.e., $\alpha ^{\prime },$ when Bob decides to
play $q^{\ast }=1$ and equals to $\beta ^{\prime }$ if Bob decides to play $%
q^{\ast }=0,$ which is higher than the worst possible payoff, i.e., $\sigma
^{\prime }$. In a similar manner for Bob the rational choice is $q^{\ast }=1$
since his payoff is maximum, i.e., $\alpha ^{\prime },$ when Alice also
plays $p^{\prime }=1$ and equals to $\beta ^{\prime }$ when Alice plays $%
p^{\prime }=0$ which better than the worst possible. Thus for the initial
quantum with parameters given by Eq. (\ref{para}), Nash equilibrium $\left(
1,1\right) $ is clearly a preferred strategy for both players giving a
unique solution to the game.

Similarly an initial quantum state, for example, with state parameters $%
\left| a\right| ^{2}=\left| d\right| ^{2}=\left| c\right| ^{2}=\frac{5}{16}%
,\left| b\right| ^{2}=\frac{1}{16}$\ can be found for which $\left(
0,0\right) $ is left as a preferred strategy for both the players giving a
unique solution for the game.

Case(c): When $(p^{\ast }=0,q^{\ast }=1),$\ then Eqs. (\ref{nash1}) and (\ref%
{nash2}) impose following set of conditions for these strategies to qualify
to be a Nash equilibrium: 
\begin{eqnarray}
(\alpha -\sigma )\left| a\right| ^{2}+(\sigma -\beta )\left| b\right|
^{2}+(\sigma -\alpha )\left| c\right| ^{2}+(\beta -\sigma )\left| d\right|
^{2} &<&0,  \notag \\
(\sigma -\alpha )\left| a\right| ^{2}+(\alpha -\sigma )\left| b\right|
^{2}+(\beta -\sigma )\left| c\right| ^{2}+(\sigma -\beta )\left| d\right|
^{2} &>&0.  \label{cond3}
\end{eqnarray}

Case (d): When $(p^{\ast }=1,q^{\ast }=0),$\ then Eqs. (\ref{nash1}) and (%
\ref{nash2}) impose following set of conditions for these strategies to
qualify to be a Nash equilibrium: 
\begin{eqnarray}
(\sigma -\beta )\left| a\right| ^{2}+(\alpha -\sigma )\left| b\right|
^{2}+(\beta -\sigma )\left| c\right| ^{2}+(\sigma -\alpha )\left| d\right|
^{2} &>&0,  \notag \\
(\beta -\sigma )\left| a\right| ^{2}+(\sigma -\beta )\left| b\right|
^{2}+(\sigma -\alpha )\left| c\right| ^{2}+(\alpha -\sigma )\left| d\right|
^{2} &<&0.  \label{cond4}
\end{eqnarray}

It is also possible to find initial quantum states for which above
conditions, i.e., inequalities (\ref{cond3}) and (\ref{cond4}) are satisfied
and either $(p^{\ast }=0,q^{\ast }=1)$\ or $(p^{\ast }=1,q^{\ast }=0)$\
remains a single preferable strategy for both the players.

Recently Cheon and Tsutsui \cite{cheon} introduced a quantization scheme%
\index{Quantization scheme!of Cheon and Tsutsui} and observed that the
dilemma can be resolved even within the full strategic space. They argued
that the Nash equilibria they obtained are truly optimal within the entire
Hilbert space%
\index{Hilbert space}. Further they also observed two types of Nash
equilibria. One which can be simulated classically even for entangled
strategies however the second that they termed as the true quantum
mechanical Nash equilibrium have no classical analogue.

\chapter{\label{QIT}Quantum Information Theory}

Quantum mechanics has witnessed a long period of philosophical debates on
issues like EPR paradox and single quantum interference\ of electrons and
photons. Quantum information theory, on the contrary, provides us with one
of the best examples for its\ real world applications where each and every
paradox of quantum mechanics offers a remarkable practical potential. Here
the discrete characteristics of quantum mechanical systems such as atoms,
electrons or photons can be exploited for encoding classical information.
Left and right circularly polarized photons, for example, can be encoded as
0 and 1 respectively. Where as a transversely polarized photon, which unlike
any classical system is a superposition of right circularly and a left
circularly polarized photon, can be used to encode both 0 and 1 at the same
time. There also exist interesting examples of entangled states where in
some sense one can encode both 00 or 11 at the same time \cite{lyod}. It is
said that quantum information theory completes its classical counterpart in
the same way as the complex numbers extend and complete the real numbers 
\cite{bennett00}. The unit of quantum information is qubit (quantum bit)
which is amount of quantum information that can be registered on a quantum
system \textrm{having two distinguishable quantum states} \cite{schumacher}.
For the transmission of quantum information the data encoded in quantum
state of a particle being emitted from a suitable quantum source is passed
through a quantum channel where it interacts with the environment of the
channel and a decohered signal is received at receiver's end. The receiver
performs measurement on the perturbed quantum states to extract useful
information. For example, individual monochromatic photons being emitted
from a highly attenuated laser can be thought as a quantum source, an
optical fibre as quantum channel and a photocell as a receiver. Similarly a
source can be a set of ions trapped in an ion trap computer prepared in
entangled state by a sequence of laser pulses \cite{cirac}; the channel in
this case is an ion trap in which the ions evolve over time and the receiver
could be a microscope to read out states of the ion by laser induced
florescence.

\begin{figure}[th]
\centering
\includegraphics[scale=.6]{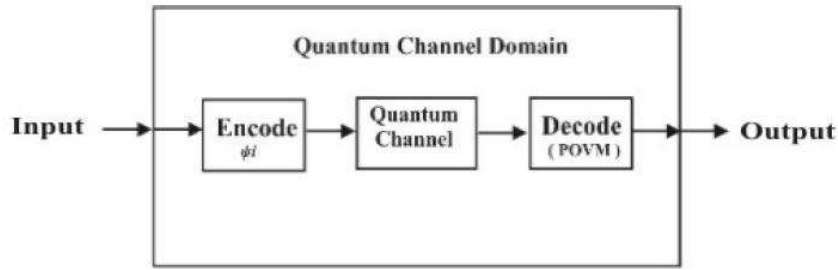}
\caption{Communication through quantum channel}
\label{quantum-channel}
\end{figure}

Figure \ref{quantum-channel} shows a schematic diagram for evolution of
quantum state\ under the action of a quantum channel. \textrm{In the this
figure POVM is a measurement strategy (see subsection \ref{POVM} for detail)}

\section{Quantum Data%
\index{Quantum data}}

Classical data%
\index{Classical data} is a string of classical bits. A classical bit
consists of many quantum systems. It is represented by $0$ and $1$\ known as
Boolean states. A bit encoded in a system can take one of the two possible
distinct values. For example, a bit on a compact disk means whether a laser
beam is reflected or not reflected from its surface; on a credit card it is
stored in the magnetization properties of a series of tiny domains; and for
a computer a bit is the presence or the absence of voltage on tiny wires. A
qubit, on the other hand is a microscopic system such as an atom or nuclear
spin or a polarization of photon. A pair of quantum states that can reliably
be distinguished are used to represent the Boolean states $0$ and $1\ $\cite%
{bennett00}. Spin up and spin down of an electron and horizontal and
vertical polarizations of a photon are among the remarkable examples.
Furthermore a qubit can also exist in superposition states. In two
dimensional Hilbert space spanned by unit vectors $\left| 0\right\rangle $\
and $\left| 1\right\rangle $, a qubit can exist in state $\alpha \left|
0\right\rangle +\beta \left| 1\right\rangle $, $(\left| \alpha \right|
^{2}+\left| \beta \right| ^{2}=1)$. Physically it means that for any
measurement that can discriminate between $\left| 0\right\rangle $ and $%
\left| 1\right\rangle $ the state $\alpha \left| 0\right\rangle +\beta
\left| 1\right\rangle $ gives $\left| 0\right\rangle $ with probability $%
\left| \alpha \right| ^{2}$ and $\left| 1\right\rangle $ with probability $%
\left| \beta \right| ^{2}$. The state of two qubit system is a vector in the
tensor product space spanned by basis $\left| 00\right\rangle $, $\left|
01\right\rangle $,\ $\left| 10\right\rangle $, $\left| 11\right\rangle .$ In
tensor product space there exist entangled states%
\index{Entangled states} which have no classical counterpart (see subsection %
\ref{postulate 4} for detail).

A $n$ bit string of classical data can exist in any $2^{n}$\ states from $%
x=00.......0$ to $11.....1.$ Similarly a string of $n$ qubits can exist in
any state of the form 
\begin{equation}
\left| \psi \right\rangle =\overset{11...1}{\underset{x=00...0}{\sum }}%
c_{x}\left| x\right\rangle ,
\end{equation}%
where $c_{x}$ are the complex numbers such that $\underset{x}{\sum }\left|
c_{x}\right| ^{2}=1.$

\section{von Neumann Entropy}

If a quantum source is emitting quantum states $\left| \psi
_{i}\right\rangle $ with probability $p_{i}$ then the minimum numbers of the
qubits into which the source can be compressed by a quantum encoder such
that it can reliably be decoded is given by von Neumann entropy of the
source. von Neumann entropy is the quantum analogue of Shannon entropy%
\index{von Neumann entropy}%
\index{Shannon entropy} and is mathematically defined as 
\begin{equation}
S\left( \rho \right) =-%
\text{Tr}\left( \rho \log _{2}\rho \right) ,
\end{equation}%
where $\rho =\underset{i}{\sum }p_{i}\left| \psi _{i}\right\rangle
\left\langle \psi _{i}\right| $. If $\lambda _{x}$ are the eigenvalues of $%
\rho $ then von Neumann entropy can be expressed as 
\begin{equation}
S\left( \rho \right) =-\underset{x}{\sum }\lambda _{x}\log _{2}\lambda _{x},
\end{equation}%
where by definition $0\log _{2}0=0.$

\section{\label{holevo-bound}The Holevo Bound}

Holevo bound is the upper bound on the accessible information from a quantum
system \cite{holevo}. Let Alice prepares a quantum system $\rho _{x}$ where $%
x=0,1,....,n$ with probabilities $p_{1,}p_{2},....p_{n}$ and Bob performs
the measurement on the system using POVM elements $\{E_{1},E_{2},...,E_{n}%
\}. $ Then the Holevo bound%
\index{Holevo bound} is given by 
\begin{equation}
H\left( X:Y\right) \leq S\left( \rho \right) -\underset{x}{\sum }%
p_{x}S\left( \rho _{x}\right) ,
\end{equation}%
where $\rho =\underset{x}{\sum }p_{x}\rho _{x},$ \textrm{and }$H\left(
X:Y\right) $\textrm{\ is mutual information of }$X$\textrm{\ and }$Y$\textrm{%
\ which measures how much information }$X$\textrm{\ and }$Y$\textrm{\ have
in common \cite{chuang}.}

\section{Quantum Channels}

A quantum channel is a completely positive trace preserving linear map from
input state density matrices to output state density matrices%
\index{Quantum channel} \cite{kraus,schumacher-1}. A positive map transforms
the matrices with non-negative eigenvalues to the matrices with non-negative
eigenvalues. On the other hand if the system of interest is a part of the
larger system $A$\ and $\varepsilon _{B}$ is a map such that $\varepsilon
_{B}(\rho _{B})\rightarrow \rho _{B}^{\prime }$ then $\varepsilon _{B}$ is
completely positive%
\index{Completely positive map} \textrm{if and only if }$\left( I_{A}\otimes
\varepsilon _{B}\right) \left( \rho _{A}\otimes \rho _{B}\right) $\textrm{\
is also a positive map \cite{kraus}.}

If $\rho $ and $\rho ^{\prime }$are the input and output density matrices,
respectively, then the channel dynamics in operator sum representation,%
\index{Operator sum representation} is described as 
\begin{equation}
\rho ^{\prime }=\varepsilon (\rho )=\underset{k}{\sum }A_{k}^{\dagger }\rho
A_{k},  \label{output}
\end{equation}%
where $\varepsilon $ is completely positive trace preserving linear map and $%
A_{k}$'s are the Kraus operators%
\index{Kraus operators} of a quantum channel. Let Alice wants to send a
message to Bob using a quantum channel. She prepares a input\emph{\ }signal
state $\rho _{k}$ with probability $p_{k}.$ Then the corresponding ensemble
of input states is given as $\rho =\underset{k}{\sum }p_{k}\rho _{k}$. On
receiving the quantum states, Bob performs the measurement by using POVM%
\index{POVM} to determine the state of the signal. According to the Holevo
bound%
\index{Holevo bound} (Sec. \ref{holevo-bound}) the\emph{\ }mutual information%
\index{Mutual information} accessible between Alice and Bob is 
\begin{equation}
I\left( p_{k},\rho _{k}^{%
%TCIMACRO{\U{b4}}%
%BeginExpansion
{\acute{}}%
%EndExpansion
}\right) =S(\underset{k}{\sum }p_{k}\rho _{k}^{%
%TCIMACRO{\U{b4}}%
%BeginExpansion
{\acute{}}%
%EndExpansion
})-\underset{k}{\sum }p_{k}S(\rho _{k}^{%
%TCIMACRO{\U{b4}}%
%BeginExpansion
{\acute{}}%
%EndExpansion
}),  \label{h-bound}
\end{equation}%
where 
\begin{equation}
S(\zeta )=-%
\text{Tr}\left( \zeta \log _{2}\zeta \right) ,
\end{equation}%
is von-Neumann entropy%
\index{von Neumann entropy} for the density matrix $\zeta .$ For the $n$
uses of a memoryless quantum channel with a given input entangled state the
output becomes: 
\begin{equation}
\rho ^{%
%TCIMACRO{\U{b4}}%
%BeginExpansion
{\acute{}}%
%EndExpansion
}=\Phi (\rho )=\underset{k_{1},....k_{n}}{\sum }(A_{k_{n}}\otimes ...\otimes
A_{k_{1}})^{\dagger }\rho _{e}(A_{k_{n}}\otimes ...\otimes A_{k_{1}}),
\label{Kraus-entangled}
\end{equation}%
where $\rho _{e}$ is some entangled state. According to the Eq. (\ref%
{h-bound}) the maximum\emph{\ }amount of reliable information that can be
transmitted along the channel is given as \cite{schumacher,holevo},%
\begin{equation}
C^{\left( n\right) }=%
\frac{1}{n}\sup_{p_{k},\rho _{k}^{\prime 
%TCIMACRO{\U{b4}}%
%BeginExpansion
{\acute{}}%
%EndExpansion
}}I^{\left( n\right) }(p_{k},\rho _{k}^{%
%TCIMACRO{\U{b4}}%
%BeginExpansion
{\acute{}}%
%EndExpansion
}),
\end{equation}%
here $n$ stands for the number of times the channel is used. The use of the
entangled states as an input is interesting since there is a possibility of
superadditivity of channel capacity%
\index{Superadditivity}, i.e., $I_{n+m}>I_{n}+I_{m}$. For the multiple uses
of the channel the classical capacity%
\index{Channel capacity} $C$ of quantum channel is defined as 
\begin{equation}
C=\underset{n\longrightarrow \infty }{\lim }C^{\left( n\right) }.
\label{capacity}
\end{equation}%
\emph{\ }The important examples of the quantum channels are depolarizing
channel, phase damping channel and the amplitude damping channel. Next we
explain these one by one.

\subsection{Depolarizing Channel}

\begin{figure}[th]
\centering
\includegraphics[scale=.8]{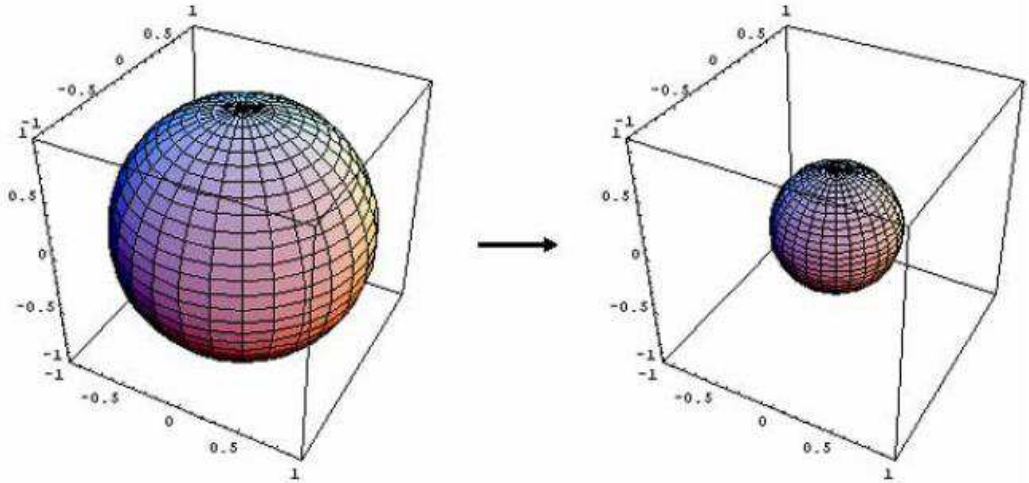}
\caption{Evolution of Bloch sphere after passing through depolarizing
channel with p=.3.}
\label{depolarizing}
\end{figure}

Depolarizing channel%
\index{Quantum channel!depolarizing} models the decohering qubit that
particularly has a nice symmetry. It can cause bit flip, phase flip or both.
Under the action of this channel pure input state $\left\vert \psi
\right\rangle $\ is transformed into $\sigma _{x}\left\vert \psi
\right\rangle $, $\sigma _{y}\left\vert \psi \right\rangle $, $\sigma
_{z}\left\vert \psi \right\rangle $ with equal probability in addition to
retaining its original form \cite{preskill}. Here $\sigma _{x}$, $\sigma
_{y} $, $\sigma _{z}$ are the Pauli matrices. The Kraus operators \cite%
{chuang}%
\index{Kraus operators!for depolarizing channel} for this channel are 
\begin{eqnarray}
A_{0} &=&\left( 
\sqrt{1-p}\right) I,  \notag \\
A_{1} &=&\sqrt{\frac{p}{3}}\sigma _{x},  \notag \\
A_{2} &=&\sqrt{\frac{p}{3}}\sigma _{y},  \notag \\
A_{3} &=&\sqrt{\frac{p}{3}}\sigma _{z}.
\end{eqnarray}%
The state of a quantum system $\rho $ after this noise operation becomes 
\begin{equation}
\acute{\rho}=\underset{k}{\sum }A_{k}\text{ }\rho \text{ }A_{k}^{\dagger
}=\left( 1-p\right) \rho +\frac{p}{3}\left( \sigma _{x}\rho \sigma
_{x}+\sigma _{y}\rho \sigma _{y}+\sigma _{z}\rho \sigma _{z}\right) .
\end{equation}%
Its effect on the Bloch sphere%
\index{Bloch sphere!depolarizing channel} is given as 
\begin{equation}
\left( r_{x},r_{y},r_{z}\right) \rightarrow \left( \left( 1-%
\frac{4}{3}p\right) r_{x},\left( 1-\frac{4}{3}p\right) r_{y},\left( 1-\frac{4%
}{3}p\right) r_{z}\right) .
\end{equation}%
Physically it means that under the action of depolarizing channel the Bloch
sphere shrinks uniformly along the x,y,z by a shrinking factor $1-\frac{4}{3}%
p.$ The effect of depolarizing channel on a Bloch sphere is shown in figure %
\ref{depolarizing} for $p=0.3.$ 

\subsection{Phase Damping Channel}

In phase damping channel%
\index{Quantum channel!phase damping} information is lost without any loss
of energy. This type of noise is unique to quantum mechanics. Kraus operators%
\index{Kraus operators!for phase damping channel} for this channel are 
\begin{eqnarray}
A_{0} &=&%
\sqrt{1-\frac{p}{2}}I,  \notag \\
A_{1} &=&\sqrt{\frac{p}{2}}\sigma _{z}.
\end{eqnarray}%
Under the action of this channel the density matrix $\rho $ transforms as%
\begin{equation}
\acute{\rho}=\left( 1-\frac{p}{2}\right) \rho +\frac{p}{2}\sigma _{z}\rho
\sigma _{z}.
\end{equation}%
Phase damping channel transforms the Bloch sphere as 
\begin{equation}
\left( r_{x},r_{y},r_{z}\right) \rightarrow \left( \left( 1-p\right)
r_{x},\left( 1-p\right) r_{y},r_{z}\right) .
\end{equation}%
The above transformation means that the phase damping channel leaves the
z-axis of the Bloch sphere unchanged whereas x-y plane is uniformly
contracted by a factor of $\left( 1-p\right) .$ The effect of phase damping
channel on Bloch sphere%
\index{Bloch sphere!phase damping channel} is shown in figure \ref%
{Phase-damping} for $p=0.2.$

\begin{figure}[th]
\centering
\includegraphics[scale=.8]{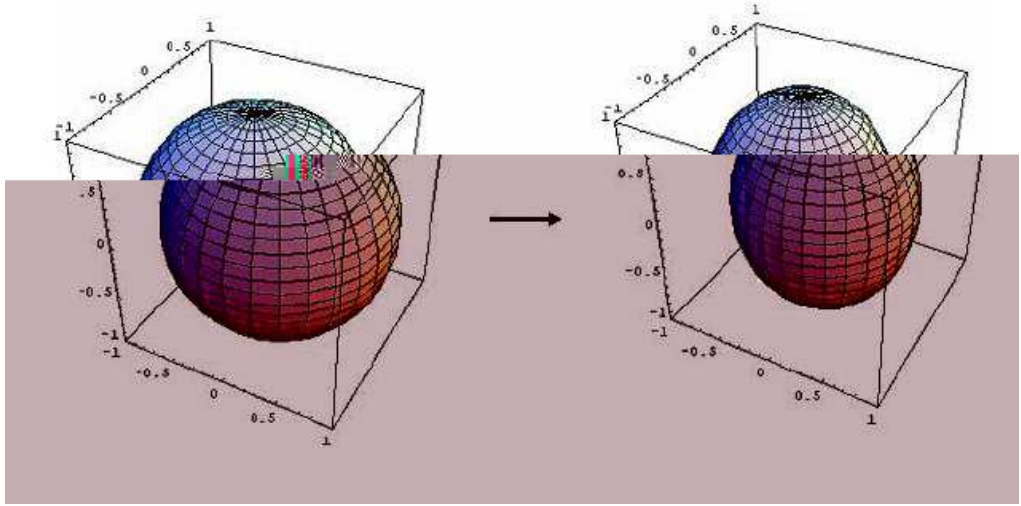}
\caption{Evolution of Bloch sphere after passing through phase daming
channel at p=.2.}
\label{Phase-damping}
\end{figure}

\subsection{Amplitude Damping Channel}

\begin{figure}[th]
\centering
\includegraphics[scale=.8]{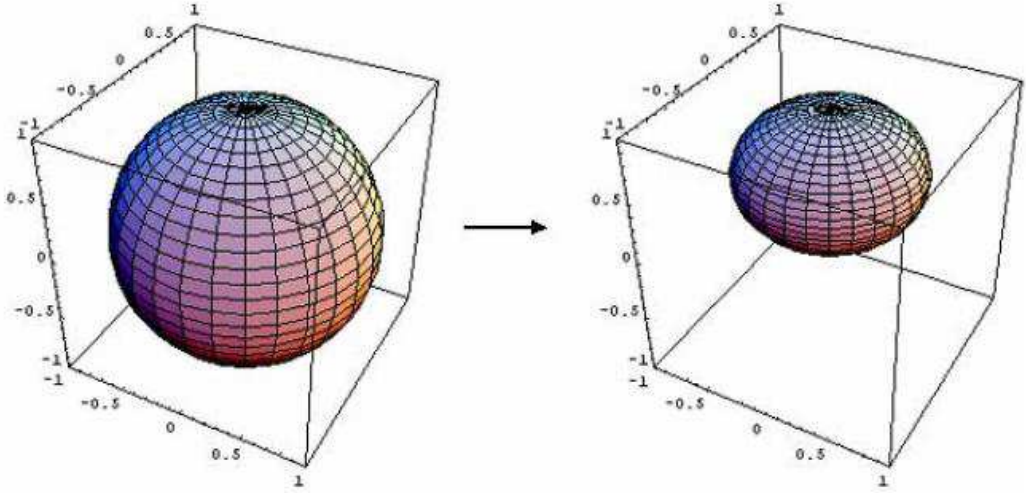}
\caption{The effect of amplitude damping channel on Bloch sphere, for p=.5.}
\label{amplitude-damping-channel}
\end{figure}

This channel%
\index{Quantum channel!amplitude damping} models the loss of energy from the
quantum system. Kraus operators%
\index{Kraus operators!for amplitude damping channel} for this channel are 
\cite{preskill} 
\begin{eqnarray}
A_{0} &=&\left[ 
\begin{array}{cc}
1 & 0 \\ 
0 & 
\sqrt{1-p}%
\end{array}%
\right] ,  \notag \\
A_{1} &=&\left[ 
\begin{array}{cc}
0 & \sqrt{p} \\ 
0 & 0%
\end{array}%
\right] .
\end{eqnarray}%
The operator $A_{1}$ changes the state from $\left\vert 1\right\rangle $ to $%
\left\vert 0\right\rangle $ which is physically a process of losing energy
to environment. The operator $A_{0}$ leaves $\left\vert 0\right\rangle $
unchanged but reduces the amplitude of $\left\vert 1\right\rangle .$
Amplitude damping channel transforms the Bloch sphere%
\index{Bloch sphere!amplitude damping channel} as 
\begin{equation}
\left( r_{x},r_{y},r_{z}\right) \rightarrow \left( \left( 
\sqrt{1-p}\right) r_{x},\left( \sqrt{1-p}\right) r_{y},p+\left( 1-p\right)
r_{z}\right) .
\end{equation}%
The effect of amplitude damping channel on Bloch sphere is shown in figure %
\ref{amplitude-damping-channel} for $p=0.5$. 

\section{Channel Capacity}

Channel capacity%
\index{Channel capacity} is the maximum reliable information that can be
transmitted across the channel. Unlike classical channel, quantum channels
can have various types of capacities. These include \cite{bennett1}

\begin{itemize}
\item Classical capacity%
\index{Quantum channel capacity!classical} C is the maximum asymptotic rate
at which classical bits can be transmitted reliably across the quantum
channel with the help of quantum encoder and decoder.

\item Quantum capacity%
\index{Quantum channel capacity!quantum} Q is the maximum asymptotic rate at
which qubits are transmitted across the channel with quantum encoder and
decoder

\item The classically assisted quantum capacity%
\index{Quantum channel capacity!classically assisted quantum@classically assisted quantum ,$Q_{2\text{ }}$}
$Q_{2\text{ }}$is the maximum asymptotic rate of reliable qubit transmission
with the help of unlimited use of two way classical side channel between
sender and receiver.

\item The entanglement assisted channel capacity%
\index{Quantum channel capacity!entanglement assisted@entanglement assisted, $C_{E}$}
$C_{E}$ is the maximum asymptotic rate of reliable bit transmission with the
help of unlimited prior entanglement between sender and receiver.
\end{itemize}

It is further to be noted that the classical channel capacity of a quantum
channel can further be subdivided into four types \cite{king} depending upon
the input quantum states and the measurement basis. These four possible
capacities are:

\begin{itemize}
\item $C_{PP}$ 
\index{Quantum channel capacity!classical!classical@$C_{PP}$}when the data
is encoded in the form of product states at the sender end and the
measurement at the receiver end is also of the product form.

\item $C_{PE%
\text{ }}$ 
\index{Quantum channel capacity!classical!classical@$C_{PE}$}is the
information capacity when the input data is encoded in the form of product
states and the measurement at the receiver end is of the entangled form.

\item $C_{EP}$ 
\index{Quantum channel capacity!classical!classical@$C_{EP}$}is the
classical channel capacity for the input data is encoded in entangled states
and the measurement at the receiver end is of the product form.

\item $C_{EE}$ 
\index{Quantum channel capacity!classical!classical@$C_{EE}$}is the channel
capacity for the case when both encoding and decoding is of entangled form.
\end{itemize}

\section{No Cloning Theorem}

One of the fundamental differences between classical information and quantum
information is that classical information can perfectly be cloned or copied
where as quantum information cannot be cloned. This is due to the reason
that we cannot measure an unknown quantum state. Therefore if we are given
two non-orthogonal quantum states $\left| \psi \right\rangle ,\left| \phi
\right\rangle $ and asked to distinguish them, there is no measurement which
could distinguish them perfectly and we always have a probability of error.
If it were possible to make many copies of the unknown states then we could
repeat the optimal measurement to make the probability of error arbitrarily
small. The no cloning theorem%
\index{No cloning theorem} \cite{wootters} states that this is not
physically possible. Only the set of mutually orthogonal quantum states can
be copied by a single unitary operator.

Let there be an unknown quantum state of the form%
\begin{equation}
\left| \psi \right\rangle =a\left| 0\right\rangle +b\left| 1\right\rangle ,
\label{unknown}
\end{equation}%
where $\left| a\right| ^{2}+\left| b\right| ^{2}=1$ and a unitary
transformation $U_{cl}$\ capable of cloning unknown states then

\begin{equation}
\left| \psi \right\rangle \left| 0\right\rangle \overset{U_{cl}}{\rightarrow 
}\left| \psi \right\rangle \left| \psi \right\rangle =\left| a\right|
^{2}\left| 00\right\rangle +\left| b\right| ^{2}\left| 11\right\rangle
+ab\left| 01\right\rangle +ab\left| 10\right\rangle .  \label{cloning-1}
\end{equation}%
On the other hand if we clone the expansion of $\left| \psi \right\rangle $
then we get 
\begin{equation}
\left( a\left| 0\right\rangle +b\left| 1\right\rangle \right) \left|
0\right\rangle \overset{U_{cl}}{\rightarrow }a\left| 00\right\rangle
+b\left| 11\right\rangle .  \label{cloning-2}
\end{equation}%
Comparing the expressions (\ref{cloning-1}) and (\ref{cloning-1}) we see a
clear contradiction. Hence quantum state cannot be copied.

\section{\label{cryptography-qi}Quantum Cryptography}

Cryptography provides the techniques of making messages unintelligible to
any undesired party. For this purpose the sender, Alice shares a secret key
with receiver, Bob. In the course of time when Alice wants to send a secret
message to Bob, she encrypts it using secret key. On receiving the encrypted
message, Bob decrypts it with the help of the secret key. Any unauthorized
party, Eve, being unaware of secret key cannot understand the message. There
have been many protocols for classical cryptography from mere transposition
and substitutions to modern\ sophisticated cryptosystems such as one time
pads%
\index{One time pads} and RSA public cryptography%
\index{RSA} \cite{menezes,welsh}. In one time pads, prior to any
communication, the sender and legitimate receiver exchange secret keys
through some physical mean and then store them at a safe and secure
location. However the security of the keys can never be guaranteed, for Eve
can copy the keys while being exchanged or from either party's possession.
In public key cryptosystems, such as, RSA%
\index{RSA}, the receiver generates a pair of keys: a \textit{public key}
and a \textit{private key} \cite{rivest}. The security of the communication
relies on determining the prime factors of a large integer. It is generally
believed that the number of steps a classical computer would need to
factorize an N decimal digit, grows exponentially with N. With recent
advances in quantum computing, it is now possible to factorize very large
numbers much faster \cite{shor1}. As a result the security of RSA will be at
risk. This problem can easily be fixed by quantum cryptography.

Security of a message in quantum cryptography%
\index{Quantum cryptography} relies on the laws of quantum physics instead
of computational complexity. The laws important to mention are

\begin{enumerate}
\item A quantum system cannot be observed without being perturbed.

\item Position and momentum of a particle or the polarization of a photon in
horizontal-vertical basis and diagonal basis cannot be measured
simultaneously.

\item An unknown quantum state cannot be duplicated.
\end{enumerate}

These unique properties of the quantum mechanical systems are used for
protecting the classical information from being tampered in a multiparty
setting where all the parties do not trust each other. The first known
property used for this task was coding secret information on non orthogonal
quantum states. The idea was floated by Stephen Wiesner by introducing the
concept of quantum money \cite{wiesner}. He assumed that let a bank issue
currency such that with each currency note there is a random quantum
sequence of non-orthogonal states. Whenever anybody tries to duplicate the
currency note he will have to perform an impossible task of cloning
non-orthogonal quantum states. Although there is a problem in this scheme as
the quantum states will have decoherence time shorter than the inflationary
half life of most of the currencies, therefore, only the issuing bank can
check the validity of the currency. Thus the counterfeiter can pass a fake
note to layman \cite{bennett}, yet the idea proved very fruitful for quantum
key distribution \ In 1984 Charles Bennett and Gilles Brassard \cite%
{bennett-0} presented a protocol for key distribution that was based on the
Wiesner idea. Since then numerous quantum cryptographic protocols have been
proposed and most of them have been implemented experimentally \cite%
{ekert-1,bennett-01,bennett-6,bennett-7,deutch-1,xyz,mayers-3}.

\section{Quantum Superdense Coding}

Quantum superdense coding, introduced by Bennett and Weisner \cite{bennett-3}%
, provides one of the best examples of the use of quantum mechanics in
information processing tasks. The basic principle of superdense coding 
\index{Quantum superdense coding} is that each member from the set of Bell
states given by Eq.%
\index{Bell states} (\ref{Bell states}) can be transformed to other member
of the set by manipulating only one qubit of a state.

\begin{figure}[th]
\centering
\includegraphics[scale=0.8]{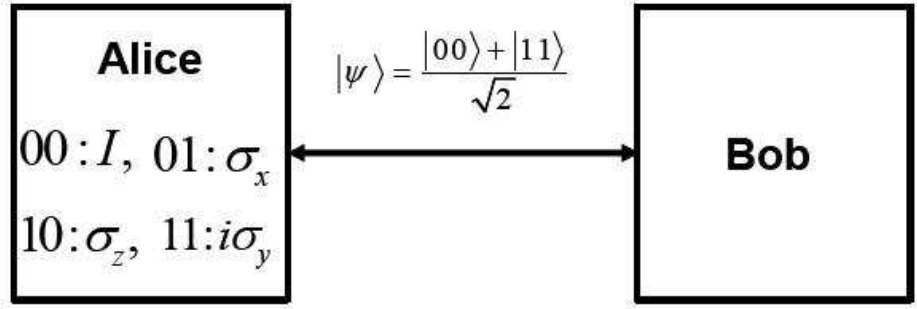}
\caption{Quantum superdense coding}
\label{dense coding}
\end{figure}

Let Alice and Bob share an entangled state of the form 
\begin{equation}
\left\vert \psi ^{+}\right\rangle =%
\frac{1}{2}\left( \left\vert 00\right\rangle +\left\vert 11\right\rangle
\right) .  \label{shared state}
\end{equation}%
Alice can encode message by applying the unitary operators $I,\sigma
_{x},i\sigma _{y}$ or $\sigma _{z}$ on her qubit$.$ For example, if she
wants to send $00$ then\ she applies the identity operator, $I,$ on her
qubit so the original state $\left\vert \psi ^{+}\right\rangle $ is
retained; if she wants to send $01$ then she applies $\sigma _{z}$\ to her
qubit and the shared Bell state transforms to $\left\vert \psi
^{-}\right\rangle ;$ for sending $10$ she applies $\sigma _{x}$ on her qubit
and the shared state becomes $\left\vert \phi ^{+}\right\rangle $ and for $%
11 $ she applies $i\sigma _{y}$ so that the shared state changes to $%
\left\vert \phi ^{-}\right\rangle $, where $\left\vert \psi
^{-}\right\rangle ,\left\vert \phi ^{+}\right\rangle $ and $\left\vert \phi
^{-}\right\rangle $ are the members of Bell states set as defined in Eq. (%
\ref{Bell states}). The resulting four Bell states are orthogonal to each
other and can easily be discriminated. In this way Alice can send two bits
of classical information to Bob while interacting only with single qubit.
Quantum superdense protocol is shown in figure \ref{dense coding}. 

\section{Teleportation}

\begin{figure}[th]
\centering
\includegraphics[scale=.8]{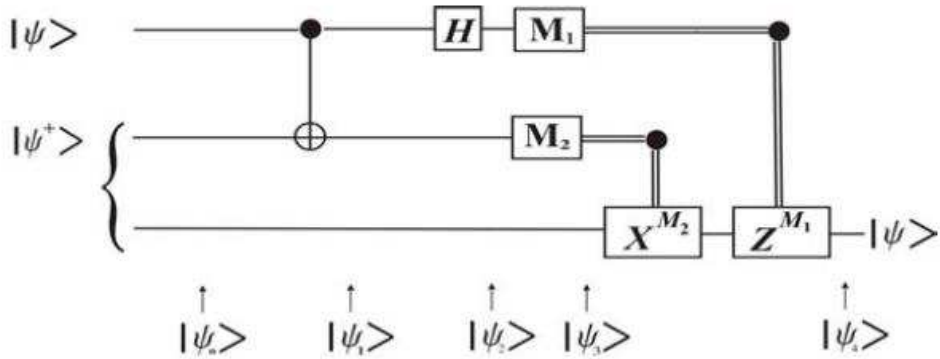}
\caption{Quantum circuit for teleporting a qubit.}
\label{teleportation}
\end{figure}

In quantum teleportation%
\index{Quantum teleportation} a sender, Alice teleports an unknown state $%
\left| \psi \right\rangle =\alpha \left| 0_{A}\right\rangle +\beta \left|
1_{A}\right\rangle $ $(\left| \alpha \right| ^{2}+\left| \beta \right|
^{2}=1)$ to a receiver Bob with whom she shares an EPR pair. Assume Alice
and Bob share an entangled state $\left| \psi ^{+}\right\rangle $. Alice
interacts the qubit to be teleported with half of her EPR pair so the input
state becomes 
\begin{equation}
\left| \psi _{0}\right\rangle =\left| \psi \right\rangle \left| \psi
^{+}\right\rangle =%
\frac{1}{\sqrt{2}}\left[ \alpha \left| 0_{A}\right\rangle \left( \left|
0_{A}0_{B}\right\rangle +\left| 1_{A}1_{B}\right\rangle \right) +\beta
\left| 1_{A}\right\rangle \left( \left| 0_{A}0_{B}\right\rangle +\left|
1_{A}1_{B}\right\rangle \right) \right] ,  \label{teleport}
\end{equation}%
where the subscripts $A$ and $B$ are for Alice and Bob respectively. Then
she sends her qubits through a CNOT gate so that Eq. (\ref{teleport}) becomes

\begin{equation}
\left| \psi _{1}\right\rangle =\frac{1}{2}\left[ \alpha \left|
0_{A}\right\rangle \left( \left| 0_{A}0_{B}\right\rangle +\left|
1_{A}1_{B}\right\rangle \right) +\beta \left| 1_{A}\right\rangle \left(
\left| 1_{A}0_{B}\right\rangle +\left| 0_{A}1_{B}\right\rangle \right) %
\right] ,  \label{teleport-1}
\end{equation}%
and then she sends her first qubit through Hadamard gate%
\index{Hadamard gate} and Eq. (\ref{teleport-1}) transforms to 
\begin{equation}
\left| \psi _{2}\right\rangle =%
\frac{1}{2}\left[ \alpha \left( \left| 0_{A}\right\rangle +\left|
1_{A}\right\rangle \right) \left( \left| 0_{A}0_{B}\right\rangle +\left|
1_{A}1_{B}\right\rangle \right) +\beta \left( \left| 0_{A}\right\rangle
-\left| 1_{A}\right\rangle \right) \left( \left| 1_{A}0_{B}\right\rangle
+\left| 0_{A}1_{B}\right\rangle \right) \right] .  \label{transform}
\end{equation}%
Rearranging Eq. (\ref{transform}) we get

\begin{eqnarray}
\left| \psi _{2}\right\rangle &=&\frac{1}{2}\left[ \left|
0_{A}0_{A}\right\rangle \left( \alpha \left| 0_{B}\right\rangle +\beta
\left| 1_{B}\right\rangle \right) +\left| 0_{A}1_{A}\right\rangle \left(
\left( \alpha \left| 1_{B}\right\rangle +\beta \left| 0_{B}\right\rangle
\right) \right) \right.  \notag \\
&&+\left. \left| 1_{A}0_{A}\right\rangle \left( \alpha \left|
0_{B}\right\rangle -\beta \left| 1_{B}\right\rangle \right) +\left|
1_{A}1_{A}\right\rangle \left( \left( \alpha \left| 1_{B}\right\rangle
-\beta \left| 0_{B}\right\rangle \right) \right) \right] .
\label{teleport-2}
\end{eqnarray}%
Now Alice performs measurement on the qubits in her possession. The
measurement gives her one of the four possible classical bits, $\left|
0_{A}0_{A}\right\rangle ,\left| 0_{A}1_{A}\right\rangle ,\left|
1_{A}0_{A}\right\rangle ,\left| 1_{A}1_{A}\right\rangle .$ If the outcome is 
$\left| 0_{A}0_{A}\right\rangle $ then Bob's state will be $\alpha \left|
0_{B}\right\rangle +\beta \left| 1_{B}\right\rangle $ and for all other
possible results Bob's states are given as follows

\begin{equation*}
\left| 0_{A}1_{A}\right\rangle \rightarrow \alpha \left| 1_{B}\right\rangle
+\beta \left| 0_{B}\right\rangle ,
\end{equation*}

\begin{equation*}
\left| 1_{A}0_{A}\right\rangle \rightarrow \alpha \left| 0_{B}\right\rangle
-\beta \left| 1_{B}\right\rangle ,
\end{equation*}

\begin{equation*}
\left\vert 1_{A}1_{A}\right\rangle \rightarrow \alpha \left\vert
1_{B}\right\rangle -\beta \left\vert 0_{B}\right\rangle .
\end{equation*}%
Alice sends these results to Bob over a classical channel. When Bob comes to
know these results then he fixes up his state to recover the original state $%
\left\vert \psi \right\rangle $ by applying the appropriate quantum gates.
For example, if Bob receives $\left\vert 0_{A}0_{A}\right\rangle $ he needs
to do nothing i.e. he will apply the identity operator $I.$\ If the outcome
is $\left\vert 0_{A}1_{A}\right\rangle $ then he will have to apply $X$ gate
to fix up the state. Similarly if he obtains $\left\vert
1_{A}0_{A}\right\rangle $ and $\left\vert 1_{A}1_{A}\right\rangle $ then he
can fix up the state by applying $Z$ gate and $XZ$\ gate respectively.
Quantum teleportation protocol is shown in the figure (\ref{teleportation}).

\section{Quantum State Discrimination}

The state of a classical system is described by its dynamical variables. For
example the state of a one dimensional point particle can be given by its
momentum $p$ and position $q$. There are no fundamental limitations on
making these values more precise by refining the measurement process. This
is because of the fact that the state variables are also the observables for
a classical system. If the values of these variables along with the
equations governing the dynamics of the system are exactly known then the
future state of the system can be predicted correctly. The state of the
quantum system is described by a normalized vector $\left| \psi
\right\rangle $\ in a complex linear vector space known as Hilbert space.
These state vectors are not observables of quantum mechanics. Therefore,
when one tries to read information stored in these state vectors after a
desired processing of the states, he faces a lot of problem. When the
information is encoded in known orthogonal states then decoding is
relatively simple but if \ the states are non orthogonal and even known,
these cannot be discriminated perfectly. Discriminating among non orthogonal
states is one of the burning issues of quantum information theory. For this
purpose various strategies have been developed. Historically the first
quantum state discrimination strategy, known as quantum hypothesis testing%
\index{Quantum state discrimination!quantum hypothesis testing}, was
introduced by Helstorm \cite{helstorm}. It works on the principle of seeking
the best guess on each trial while minimizing the rate of incorrect guesses.
Hence one can always find an optimal strategy for discriminating between two
non orthogonal quantum states using von Neumann projective measurement \cite%
{helstorm}.\textrm{\ }This strategy has was implemented experimentally by
Barnett and Riis \cite{barnet} using photon polarizing states as
non-orthogonal quantum states. Another interesting strategy for
discriminating among non-orthogonal quantum states, known as unambiguous
state discrimination%
\index{Quantum state discrimination!unambiguous}, was introduced by I. D.
Ivanovic \cite{ivanovic}.\ For the case of two non orthogonal quantum states
Ivanovic studied the following problem. A preparator prepares a collection
of quantum systems in a set of two known non-orthogonal quantum states and
hands them over to an observer one by one for discrimination. He showed that
if the observer is allowed to obtain inconclusive results occasionally then
for the other cases he can perform error free discrimination between the
given non-orthogonal states.\ Since then various strategies have been
developed for optimal state discrimination \cite%
{chelfes-1,hillery,dieks,peres1,hillery1}.

\section{Quantum State Tomography}

All information about a quantum system is encoded in the state of a system
but it is one of the great challenges for experimentalists to measure the
state of the quantum system perfectly \cite{schleich}. This is because of
the fact that the state is not an observable in quantum mechanics \cite%
{peres} and therefore, it is not possible to perform all measurements on the
single state to extract the whole information about the system. It is also
impossible to create a perfect copy of an unknown quantum state \cite%
{wootters}. Therefore, there is no way, even in principle, to infer the
quantum state of a single system without some prior knowledge about it \cite%
{ariano}. However it becomes possible to estimate the unknown quantum state
of a system when many identical copies of the system are available. This
procedure of reconstructing an unknown quantum state through a series of
measurements on a number of identical copies of the system is called quantum
state tomography%
\index{Quantum state tomography}. Each measurement gives a new dimension of
system. To reconstruct the exact state of the system infinite number of
copies are required. This type of procedure was first addressed by Fano \cite%
{fano} and remained mere speculation until original proposal for quantum
tomography and its experimental verification \cite{ariano,vegal,raymer}.
Since than it has been applied \textrm{successfully}\ to the measurement of
photon statistics of a semiconductor laser \cite{munroe}, reconstruction of
density matrix of squeezed vacuum \cite{schiller} and\ probing the entangled
states of light and ions \cite{paris}.

In the following we present a brief introduction to single qubit tomography
following Refs. \cite{chuang,altepeter}.

\subsection{\label{stokes}The Stokes Parameters Representation of Qubit}

Any single qubit density matrix $\rho $ can uniquely be represented with the
help of three parameters $\left\{ S_{1},S_{2},S_{3}\right\} $ and Pauli
matrices $\sigma _{i}^{\prime }s$\ by the expression 
\begin{equation}
\rho =%
\frac{1}{2}\underset{i=0}{\overset{3}{\tsum }}S_{i}\sigma _{i},
\label{stokes representation}
\end{equation}%
where $S_{0}=1$ and the other parameters obey the relation $\underset{i=0}{%
\overset{3}{\tsum }}S_{i}^{2}\leq 1$. The parameters, $S_{i}$\ are called
Stokes parameters%
\index{Stokes parameters} and for a quantum state $\rho $\ these can be
calculated as 
\begin{equation}
S_{i}=%
\text{Tr}\left( \sigma _{i}\rho \right) .
\end{equation}%
Physically these parameters give the outcome of a projective measurements%
\index{Measurement!projective measurement} as

\begin{eqnarray}
S_{0} &=&P_{\left| 0\right\rangle }+P_{\left| 1\right\rangle }  \notag \\
S_{1} &=&P_{%
\frac{1}{\sqrt{2}}\left( \left| 0\right\rangle +\left| 1\right\rangle
\right) }-P_{\frac{1}{\sqrt{2}}\left( \left| 0\right\rangle -\left|
1\right\rangle \right) }  \notag \\
S_{2} &=&P_{\frac{1}{\sqrt{2}}\left( \left| 0\right\rangle +i\left|
1\right\rangle \right) }-P_{\frac{1}{\sqrt{2}}\left( \left| 0\right\rangle
-i\left| 1\right\rangle \right) }  \notag \\
S_{3} &=&P_{\left| 0\right\rangle }-P_{\left| 1\right\rangle }
\end{eqnarray}%
where $P_{\left| i\right\rangle }$ is the probability to measure state $%
\left| i\right\rangle $ given by 
\begin{eqnarray}
P_{\left| i\right\rangle } &=&\left\langle i\right| \rho \left|
i\right\rangle  \notag \\
&=&\text{Tr}\left( \left| i\right\rangle \left\langle i\right| \rho \right) .
\end{eqnarray}%
If we are provided with many copies of a quantum state then with the help of
orthogonal set of matrices $\frac{\sigma _{0}}{\sqrt{2}},\frac{\sigma _{1}}{%
\sqrt{2}},\frac{\sigma _{2}}{\sqrt{2}},\frac{\sigma _{3}}{\sqrt{2}}$ the
density matrix (\ref{stokes representation}) can be written as 
\begin{equation}
\rho =\frac{\text{\textrm{Tr}}(\rho )\sigma _{0}+\text{\textrm{Tr}}(\rho
\sigma _{1})\sigma _{1}+\text{\textrm{Tr}}(\rho \sigma _{2})\sigma _{2}+%
\text{\textrm{Tr}}(\rho \sigma _{3})\sigma _{3}}{2}.  \label{rho}
\end{equation}%
\textrm{where the} expression like \textrm{Tr}$(\rho \sigma _{i})$\
represents the expectation value of the observable. For example to estimate 
\textrm{Tr}$(\rho \sigma _{3})$ we measure $\sigma _{3}$ for $m$ numbers of
time giving the values $z_{1},z_{2,}.....,z_{m}$ all equal to +1 or -1. The
average $\tsum \frac{z_{i}}{m_{i}}$ is an estimate to true value of the
quantity \textrm{Tr}$(\rho \sigma _{3}).$ By central limit theorem this
estimate has standard deviation%
\index{Standard deviation} $%
\frac{\Delta \sigma _{3}}{m}$ where $\Delta \sigma _{3}$\ is the standard
deviation for single measurement of $\sigma _{3}$ that is upper bounded by
1. Therefore, the standard deviation for estimate $\tsum \frac{z_{i}}{m_{i}}$
is at most $\frac{i}{\sqrt{m}}.$The standard deviation for each of the
measurement in Eq. (\ref{rho}) is the same \cite{chuang}. In this way with
the help of Eq. (\ref{rho}) tomography can be performed for an unknown
single qubit state.

\subsection{Single Qubit Tomography}

A single qubit state can very conveniently be represented by a vector in
three dimensional vector space spanned by Pauli matrices. This
representation provides very helpful way for geometrical visualization of
single qubit state, where all the legal states fall within a unit sphere
(Bloch sphere)%
\index{Bloch sphere}. In this representation all the pure states lie on the
surface of the sphere and mixed states fall inside the sphere. The pure
states can be written as 
\begin{equation}
\left| \psi \right\rangle =\cos 
\frac{\theta }{2}\left| 0\right\rangle +e^{i\phi }\sin \frac{\theta }{2}%
\left| 1\right\rangle  \label{state}
\end{equation}%
where\textrm{\ }$\theta $\textrm{\ }and\textrm{\ }$\phi $\textrm{\ }map them
on the surface of the sphere. Any state $\left| \psi \right\rangle $ and its
orthogonal component $\left| \psi ^{\perp }\right\rangle $ fall on two
opposite points on the surface of the sphere such that the line connecting
these points form the axis of the sphere.

\begin{figure}[th]
\centering
\includegraphics[scale=.6]{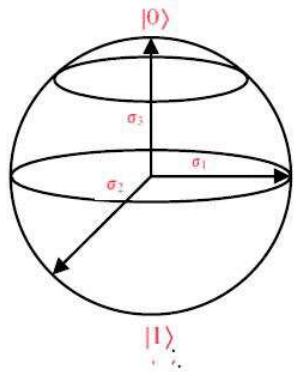}
\caption{The measuremsnt in $\protect\sigma _{3}$ basis confines the unkown
quantum state to a plane $z=\cos \protect\theta .$}
\label{tomography-1}
\end{figure}

For the tomography of an unknown single qubit state three consecutive
measurements are required. Each measurement gives one dimension of the
system until one becomes aware of all dimensions after the complete set of
measurement. For example, a single qubit state $\rho =\left| \psi
\right\rangle \left\langle \psi \right| $ where $\left| \psi \right\rangle $
is defined in Eq. (\ref{state}), can be expressed as 
\begin{equation}
\rho =\frac{1}{2}\left( \sigma _{0}+\sin \theta \cos \phi \text{ }\sigma
_{1}+\sin \theta \sin \phi \text{ }\sigma _{2}+\cos \theta \text{ }\sigma
_{3}\right)  \label{Stokes}
\end{equation}%
Comparing Eqs. (\ref{stokes representation}) and (\ref{Stokes}) the Stokes
parameters for this state become 
\begin{equation}
S_{1}=\sin \theta \cos \phi ,\text{ }S_{2}=\sin \theta \sin \phi ,\text{ }%
S_{3}=\cos \theta .  \label{stokes parameters}
\end{equation}%
For an unknown state of the form Eq. (\ref{Stokes}) when a measurement is
performed in $\sigma _{3}$ basis it confines the state to a plane $z=\cos
\theta $; as shown in Fig. (\ref{tomography-1}).

\begin{figure}[th]
\centering
\includegraphics[scale=.6]{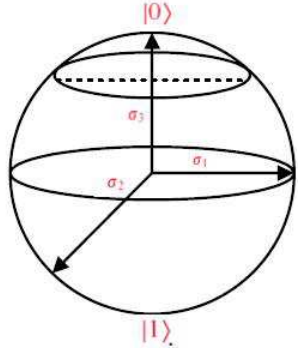}
\caption{The measurement in $\protect\sigma _{2}$ basis confines the state
to $y=\sin \protect\theta \sin \protect\phi $ plane. When this measurement
is combined with first measurement the unknown state reduces to a line
parrallel to x-axis.}
\label{tomography-2}
\end{figure}

Then a measurement in $\sigma _{2}$\ basis is performed that further
confines it to the plane $y=\sin \theta \sin \phi $. The combined effect of
both these measurements restricts the unknown quantum state to a line
parallel to x-axis as shown in Fig. (\ref{tomography-2}). 
At last the measurement in $\sigma _{1}$ basis pinpoints the state as point
lying on this line (resulting from the intersection of $y$ and $z$ planes)
at distance $x=\sin \theta \cos \phi $; as illustrated in Fig. (\ref%
{tomography-3}).

\begin{figure}[th]
\centering
\includegraphics[scale=.6]{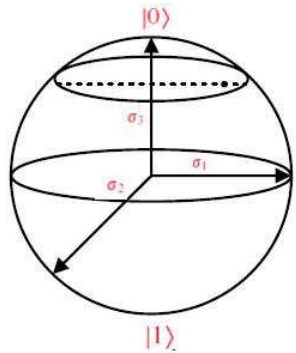}
\caption{The last measurement in $\protect\sigma _{1}$ basis pinpoints the
state as point \ that results from the intersection of three orthogonal
planes.}
\label{tomography-3}
\end{figure}

Since the resultant state is due the intersection of three orthogonal planes
therefore the order these measurements is immaterial in the whole process.
In experimental analysis one usually faces three types of most prominent
errors.

\begin{enumerate}
\item Errors due to measurement basis:- This type of errors appear by an
accidental use of different measurement basis. It can be reduced by
increasing the accuracy of the apparatus.

\item Errors due to counting statistics:- To extract full information about
an unknown quantum state it requires infinite number of measurements. On the
other hand all real life measurements can be performed on limited size of
ensembles which is a source for this type of errors. This can be overcome by
performing measurement on a larger ensemble.

\item Errors from experimental stability:- The drift can occur either in the
state produces or due the efficiency of the detection system that can
constrain the data collection time.
\end{enumerate}

\chapter{\label{general}Generalized Quantization Scheme for Two-Person
Non-zero-Sum Games}

There have been two well known quantization schemes for two person non-zero
sum games. The first was presented by Eisert \textit{et al.} \cite{eisert}
and the second by Marinatto and Weber \cite{marinatto}. The main purpose of
this endeavour was to find a way for resolving dilemmas in games like
Prisoners' Dilemma%
\index{Prisoners' dilemma!.} and Battle of Sexes%
\index{Battle of sexes!-}. These\emph{\ }quantization\emph{\ }schemes gave
very interesting results when applied to different games \cite%
{flitney,azhar,azhar1,azhar2,jiang,rosero}. A detailed description of these
schemes has been provided in section (\ref{eisert-scheme}) and section (\ref%
{marinatto-scheme}) respectively.

In this chapter we introduce the generalized quantization%
\index{Quantization scheme!generalized} scheme for two person non-zero sum
games that gives a relationship between these two apparently different
quantization schemes. The game of Battle of Sexes%
\index{Battle of sexes!-} has been used as an example to introduce this
quantization scheme but this scheme is applicable to other games as well.
Separate set of parameters are identified for which this scheme reduces to
that of Marinatto and Weber and Eisert \textit{et al}. quantization schemes.
Furthermore some other interesting situations are identified which are not
apparent within the existing two quantizations schemes.

\section{Comparison of Quantization Schemes}

A straight forward comparison of Eisert \textit{et al}. \cite{eisert}\ and
Marinatto and Weber \cite{marinatto} quantization schemes%
\index{Quantization scheme!comparison of} can be performed by making use of
entanglement operator $%
\hat{J}$. It was pointed out\ by Eisert \textit{et al}. \cite{eisert}\ that
the initial entangled state for quantum games can be prepared with the
application of an entanglement operator $\hat{J}$ as 
\begin{equation}
\left| \psi _{in}\right\rangle =\hat{J}(\frac{\gamma }{2})\left|
CC\right\rangle ,
\end{equation}%
where $\hat{J}$\ is defined as 
\begin{equation}
\hat{J}(\frac{\gamma }{2})=\exp (-i\frac{\gamma }{2}D\otimes D).
\end{equation}%
The qubits are forwarded one to the each player. The strategic moves of
Alice and Bob are associated with the unitary operators $U_{1}(\theta
_{1},\phi _{1})$ and $U_{2}(\theta _{2},\phi _{2})$, respectively. After the
application of players moves the state of game is 
\begin{equation}
\left| \psi _{f}\right\rangle =\left( U_{1}\otimes U_{2}\right) \hat{J}(%
\frac{\gamma }{2})\left| CC\right\rangle .
\end{equation}%
Then Alice and Bob return back their qubits to arbiter for measurement, the
final state of the game prior to the measurement is 
\begin{equation}
\left| \psi _{f}\right\rangle =\hat{J}^{\dagger }(\frac{\delta }{2})\left(
U_{1}\otimes U_{2}\right) \hat{J}(\frac{\gamma }{2})\left| CC\right\rangle ,
\label{comparison-eisert-marinatto}
\end{equation}%
where 
\begin{equation}
\hat{J}(\frac{\delta }{2})=\exp (-i\frac{\delta }{2}D\otimes D),
\end{equation}%
is the disentanglement operator. Putting$\ \delta =\gamma $ in (\ref%
{comparison-eisert-marinatto})\ the original scheme of Eisert \textit{et al.}
\cite{eisert} is reproduced and letting $\delta =0$ with restriction of $%
\hat{U}_{1}$ and $\hat{U}_{2}$ as a linear combination of identity operator $%
\hat{I},$ and the flip operator $\hat{C}$, the scheme of Marinatto and Weber 
\cite{marinatto} is retrieved.

\section{\label{GQS}Generalized Quantization Scheme}

To introduce the generalized quantization scheme we take Battle of Sexes%
\index{Battle of sexes!generalized quantization scheme!-} as an example
which is an interesting static game of complete information with payoff
matrix (\ref{matrix-BoS}). For the quantization of this game we suppose that
Alice and Bob are given the following initial state 
\begin{equation}
\left| \psi _{in}\right\rangle =\cos 
\frac{\gamma }{2}\left| 00\right\rangle +i\sin \frac{\gamma }{2}\left|
11\right\rangle .  \label{state in}
\end{equation}%
Here $\left| 0\right\rangle $ and $\left| 1\right\rangle $ represent the
vectors in the strategy space corresponding to Opera and TV, respectively%
\emph{\ }with\emph{\ }$\gamma \in \left[ 0,\frac{\pi }{2}\right] $. Here $%
\gamma $ is entanglement of initial quantum state.\emph{\ }The strategy of
each of the\ players is represented by the unitary operator $U_{i}$\ of the
form\emph{\ } 
\begin{equation}
U_{i}=\cos \frac{\theta _{i}}{2}R_{i}+\sin \frac{\theta _{i}}{2}C_{i},\text{
\ \ \ }  \label{combination}
\end{equation}%
where $i=1$\ or $2$\ and $R_{i}$, $C_{i}$\emph{\ }are the unitary operators
defined as%
\begin{align}
R_{i}\left| 0\right\rangle & =e^{i\phi _{i}}\left| 0\right\rangle ,\text{ \
\ }R_{i}\left| 1\right\rangle =e^{-i\phi _{i}}\left| 1\right\rangle ,  \notag
\\
C_{i}\left| 0\right\rangle & =-\left| 1\right\rangle ,\text{ \ \ \ \ \ }%
C_{i}\left| 1\right\rangle =\left| 0\right\rangle .  \label{oper}
\end{align}%
Here we restrict our treatment to two parameter set of strategies for
mathematical simplicity in accordance with Ref. \cite{eisert}.\emph{\ }After
the application of the strategies, the initial state given by Eq. (\ref%
{state in}) transforms into 
\begin{equation}
\left| \psi _{f}\right\rangle =(U_{1}\otimes U_{2})\left| \psi
_{in}\right\rangle .  \label{final}
\end{equation}%
Using Eqs. (\ref{oper}) and (\ref{final}) the above expression becomes 
\begin{align}
\left| \psi _{f}\right\rangle & =\cos \frac{\gamma }{2}[\cos \frac{\theta
_{1}}{2}\cos \frac{\theta _{2}}{2}e^{i(\phi _{1}+\phi _{2})}\left|
00\right\rangle -\cos \frac{\theta _{1}}{2}\sin \frac{\theta _{2}}{2}%
e^{i\phi _{1}}\left| 01\right\rangle  \notag \\
& -\cos \frac{\theta _{2}}{2}\sin \frac{\theta _{1}}{2}e^{i\phi _{2}}\left|
10\right\rangle +\sin \frac{\theta _{1}}{2}\sin \frac{\theta _{2}}{2}\left|
11\right\rangle ]  \notag \\
& +i\sin \frac{\gamma }{2}[\cos \frac{\theta _{1}}{2}\cos \frac{\theta _{2}}{%
2}e^{-i(\phi _{1}+\phi _{2})}\left| 11\right\rangle +\cos \frac{\theta _{1}}{%
2}\sin \frac{\theta _{2}}{2}e^{-i\phi _{1}}\left| 10\right\rangle  \notag \\
& +\cos \frac{\theta _{2}}{2}\sin \frac{\theta _{1}}{2}e^{-i\phi _{2}}\left|
01\right\rangle +\sin \frac{\theta _{1}}{2}\sin \frac{\theta _{2}}{2}\left|
00\right\rangle ].  \label{state fin}
\end{align}%
The payoff operators for Alice and Bob corresponding to payoff matrix (\ref%
{matrix-BoS}) are

\begin{align}
P_{A}& =\alpha P_{00}+\beta P_{11}+\sigma (P_{01}+P_{10}),  \notag \\
P_{B}& =\alpha P_{11}+\beta P_{00}+\sigma (P_{01}+P_{10}),
\label{pay-operator}
\end{align}%
where 
\begin{subequations}
\label{oper d}
\begin{align}
P_{00}& =\left\vert \psi _{00}\right\rangle \left\langle \psi
_{00}\right\vert \text{, \ }\left\vert \psi _{00}\right\rangle =\cos \frac{%
\delta }{2}\left\vert 00\right\rangle +i\sin \frac{\delta }{2}\left\vert
11\right\rangle ,  \label{oper 1} \\
P_{11}& =\left\vert \psi _{11}\right\rangle \left\langle \psi
_{11}\right\vert ,\text{ \ }\left\vert \psi _{11}\right\rangle =\cos \frac{%
\delta }{2}\left\vert 11\right\rangle +i\sin \frac{\delta }{2}\left\vert
00\right\rangle ,  \label{oper 2} \\
P_{10}& =\left\vert \psi _{10}\right\rangle \left\langle \psi
_{10}\right\vert \text{, \ }\left\vert \psi _{10}\right\rangle =\cos \frac{%
\delta }{2}\left\vert 10\right\rangle -i\sin \frac{\delta }{2}\left\vert
01\right\rangle ,  \label{oper 3} \\
P_{01}& =\left\vert \psi _{01}\right\rangle \left\langle \psi
_{01}\right\vert \text{, \ }\left\vert \psi _{01}\right\rangle =\cos \frac{%
\delta }{2}\left\vert 01\right\rangle -i\sin \frac{\delta }{2}\left\vert
10\right\rangle ,  \label{oper 4}
\end{align}%
and\emph{\ }$\delta \in \left[ 0,\frac{\pi }{2}\right] $ refers to the
entanglement of the measurement basis. Above payoff operators reduce to that
of Eisert's scheme for $\delta $ equal to $\gamma ,$ which represents the
entanglement of the initial state. For $\delta =0$ above operators transform
into that of Marinatto and Weber's scheme. In generalized quantization
scheme payoffs for the players are calculated as 
\end{subequations}
\begin{eqnarray}
\$_{A}(\theta _{1},\phi _{1},\theta _{2},\phi _{2}) &=&\text{Tr}(P_{A}\rho
_{f})\text{,}  \notag \\
\$_{B}(\theta _{1},\phi _{1},\theta _{2},\phi _{2}) &=&\text{Tr}(P_{B}\rho
_{f}),  \label{payoff-generalized}
\end{eqnarray}%
where $\rho _{f}=\left\vert \psi _{f}\right\rangle \left\langle \psi
_{f}\right\vert $ is the density matrix for the quantum state given by Eq. (%
\ref{state fin}) and Tr represents the trace of a\emph{\ }matrix. Using Eqs.
(\ref{state fin}), (\ref{pay-operator}) and (\ref{payoff-generalized}) the
payoffs for players are obtained as 
\begin{subequations}
\label{oper a}
\begin{align}
\$_{A}(\theta _{1},\phi _{1},\theta _{2},\phi _{2})& =\cos ^{2}\frac{\theta
_{1}}{2}\cos ^{2}\frac{\theta _{2}}{2}\left[ \eta \sin ^{2}\frac{\gamma }{2}%
+\xi \cos ^{2}\frac{\gamma }{2}+\chi \cos 2(\phi _{1}+\phi _{2})\sin \gamma
\right.  \notag \\
& \left. -\sigma \right] +\sin ^{2}\frac{\theta _{1}}{2}\sin ^{2}\frac{%
\theta _{2}}{2}(\eta \cos ^{2}\frac{\gamma }{2}+\xi \sin ^{2}\frac{\gamma }{2%
}-\chi \sin \gamma -\sigma )  \notag \\
& +\frac{(\alpha +\beta -2\sigma )\sin \gamma -2\chi }{4}\sin \theta
_{1}\sin \theta _{2}\sin (\phi _{1}+\phi _{2})+\sigma ,  \notag \\
&  \label{GPA} \\
\$_{B}(\theta _{1},\phi _{1},\theta _{2},\phi _{2})& =\cos ^{2}\frac{\theta
_{1}}{2}\cos ^{2}\frac{\theta _{2}}{2}\left[ \xi \sin ^{2}\frac{\gamma }{2}%
+\eta \cos ^{2}\frac{\gamma }{2}-\chi \cos 2(\phi _{1}+\phi _{2})\sin \gamma
\right.  \notag \\
& \left. -\sigma \right] +\sin ^{2}\frac{\theta _{1}}{2}\sin ^{2}\frac{%
\theta _{2}}{2}(\xi \cos ^{2}\frac{\gamma }{2}+\eta \sin ^{2}\frac{\gamma }{2%
}+\chi \sin \gamma -\sigma )+  \notag \\
& \frac{(\alpha +\beta -2\sigma )\sin \gamma +2\chi }{4}\sin \theta _{1}\sin
\theta _{2}\sin (\phi _{1}+\phi _{2})+\sigma ,  \notag \\
&  \label{GPB}
\end{align}%
where 
\end{subequations}
\begin{equation}
\xi =\alpha \cos ^{2}\frac{\delta }{2}+\beta \sin ^{2}\frac{\delta }{2},
\end{equation}%
\begin{equation}
\eta =\alpha \sin ^{2}\frac{\delta }{2}+\beta \cos ^{2}\frac{\delta }{2},
\end{equation}%
\begin{equation}
\chi =\frac{(\alpha -\beta )}{2}\sin \delta .
\end{equation}%
Classical results can easily be found from Eqs. (\ref{GPA}) and (\ref{GPB})
by simply unentangling, the initial quantum state of the game i.e. letting $%
\gamma =0$\emph{.} Furthermore all the results found by Marinatto and Weber 
\cite{marinatto} and Eisert \textit{et al}. \cite{eisert} are also embedded
in these payoffs.

\subsection{Reduction to Marinatto and Weber Quantization Scheme}

The generalized quantization scheme reduces to Marinatto and Weber \cite%
{marinatto} quantization scheme for $\delta =0.$\ In this situation we have
following two cases of interest.

\textbf{Case(a): }When $\delta =0$ and $\phi _{1}=0$, $\phi _{2}=0.$ then
the payoffs for the players from Eqs. (\ref{GPA}) and (\ref{GPB}) reduce to 
\begin{subequations}
\label{marinto}
\begin{align}
\$_{A}(\theta _{1},\phi _{1,}\theta _{2},\phi _{2})& =\cos ^{2}\frac{\theta
_{1}}{2}[\cos ^{2}\frac{\theta _{2}}{2}(\alpha +\beta -2\sigma )-\alpha \sin
^{2}\frac{\gamma }{2}-\beta \cos ^{2}\frac{\gamma }{2}+\sigma ]  \notag \\
& +\cos ^{2}\frac{\theta _{2}}{2}(-\alpha \sin ^{2}\frac{\gamma }{2}-\beta
\cos ^{2}\frac{\gamma }{2}+\sigma )+\alpha \sin ^{2}\frac{\gamma }{2}+\beta
\cos ^{2}\frac{\gamma }{2},  \label{marintoa} \\
\$_{B}(\theta _{1},\phi _{1},\theta _{2},\phi _{2})& =\cos ^{2}\frac{\theta
_{2}}{2}[\cos ^{2}\frac{\theta _{1}}{2}(\alpha +\beta -2\sigma )-\beta \sin
^{2}\frac{\gamma }{2}-\alpha \cos ^{2}\frac{\gamma }{2}+\sigma ]  \notag \\
& +\cos ^{2}\frac{\theta _{1}}{2}(-\beta \sin ^{2}\frac{\gamma }{2}-\alpha
\cos ^{2}\frac{\gamma }{2}+\sigma )+\beta \sin ^{2}\frac{\gamma }{2}+\alpha
\cos ^{2}\frac{\gamma }{2}.  \label{marinatob}
\end{align}%
These payoffs are the same as found by Marinatto and Weber \cite{marinatto}
where the players apply the identity operators $I_{1}$ and $I_{2}$ with
probabilities $\cos ^{2}\frac{\theta _{1}}{2}$ and $\cos ^{2}\frac{\theta
_{2}}{2}$ respectively on the given initial quantum state of the form of Eq.
(\ref{state in}).

\textbf{Case(b): }When $\delta =0$ and $\phi _{1}+\phi _{2}=\frac{\pi }{2}$
then Eqs. (\ref{GPA}) and (\ref{GPB}) reduce to

\end{subequations}
\begin{subequations}
\label{marin+chinese b}
\begin{align}
\$_{A}(\theta _{1},\phi _{1},\theta _{2},\phi _{2})& =\cos ^{2}\frac{\theta
_{1}}{2}\left[ \cos ^{2}\frac{\theta _{2}}{2}(\alpha +\beta -2\sigma
)-\alpha \sin ^{2}\frac{\gamma }{2}-\beta \cos ^{2}\frac{\gamma }{2}+\sigma %
\right]  \notag \\
& +\cos ^{2}\frac{\theta _{2}}{2}\left( -\alpha \sin ^{2}\frac{\gamma }{2}%
-\beta \cos ^{2}\frac{\gamma }{2}+\sigma \right) +\alpha \sin ^{2}\frac{%
\gamma }{2}+\beta \cos ^{2}\frac{\gamma }{2}  \notag \\
& +\frac{\left( \alpha +\beta -2\sigma \right) }{4}\sin \gamma \sin \theta
_{1}\sin \theta _{2},  \label{marin+chinese1a} \\
\$_{B}(\theta _{1},\phi _{1},\theta _{2},\phi _{2})& =\cos ^{2}\frac{\theta
_{2}}{2}\left[ \cos ^{2}\frac{\theta _{1}}{2}(\alpha +\beta -2\sigma )-\beta
\sin ^{2}\frac{\gamma }{2}-\alpha \cos ^{2}\frac{\gamma }{2}+\sigma \right] 
\notag \\
& +\cos ^{2}\frac{\theta _{1}}{2}\left( -\beta \sin ^{2}\frac{\gamma }{2}%
-\alpha \cos ^{2}\frac{\gamma }{2}+\sigma \right) +\beta \sin ^{2}\frac{%
\gamma }{2}+\alpha \cos ^{2}\frac{\gamma }{2}  \notag \\
& +\frac{\left( \alpha +\beta -2\sigma \right) }{4}\sin \gamma \sin \theta
_{1}\sin \theta _{2}.  \label{marin+chinese1b}
\end{align}%
In the context of Marinatto and Weber scheme \cite{marinatto,ma} the above
payoffs for the two players correspond to a situation when the strategies of
the players are linear combination of operators $I$ and flip operator $\hat{C%
}$ of the form $\hat{O}_{i}=\sqrt{p_{i}}\hat{I}+\sqrt{1-p_{i}}\hat{C}$ with
probabilities $p_{i}=\cos ^{2}\frac{\theta _{i}}{2}$, $i=1$ or $2$ and
initial entangled state of the form given by Eq. (\ref{state in}).

\subsection{Reduction to Eisert Quantization Scheme}

The results of Eisert \textit{et al. }\cite{eisert} quantization scheme can
be retrieved restricting $\delta =\gamma $ in the generalized quantization
scheme. Here again we have two cases of interest.

\textbf{Case (a)} When $\delta =\gamma $ and$\ \phi _{1}\neq 0$, $\phi
_{2}\neq 0$ then payoffs given by the Eqs. (\ref{GPA}) and (\ref{GPB}) very
interestingly change to the payoffs as if the game has been quantized using
Eisert \textit{et al}. \cite{eisert} scheme for the initial quantum state of
the form (\ref{state in}). In this situation the payoffs for both the
players are 
\end{subequations}
\begin{subequations}
\label{marin+chinese1}
\begin{align}
\$_{A}(\theta _{1},\phi _{1},\theta _{2},\phi _{2})& =\cos ^{2}\frac{\theta
_{1}}{2}\cos ^{2}\frac{\theta _{2}}{2}\left[ \eta _{1}\sin ^{2}\frac{\gamma 
}{2}+\xi _{1}\cos ^{2}\frac{\gamma }{2}+\chi _{1}\cos 2(\phi _{1}+\phi
_{2})\right.  \notag \\
& \left. -\sigma \right] +\sin ^{2}\frac{\theta _{1}}{2}\sin ^{2}\frac{%
\theta _{2}}{2}\left( \eta _{1}\cos ^{2}\frac{\gamma }{2}+\xi _{1}\sin ^{2}%
\frac{\gamma }{2}-\chi _{1}-\sigma \right)  \notag \\
& +\frac{\left( \beta -\sigma \right) }{2}\sin \gamma \sin \theta _{1}\sin
\theta _{2}\sin \left( \phi _{1}+\phi _{2}\right) +\sigma ,
\label{payoff-general1} \\
\$_{B}(\theta _{1},\phi _{1},\theta _{2},\phi _{2})& =\cos ^{2}\frac{\theta
_{1}}{2}\cos ^{2}\frac{\theta _{2}}{2}\left[ \xi _{1}\sin ^{2}\frac{\gamma }{%
2}+\eta _{1}\cos ^{2}\frac{\gamma }{2}-\chi _{1}\cos 2\left( \phi _{1}+\phi
_{2}\right) \right.  \notag \\
& \left. -\sigma \right] +\sin ^{2}\frac{\theta _{1}}{2}\sin ^{2}\frac{%
\theta _{2}}{2}\left( \xi _{1}\cos ^{2}\frac{\gamma }{2}+\eta _{1}\sin ^{2}%
\frac{\gamma }{2}+\chi _{1}-\sigma \right)  \notag \\
& +\frac{\left( \alpha -\sigma \right) }{2}\sin \gamma \sin \theta _{1}\sin
\theta _{2}\sin \left( \phi _{1}+\phi _{2}\right) +\sigma ,
\label{payoff-general2}
\end{align}%
where 
\end{subequations}
\begin{equation*}
\xi _{1}=\alpha \cos ^{2}\frac{\gamma }{2}+\beta \sin ^{2}\frac{\gamma }{2},
\end{equation*}%
\begin{equation*}
\eta _{1}=\alpha \sin ^{2}\frac{\gamma }{2}+\beta \cos ^{2}\frac{\gamma }{2},
\end{equation*}%
\begin{equation*}
\chi _{1}=\frac{(\alpha -\beta )}{2}\sin ^{2}\gamma .
\end{equation*}%
To draw a better comparison we take $\delta =\gamma =\frac{\pi }{2}$ then
the payoffs given by Eqs. (\ref{marin+chinese1}) reduce to 
\begin{subequations}
\label{payoffs}
\begin{align}
\$_{A}(\theta _{1},\phi _{1},\theta _{2},\phi _{2})& =\left( \alpha -\sigma
\right) \cos ^{2}\frac{\theta _{1}}{2}\cos ^{2}\frac{\theta _{2}}{2}\sin
^{2}\left( \phi _{1}+\phi _{2}\right)  \notag \\
& +\left( \beta -\sigma \right) \left[ \cos \frac{\theta _{1}}{2}\cos \frac{%
\theta _{2}}{2}\sin (\phi _{1}+\phi _{2})+\sin \frac{\theta _{1}}{2}\sin 
\frac{\theta _{2}}{2}\right] ^{2}+\sigma ,  \label{payoff-A1} \\
\$_{B}(\theta _{1},\phi _{1},\theta _{2},\phi _{2})& =\left( \alpha -\sigma
\right) \left[ \cos \frac{\theta _{1}}{2}\cos \frac{\theta _{2}}{2}\sin
(\phi _{1}+\phi _{2})+\sin \frac{\theta _{1}}{2}\sin \frac{\theta _{2}}{2}%
\right] ^{2}  \notag \\
& +\left( \beta -\sigma \right) \cos ^{2}\frac{\theta _{1}}{2}\cos ^{2}\frac{%
\theta _{2}}{2}\sin ^{2}\left( \phi _{1}+\phi _{2}\right) +\sigma .
\label{payoff-B1}
\end{align}%
The payoffs given in the Eqs. (\ref{payoffs})\ have already been found by Du 
\textit{et al}. \cite{du-3} through Eisert \textit{et al}. scheme \cite%
{eisert}.

\textbf{Case (b)} When $\delta =\gamma $ and$\ \phi _{1}=\phi _{2}=0$, then
as shown by Eisert \textit{et al}. \cite{eisert,eisert1}, one gets classical
payoffs with mixed strategies. For a better comparison putting $\gamma
=\delta =\frac{\pi }{2}$ and $\phi _{1}=\phi _{2}=0$ in the Eqs. (\ref%
{payoff-general1}) and (\ref{payoff-general2}) the same situation occurs and
the payoffs reduce to 
\end{subequations}
\begin{subequations}
\label{one-parameter b}
\begin{eqnarray}
\$_{A}(\theta _{1},\phi _{1},\theta _{2},\phi _{2}) &=&\alpha \cos ^{2}\frac{%
\theta _{1}}{2}\cos ^{2}\frac{\theta _{2}}{2}+\beta \sin ^{2}\frac{\theta
_{1}}{2}\sin ^{2}\frac{\theta _{2}}{2}  \notag \\
&&+\sigma (\cos ^{2}\frac{\theta _{1}}{2}\sin ^{2}\frac{\theta _{2}}{2}+\sin
^{2}\frac{\theta _{1}}{2}\cos ^{2}\frac{\theta _{2}}{2}),
\label{one-parametera} \\
\$_{B}(\theta _{1},\phi _{1},\theta _{2},\phi _{2}) &=&\beta \cos ^{2}\frac{%
\theta _{1}}{2}\cos ^{2}\frac{\theta _{2}}{2}+\alpha \sin ^{2}\frac{\theta
_{1}}{2}\sin ^{2}\frac{\theta _{2}}{2}  \notag \\
&&+\sigma (\cos ^{2}\frac{\theta _{1}}{2}\sin ^{2}\frac{\theta _{2}}{2}+\sin
^{2}\frac{\theta _{1}}{2}\cos ^{2}\frac{\theta _{2}}{2}).
\label{one-parameterb}
\end{eqnarray}%
In this case the game behaves just like classical game where the players are
playing mixed strategies with probabilities $\cos ^{2}\frac{\theta _{1}}{2}$
and $\cos ^{2}\frac{\theta _{2}}{2}$ respectively.

\subsection{New Explorations}

These are the situations which do not arise in the original versions of
Eisert \textit{et al}. scheme \cite{eisert} and Marinatto and Weber scheme 
\cite{marinatto}.

\textbf{Case (a}): When $\delta \neq \gamma $\ and $\phi _{1}=0$, $\phi
_{2}=0$\ the payoffs given by the Eqs. (\ref{GPA}) and (\ref{GPB}) reduce to 
\end{subequations}
\begin{subequations}
\label{one-parameter d}
\begin{align}
\$_{A}(\theta _{1},\phi _{1,}\theta _{2},\phi _{2}& =\cos ^{2}\frac{\theta
_{1}}{2}\left[ \cos ^{2}\frac{\theta _{2}}{2}\left( \alpha +\beta -2\sigma
\right) -\alpha \sin ^{2}\frac{\left( \gamma -\delta \right) }{2}\right. 
\notag \\
& \left. -\beta \cos ^{2}\frac{(\gamma -\delta )}{2}+\sigma \right] +\cos
^{2}\frac{\theta _{2}}{2}\left[ -\alpha \sin ^{2}\frac{\left( \gamma -\delta
\right) }{2}\right.  \notag \\
& \left. -\beta \cos ^{2}\frac{\left( \gamma -\delta \right) }{2}+\sigma %
\right] +\alpha \sin ^{2}\frac{\left( \gamma -\delta \right) }{2}+\beta \cos
^{2}\frac{\left( \gamma -\delta \right) }{2},  \label{delta0} \\
\$_{B}\left( \theta _{1},\phi _{1,}\theta _{2},\phi _{2}\right) & =\cos ^{2}%
\frac{\theta _{2}}{2}\left[ \cos ^{2}\frac{\theta _{1}}{2}\left( \alpha
+\beta -2\sigma \right) -\beta \sin ^{2}\frac{\left( \gamma -\delta \right) 
}{2}\right.  \notag \\
& \left. -\alpha \cos ^{2}\frac{\left( \gamma -\delta \right) }{2}+\sigma %
\right] +\cos ^{2}\frac{\theta _{1}}{2}\left[ -\beta \sin ^{2}\frac{\left(
\gamma -\delta \right) }{2}\right.  \notag \\
& \left. -\alpha \cos ^{2}\frac{\left( \gamma -\delta \right) }{2}+\sigma %
\right] +\beta \sin ^{2}\frac{\left( \gamma -\delta \right) }{2}+\alpha \cos
^{2}\frac{\left( \gamma -\delta \right) }{2}.  \label{delta}
\end{align}

These payoffs are equivalent to Marinatto and Weber \cite{marinatto} when $%
\gamma $ is replaced with $\gamma -\delta .$

\textbf{Case (b): }When $\delta \neq 0$ and $\gamma =0$ then from Eqs. (\ref%
{payoff-general1}) and (\ref{payoff-general2}) the payoffs of the players
reduce to 
\end{subequations}
\begin{subequations}
\label{one-parameter}
\begin{align}
\$_{A}(\theta _{1},\phi _{1},\phi _{2},\theta _{2})& =\cos ^{2}\frac{\theta
_{1}}{2}\left[ \cos ^{2}\frac{\theta _{2}}{2}\left( \alpha +\beta -2\sigma
\right) -\alpha \sin ^{2}\frac{\delta }{2}-\beta \cos ^{2}\frac{\delta }{2}%
+\sigma \right]  \notag \\
& +\cos ^{2}\frac{\theta _{2}}{2}\left( -\alpha \sin ^{2}\frac{\delta }{2}%
-\beta \cos ^{2}\frac{\delta }{2}+\sigma \right) +\alpha \sin ^{2}\frac{%
\delta }{2}+\beta \cos ^{2}\frac{\delta }{2}  \notag \\
& -\frac{\left( \alpha -\beta \right) }{2}\sin \delta \sin \theta _{1}\sin
\theta _{2}\sin \left( \phi _{1}+\phi _{2}\right) , \\
\$_{B}(\theta _{1},\phi _{1},\phi _{2},\theta _{2})& =\cos ^{2}\frac{\theta
_{2}}{2}\left[ \cos ^{2}\frac{\theta _{1}}{2}\left( \alpha +\beta -2\sigma
\right) -\beta \sin ^{2}\frac{\delta }{2}-\alpha \cos ^{2}\frac{\delta }{2}%
+\sigma \right]  \notag \\
& +\cos ^{2}\frac{\theta _{1}}{2}\left( -\beta \sin ^{2}\frac{\delta }{2}%
-\alpha \cos ^{2}\frac{\delta }{2}+\sigma \right) +\beta \sin ^{2}\frac{%
\delta }{2}+\alpha \cos ^{2}\frac{\delta }{2}  \notag \\
& +\frac{\left( \alpha -\beta \right) }{2}\sin \delta \sin \theta _{1}\sin
\theta _{2}\sin \left( \phi _{1}+\phi _{2}\right) .  \label{un-entangled}
\end{align}%
This shows that the measurement plays a crucial role in quantum games as if
initial state is unentangled, i.e., $\gamma =0,$ arbiter can still apply
entangled basis for the measurement to obtain quantum mechanical results.
Above payoff's are similar to that of Marinatto and Weber for the Battle of
Sexes%
\index{Battle of sexes!-} games if $\delta $ is replaced by $\gamma .$

This encourages us to investigate the role of measurement in quantum games
in more detail.

\section{The Role of Measurement in Quantum Games}

In two players quantum game arbiter prepares a two qubit initial quantum
state and passes on one qubit to each of the players (generally referred to
as Alice and Bob). After applying their local operators (or strategies) the
players return the respective qubits back to arbiter who announces the
payoffs after\ performing measurement by applying the suitable payoff
operators depending on the payoff matrix of the game. The role of the initial%
\index{Quantum game!role of initial state} quantum state remained an
interesting issue in quantum games \cite{eisert,marinatto,azhar} see also
section (\ref{dilemma}). However, the importance of the payoff operators,
used by arbiter to perform measurement to determine the payoffs of the
players, remained unnoticed. In chapter (\ref{general}) we pointed out the
importance of measurement basis in quantum games. It was shown that if the
arbiter is allowed to perform the measurement in the entangled basis some
interesting situations could arise which were not possible in the frame work
of Eisert \textit{et al}. \cite{eisert} and Marinatto and Weber \cite%
{marinatto} schemes. Here we further extend this notion to investigate the
role of measurement%
\index{Quantum game!role of measurement} basis in quantum games by taking
Prisoners' Dilemma%
\index{Prisoners' dilemma!role of measurement} as an example. It is observed
that the quantum payoffs can be divided into four different categories on
the basis of initial state and measurement basis. These different situations
arise due the possibility of having product or entangled initial state and
then applying product or entangled basis for the measurement \cite{pati,xyz1}%
. In the context of our generalized framework for quantum games, the four
different types of payoffs are\emph{\ }

(i) $\$_{PP}$\ is the payoff%
\index{Payoff!product input/product measurement} when the initial quantum
state is of the product form and product basis are used for measurement to
determine the payoffs.

(ii) $\$_{PE}$\ is the payoff%
\index{Payoff!product input/entangled measurement} when the initial quantum
state is of the product form and entangled basis are used for measurement to
determine the payoffs.

(iii) $\$_{EP}$\ is the payoff%
\index{Payoff!entangled input/product measurement} when the initial quantum
state is entangled and product basis are used for measurement to determine
the payoffs.

(iv) $\$_{EE}$\ is the payoff%
\index{Payoff!entangled input/entangled measurement} when the initial
quantum state is entangled and entangled basis are used for measurement to
determine the payoffs.

Our results show that these payoffs obey a relation, $\$_{PP}<\$_{PE}=%
\$_{EP}<\$_{EE}$ at the Nash equilibrium (NE)%
\index{Nash equilibrium (NE)!role of measurement}.

\subsection{Quantization of Prisoners' Dilemma}

The payoff matrix for Prisoners' Dilemma%
\index{Prisoners' dilemma!.} game is of the form (\ref{matrix-prisoner}). In
our generalized version of quantum games the arbiter prepares the initial
state of the form of Eq. (\ref{state in}) where $\left| 0\right\rangle $ and 
$\left| 1\right\rangle $, represent vectors in the strategy space
corresponding to Cooperate and Defect, respectively\emph{\ }with\emph{\ }$%
\gamma \in \left[ 0,\pi \right] $\textbf{.}\emph{\ }Usually this range is
set as $\gamma \in \left[ 0,\pi /2\right] $\ but as we will see later in
case (c) below that the game has two Nash equilibria one at $\theta
_{1}=\theta _{2}=\pi $ when $\sin ^{2}\left( \gamma /2\right) \leq 
\frac{1}{3}$ and the other at$\ \theta _{1}=\theta _{2}=0$\ when $\sin
^{2}\left( \gamma /2\right) \geq \frac{2}{3}.$\ The latter possibility
exists if $\gamma \in \left[ 0,\pi \right] $\ otherwise only the first Nash (%
$\theta _{1}=\theta _{2}=\pi )$ will exist. Therefore, we set this range so
that both the Nash Equilibria could be analyzed.

The strategy of each of the\ players can be represented by the unitary
operator $U_{i}$\ of the form of Eq. (\ref{combination}).\textbf{\ }Here we
restrict our treatment to two parameter set of strategies $(\theta _{i},\phi
_{i})$ for mathematical simplicity in accordance with the Ref. \cite{eisert}%
. After the application of the strategies, the initial state Eq. (\ref{state
in}) transforms to 
\end{subequations}
\begin{align}
\left| \psi _{f}\right\rangle & =\cos \left( \gamma /2\right) \left[ \cos
\left( \theta _{1}/2\right) \cos \left( \theta _{2}/2\right) e^{i\left( \phi
_{1}+\phi _{2}\right) }\left| 00\right\rangle -\cos \left( \theta
_{1}/2\right) \sin \left( \theta _{2}/2\right) e^{i\phi _{1}}\left|
01\right\rangle \right.  \notag \\
& -\left. \cos \left( \theta _{2}/2\right) \sin \left( \theta _{1}/2\right)
e^{i\phi _{2}}\left| 10\right\rangle +\sin \left( \theta _{1}/2\right) \sin
\left( \theta _{2}/2\right) \left| 11\right\rangle \right]  \notag \\
& +i\sin \left( \gamma /2\right) \left[ \cos \left( \theta _{1}/2\right)
\cos \left( \theta _{2}/2\right) e^{-i(\phi _{1}+\phi _{2})}\left|
11\right\rangle +\cos \left( \theta _{1}/2\right) \sin \left( \theta
_{2}/2\right) e^{-i\phi _{1}}\left| 10\right\rangle \right.  \notag \\
& +\left. \cos \left( \theta _{2}/2\right) \sin \left( \theta _{1}/2\right)
e^{-i\phi _{2}}\left| 01\right\rangle +\sin \left( \theta _{1}/2\right) \sin
\left( \theta _{2}/2\right) \left| 00\right\rangle \right] .
\label{state fin role}
\end{align}%
The operators used by the arbiter to determine the payoffs for Alice and Bob
are for the case of Prisoners' Dilemma%
\index{Prisoners' dilemma!.} with payoff matrix of the form (\ref%
{matrix-prisoner}) become

\begin{align}
P_{A}& =3P_{00}+P_{11}+5P_{10},  \notag \\
P_{B}& =3P_{00}+P_{11}+5P_{01},  \label{pay-operator role}
\end{align}%
where for $m,n=0,1$ and the operators $P_{mn}=\left| \psi _{mn}\right\rangle
\left\langle \psi _{mn}\right| $ are given by Eq. (\ref{oper a}) with\emph{\ 
}$\delta \in \left[ 0,\pi \right] $ (the explanation for this range is the
same as for $\gamma $ above). Using Eqs. (\ref{state fin role}), (\ref%
{pay-operator role}) and (\ref{payoff}), we get the following payoffs 
\begin{eqnarray}
\$^{A}\left( \theta _{1},\theta _{2},\phi _{1},\phi _{2}\right) &=&\sin
^{2}\left( \theta _{1}/2\right) \sin ^{2}\left( \theta _{2}/2\right) \left[
\cos ^{2}\left( 
\frac{\gamma +\delta }{2}\right) +3\sin ^{2}\left( \frac{\gamma -\delta }{2}%
\right) \right]  \notag \\
&&+\cos ^{2}\left( \theta _{1}/2\right) \cos ^{2}\left( \theta _{2}/2\right) 
\left[ 2+\cos \gamma \cos \delta +2\cos \left( 2\delta \left( \phi _{1}+\phi
_{2}\right) \right) \sin \gamma \sin \delta \right]  \notag \\
&&-\sin \theta _{1}\sin \theta _{2}\sin \left( \phi _{1}+\phi _{2}\right) 
\left[ \sin \gamma -\sin \delta \right] +\frac{5}{4}\left[ 1-\cos \theta
_{1}\cos \theta _{2}\right]  \notag \\
&&+\frac{5}{4}\left( \cos \theta _{2}-\cos \theta _{1}\right) \left[ \cos
\gamma \cos \delta +\cos \left( 2\phi _{1}\right) \sin \gamma \sin \delta %
\right] .  \label{payoff-general-role}
\end{eqnarray}%
The payoff of player $B$ can be found by interchanging $\theta
_{1}\longleftrightarrow $\ $\theta _{2}$\ and $\phi _{1}\longleftrightarrow
\phi _{2}$ in Eq. (\ref{payoff-general-role}). There can be four types of
payoffs for each player\ for different combinations of $\delta $ and $\gamma 
$. In the following $\$_{PP}\left( \theta _{1},\theta _{2}\right) $ means
payoffs of the players when the initial state of the game is product state
and payoff operator used by arbiter for measurement is also in the product
form $(\gamma =0,\delta =0)$ and $\$_{EP}\left( \theta _{1},\theta _{2},\phi
_{1},\phi _{2}\right) $ means the payoffs for entangled input state when the
payoff operator used for measurement is in the product form, i.e., $(\gamma
\neq 0,\delta =0)$. Similarly $\$_{PE}\left( \theta _{1},\theta _{2},\phi
_{1},\phi _{2}\right) $ and $\$_{EE}\left( \theta _{1},\theta _{2},\phi
_{1},\phi _{2}\right) $ can also be interpreted. Therefore, for different
values of $\delta $ and $\gamma $ the\emph{\ }following four cases can be
identified:

\textbf{Case (a) }When\textbf{\ }$\delta $\textbf{\ }$=$\textbf{\ }$\gamma
=0,$ Eq. (\ref{payoff-general-role}), becomes 
\begin{subequations}
\label{SPP-prisoner}
\begin{equation}
\$_{PP}^{A}\left( \theta _{1},\theta _{2}\right) =3\cos ^{2}\left( \theta
_{1}/2\right) \cos ^{2}\left( \theta _{2}/2\right) +\sin ^{2}\left( \theta
_{1}/2\right) \sin ^{2}\left( \theta _{2}/2\right) +5\sin ^{2}\left( \theta
_{1}/2\right) \cos ^{2}\left( \theta _{2}/2\right) .  \label{SPP-prisoner-a}
\end{equation}%
This situation corresponds to the classical game \cite{eisert1} where each
player play,\ $C,$ with probability $\cos ^{2}\left( \theta _{i}/2\right) $
with $i=1,2.$ The Nash equilibrium%
\index{Nash equilibrium (NE)!-} corresponds to $\theta _{1}=\theta _{2}=\pi
, $ i.e., $(D,D)$ with payoffs for both the players as

\end{subequations}
\begin{equation}
\$_{PP}^{A}(\theta _{1}=\pi ,\theta _{2}=\pi )=\$_{PP}^{B}(\theta _{1}=\pi
,\theta _{2}=\pi )=1.  \label{SPP-Nash}
\end{equation}

\textbf{Case (b) }When $\gamma =0,\delta $\textbf{\ }$\neq 0,$ in Eq. (\ref%
{payoff-general-role}), then the game has two Nash equilibria one at $\theta
_{1}=\theta _{2}=0$ when $\sin ^{2}\left( \delta /2\right) \geq 
\frac{2}{3}$ \ and the other at $\theta _{1}=\theta _{2}=\pi $ when $\sin
^{2}\left( \delta /2\right) \leq \frac{1}{3}$. The corresponding payoffs for
these Nash equilibria are

\begin{eqnarray}
\$_{PE}^{A}(\theta _{1} &=&0,\theta _{2}=0)=\$_{PE}^{B}(\theta _{1}=0,\theta
_{2}=0)=3-2\sin ^{2}\left( \delta /2\right) ,  \notag \\
\$_{PE}^{A}(\theta _{1} &=&\pi ,\theta _{2}=\pi )=\$_{PE}^{B}(\theta
_{1}=\pi ,\theta _{2}=\pi )=1+2\sin ^{2}\left( \delta /2\right) .
\label{SPE-nash}
\end{eqnarray}%
Here in this case at Nash equilibrium%
\index{Nash equilibrium (NE)!-} the payoffs are independent of $\phi
_{1},\phi _{2}.$ Furthermore it is clear that the above payoffs for all the
allowed values of $\delta $ remain less than 3, which is the optimal payoff
for the two players if they cooperate.

\textbf{Case (c) }For $\gamma \neq 0,$ and $\delta $\textbf{\ }$=0,$ Eq. (%
\ref{payoff-general-role}) again gives two Nash equilibria one at $\theta
_{1}=\theta _{2}=0$ when $\sin ^{2}\left( \gamma /2\right) \geq 
\frac{2}{3}$ \ and the other at $\theta _{1}=\theta _{2}=\pi $ when $\sin
^{2}\left( \gamma /2\right) \leq \frac{1}{3}$. The corresponding payoffs are

\begin{eqnarray}
\$_{EP}^{A}(\theta _{1} &=&0,\theta _{2}=0)=\$_{EP}^{B}(\theta _{1}=0,\theta
_{2}=0)=3-2\sin ^{2}\left( \gamma /2\right) ,  \notag \\
\$_{EP}^{A}(\theta _{1} &=&\pi ,\theta _{2}=\pi )=\$_{EP}^{B}(\theta
_{1}=\pi ,\theta _{2}=\pi )=1+2\sin ^{2}\left( \gamma /2\right) .
\label{SEP-nash}
\end{eqnarray}%
It can be seen that the payoffs at both Nash equilibrium%
\index{Nash equilibrium (NE)!-} for allowed values of $\sin ^{2}%
\frac{\gamma }{2}$\ remain less than 3. From equations (\ref{SPE-nash}) and (%
\ref{SEP-nash}), it is also clear that $\$_{EP}^{A}(0,0)=\$_{PE}^{A}(\pi
,\pi )$ only for $\delta =\gamma .$

\textbf{Case (d) }When$\ \gamma =\delta $\textbf{\ }$=\pi /2,$ Eq. (\ref%
{payoff-general-role}) becomes 
\begin{align}
\$_{EE}^{A}\left( \theta _{1},\theta _{2},\phi _{1},\phi _{2}\right) & =3 
\left[ \cos \left( \theta _{1}/2\right) \cos \left( \theta _{2}/2\right)
\cos \left( \phi _{1}+\phi _{2}\right) \right] ^{2}  \notag \\
& +\left[ \sin \left( \theta _{1}/2\right) \sin \left( \theta _{2}/2\right)
+\cos \left( \theta _{1}/2\right) \cos \left( \theta _{2}/2\right) \sin
\left( \phi _{1}+\phi _{2}\right) \right] ^{2}  \notag \\
& +5\left[ \sin \left( \theta _{1}/2\right) \cos \frac{\theta _{2}}{2}\cos
\phi _{2}-\cos \left( \theta _{1}/2\right) \sin \left( \theta _{2}/2\right)
\sin \phi _{1}\right] ^{2}.  \notag \\
&  \label{SEE-prisoner-b}
\end{align}%
This payoff is same\emph{\ }as found by Eisert \textit{et al}. \cite{eisert}
and $\theta _{1}=\theta _{2}=0,\phi _{1}=\phi _{2}=\frac{\pi }{2}$ is the
Nash equilibrium%
\index{Nash equilibrium (NE)!-} of the game that gives the payoffs for both
players as 
\begin{equation}
\$_{EE}^{A}(0,0,%
\frac{\pi }{2},\frac{\pi }{2})=\$_{EE}^{B}(0,0,\frac{\pi }{2},\frac{\pi }{2}%
)=3.  \label{SEE-nash}
\end{equation}%
Comparing Eqs. (\ref{SPP-Nash}), (\ref{SPE-nash}), (\ref{SEP-nash}) and (\ref%
{SEE-nash}) it is evident that 
\begin{equation}
\$_{EE}^{l}(0,0,\frac{\pi }{2},\frac{\pi }{2})>\left( \$_{PE}^{l}(\theta
_{1}=k,\theta _{2}=k),\$_{EP}^{l}(\theta _{1}=k,\theta _{2}=k)\right)
>\$_{PP}^{l}(\theta _{1}=\pi ,\theta _{2}=\pi ),  \label{payoffs-relation}
\end{equation}%
and 
\begin{equation}
\$_{PE}^{l}(\theta _{1}=k,\theta _{2}=k)=\$_{EP}^{l}(\theta _{1}=k,\theta
_{2}=k)\text{ for }\gamma =\delta ,
\end{equation}%
with $k=0,\pi $ and $l=A,B$. \textrm{The expression (\ref{payoffs-relation})
shows that entanglement plays a crucial role in quantum games}. The
combination of initial entangled state with entangled payoff operators gives
higher payoffs as compared to all other combinations of $\gamma $\ and $%
\delta $.

\section{Extension to Three Parameter Set of Strategies}

Generalized quantization scheme%
\index{Three parameters set of strategies!generalized} can be extended to
three parameter set of strategies by introducing the unitary operator of the
form

\begin{equation}
\hat{U}(\theta ,\phi ,\psi )=\left[ 
\begin{array}{cc}
e^{i\phi }\cos \frac{\theta }{2} & ie^{i\psi }\sin \frac{\theta }{2} \\ 
ie^{-i\psi }\sin \frac{\theta }{2} & e^{-i\phi }\cos \frac{\theta }{2}%
\end{array}%
\right] ,
\end{equation}%
by replacing operators (\ref{oper}) by 
\begin{align}
R\left| 0\right\rangle & =e^{i\phi _{j}}\left| 0\right\rangle ,\text{ \ \ }%
R\left| 1\right\rangle =e^{-i\phi _{j}}\left| 1\right\rangle ,  \notag \\
C\left| 0\right\rangle & =e^{i(\frac{\pi }{2}+\psi _{j})}\left|
1\right\rangle ,\text{ \ \ \ \ \ }C\left| 1\right\rangle =e^{i(\frac{\pi }{2}%
-\psi _{j})}\left| 0\right\rangle ,
\end{align}%
where $-\pi \leq \phi _{j},\psi _{j}\leq \pi ,0\leq \theta \leq \pi .$

In this case the payoffs for any game with $\$_{mn}$\ as the elements of the
payoff matrix come out to be 
\begin{eqnarray}
\$(\theta _{j},\alpha _{j},\beta _{j}) &=&\cos ^{2}\frac{\theta _{1}}{2}\cos
^{2}\frac{\theta _{2}}{2}\left[ \eta \$_{00}+\chi \$_{11}+\left(
\$_{00}-\$_{11}\right) \xi \cos 2(\alpha _{1}+\alpha _{2})\right]  \notag \\
&&+\sin ^{2}\frac{\theta _{1}}{2}\sin ^{2}\frac{\theta _{2}}{2}\left[ \eta
\$_{11}+\chi \$_{00}-\left( \$_{00}-\$_{11}\right) \xi \cos 2(\beta
_{1}+\beta _{2})\right]  \notag \\
&&+\cos ^{2}\frac{\theta _{1}}{2}\sin ^{2}\frac{\theta _{2}}{2}\left[ \eta
\$_{01}+\chi \$_{10}+\left( \$_{01}-\$_{10}\right) \xi \cos 2(\alpha
_{1}-\beta _{2})\right]  \notag \\
&&+\sin ^{2}\frac{\theta _{1}}{2}\cos ^{2}\frac{\theta _{2}}{2}\left[ \eta
\$_{10}+\chi \$_{01}-\left( \$_{01}-\$_{10}\right) \xi \cos 2(\alpha
_{2}-\beta _{1})\right]  \notag \\
&&+\frac{\left( \$_{00}-\$_{11}\right) }{4}\sin \theta _{1}\sin \theta
_{2}\sin \delta \sin \left( \alpha _{1}+\alpha _{2}+\beta _{1}+\beta
_{2}\right)  \notag \\
&&+\frac{\left( \NEG{\$}_{10}-\$_{01}\right) }{4}\sin \theta _{1}\sin \theta
_{2}\sin \delta \sin \left( \alpha _{1}-\alpha _{2}+\beta _{1}-\beta
_{2}\right)  \notag \\
&&+\frac{\left( -\$_{00}-\$_{11}+\$_{01}+\$_{10}\right) }{4}\sin \theta
_{1}\sin \theta _{2}\sin \gamma \sin \left( \alpha _{1}+\alpha _{2}-\beta
_{1}-\beta _{2}\right) ,  \notag \\
&&  \label{three-strategy}
\end{eqnarray}%
where $j=1,2$. When $\delta =0$\ and $\gamma =\frac{\pi }{2}$\ then the
generalized quantization scheme reduces to the Marinatto and Weber
quantization scheme \cite{marinatto}. For Battle of Sexes%
\index{Battle of sexes!three parameters set of strategies} (\ref{matrix-BoS}%
)\ the payoffs (\ref{three-strategy}) with $\delta =0$\ and $\gamma =%
\frac{\pi }{2}$\ become 
\begin{subequations}
\label{BoS}
\begin{eqnarray}
\$^{A}(\theta _{j},\alpha _{j},\beta _{j}) &=&\frac{\left( \alpha +\beta
\right) }{2}\cos ^{2}\frac{\theta _{1}}{2}\cos ^{2}\frac{\theta _{2}}{2}+%
\frac{\left( \alpha +\beta \right) }{2}\sin ^{2}\frac{\theta _{1}}{2}\sin
^{2}\frac{\theta _{2}}{2}+  \notag \\
&&\left( \cos ^{2}\frac{\theta _{1}}{2}\sin ^{2}\frac{\theta _{2}}{2}+\sin
^{2}\frac{\theta _{1}}{2}\cos ^{2}\frac{\theta _{2}}{2}\right) \sigma - 
\notag \\
&&\frac{\left( \alpha +\beta -2\sigma \right) }{4}\sin \theta _{1}\sin
\theta _{2}\sin (\alpha _{1}-\beta _{1}+\alpha _{2}-\beta _{2})  \notag \\
&&  \label{BoS-3pa} \\
\$^{B}(\theta _{j},\alpha _{j},\beta _{j}) &=&\frac{\left( \alpha +\beta
\right) }{2}\cos ^{2}\frac{\theta _{1}}{2}\cos ^{2}\frac{\theta _{2}}{2}+%
\frac{\left( \alpha +\beta \right) }{2}\sin ^{2}\frac{\theta _{1}}{2}\sin
^{2}\frac{\theta _{2}}{2}+  \notag \\
&&\left( \cos ^{2}\frac{\theta _{1}}{2}\sin ^{2}\frac{\theta _{2}}{2}+\sin
^{2}\frac{\theta _{1}}{2}\cos ^{2}\frac{\theta _{2}}{2}\right) \sigma - 
\notag \\
&&\frac{\left( \alpha +\beta -2\sigma \right) }{4}\sin \theta _{1}\sin
\theta _{2}\sin (\alpha _{1}-\beta _{1}+\alpha _{2}-\beta _{2})  \notag \\
&&  \label{BoS-3pb}
\end{eqnarray}

\section{Summary}

A generalized quantization scheme for non zero sum games is proposed. The
game of Battle of Sexes%
\index{Battle of sexes!-} has been used as an example to introduce this
quantization scheme. However our quantization scheme is applicable to other
games as well. This new scheme reduces to Eisert's \textit{et al.} \cite%
{eisert} scheme under the condition 
\end{subequations}
\begin{equation*}
\delta =\gamma ,\phi _{1}+\phi _{2}=\pi /2
\end{equation*}%
and to Marinatto and Weber \cite{marinatto} scheme when 
\begin{equation*}
\delta =0,\phi _{1}=0,\phi _{2}=0.
\end{equation*}%
In the above conditions $\gamma $ is a measure of entanglement of the
initial state. For $\gamma =0,$ classical results are obtained when $\delta
=0,\phi _{1}=0,\phi _{2}=0$. Furthermore, some interesting situations are
identified which are not apparent within the exiting two quantizations
schemes. For example, with $\delta \neq 0,$\ nonclassical results are
obtained for initially unentangled state. This shows that the measurement
plays a crucial role in quantum%
\index{Role of measurement} games. Under the context of generalized
quantization scheme,\ by taking Prisoners' Dilemma game as an example we
showed that depending on the initial state and type of measurement (product
or entangled), quantum payoffs in games can be categorized in to four
different types. These four categories are $\$_{PP},\$_{PE},\$_{EP},\$_{EE}$
where $P,$ and $E$\ are abbreviations for the product and entanglement at
input and output. It is shown that there exists a relation of the form $%
\$_{PP}<\$_{PE}=\$_{EP}<\$_{EE}$\ among different payoffs at Nash
equilibrium.

\chapter{\label{correlated noise}Quantum Games with Correlated Noise}

It requires exchange of qubits between arbiter and players to play quantum
games. The\ transmission of qubit\ through a channel is generally prone to
decoherence due to its interaction with the environment. In the game
theoretic sense this situation can be imagined as if a demon is present
between the arbiter and the players who corrupts the qubits. The players are
not necessarily aware of the actions of the demon \cite{lee}. This type of
protocol was first applied to quantum games to show that above a certain
level of decoherence the quantum player has no advantage over a classical
player \cite{johnson}.\emph{\ }Later quantum version of Prisoners' Dilemma%
\index{Prisoners' dilemma!in presence of decoherence} was analyzed in
presence of decoherence to prove that Nash equilibrium%
\index{Nash equilibrium (NE)!-} is not affected by decoherence%
\index{Decoherence} \cite{chen}. Recently, Flitney and Abbott \cite%
{flitney-3} showed for the quantum games in presence dephasing quantum
channel%
\index{Quantum channel!dephasing} that the advantage that a quantum player
enjoys over a classical player diminishes as decoherence%
\index{Decoherence} increases and it vanishes for the maximum decoherence.

In this chapter we analyze quantum games in presence of quantum correlated
dephasing channel%
\index{Quantum channel!dephasing!cerrelated} in the context of our
generalized quantization scheme%
\index{Quantization scheme!generalized} for non-zero sum games. We identify
four different combinations on the basis of initial state entanglement
parameter, $\gamma ,$\ and the measurement parameter, $\delta ,$\ for three
quantum games.\emph{\ }It is shown that for $\gamma =\delta =0$ the games
reduce to the classical and\emph{\ }become independent of decoherence%
\index{Decoherence} and memory effects. For the case when $\gamma \neq
0,\delta =0$ the scheme reduces to Marinatto and Weber quantization scheme%
\index{Quantization scheme!of Marinatto and Weber} \cite{marinatto}. It is
interesting to note that though the initial state is entangled, quantum
player has no advantage over the classical player.\emph{\ }Same happens for
the case of $\gamma =0,\delta \neq 0$. However, for the case when $\gamma
=\delta =%
\frac{\pi }{2}$ the scheme transforms to the Eisert's quantization scheme%
\index{Quantization scheme!of Eisert et al.|textit} \cite{eisert} and
quantum player always remains better off against a player restricted to
classical strategies. Furthermore, in the limit of maximum correlation the
effect of decoherence%
\index{Decoherence} vanishes and the quantum games behave as noiseless games.

\section{Classical Noise}

To understand classical noise%
\index{Classical noise} take an example of a storage device that stores
information in form of a string of $0$ and $1$. The bits interact with
environment and therefore, with the passage of time each of the bit has a
probability $p$ to flip. This situation is illustrated in Fig. \ref%
{classical noise}. 
\begin{figure}[th]
\centering
\includegraphics[scale=1]{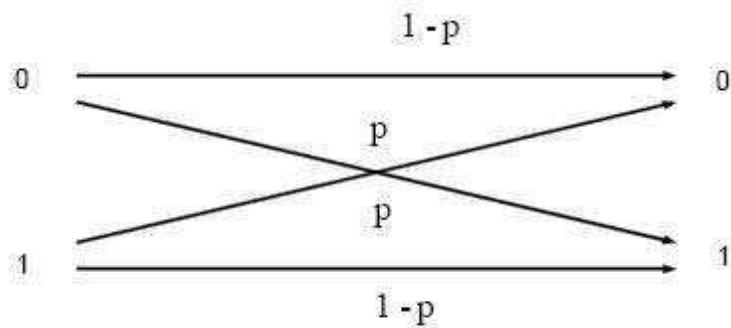}
\caption{Classical noise model}
\label{classical noise}
\end{figure}

To describe the process mathematically suppose that $\left(
p_{0},p_{1}\right) $ and $\left( q_{0},q_{1}\right) $ are the probabilities
of bits of being in state $\left( 0,1\right) $ before and after the
interaction respectively. Therefore, the whole process can be expressed as 
\begin{equation}
\left[ 
\begin{array}{c}
q_{0} \\ 
q_{1}%
\end{array}%
\right] =\left[ 
\begin{array}{cc}
1-p & p \\ 
p & 1-p%
\end{array}%
\right] \left[ 
\begin{array}{c}
p_{0} \\ 
p_{1}%
\end{array}%
\right] ,
\end{equation}%
which can be written as 
\begin{equation}
\vec{q}=E\vec{p},  \label{distribution}
\end{equation}%
where $E$ is known as evolution matrix. For any valid probability
distribution $\vec{p},$ the evolution matrix $E$ must fulfill the conditions
of positivity and completeness. Where the positivity means that all the
elements of matrix $E$ are positive and the completeness reflects the fact
that the sum of elements in each column of evolution matrix $E$\ is one \cite%
{chuang}.

\section{Quantum Noise}

A quantum system that is totally isolated from unwanted interactions of the
outside world is called a closed system%
\index{Closed system}; but in real world there is no perfect closed system.
For it may suffer from unwanted interactions with the outside world -\textit{%
the environment}- that can introduce noise in it. Therefore, such a system
is termed as open quantum system%
\index{Open system}. A nice example in this regard is that of a qubit
prepared by two positions of an electron that is being interacted by a
charge particle acting as a source of uncontrolled noise.

Quantum operations formalism%
\index{Quantum operations formalism} is a mathematical tool that is used to
study the behavior of an open system \cite{chuang}. Let a physical process $%
\varepsilon $\ transform a quantum state $\rho $ as 
\begin{equation}
\acute{\rho}=\varepsilon (\rho ).  \label{dynamics}
\end{equation}%
For a closed system $\varepsilon $\ in Eq. (\ref{dynamics}) is a unitary
transformation $U$ and $\varepsilon (\rho )=U\rho U^{\dagger }.$ Whereas the
dynamics of an open quantum system is considered to be arising from the
interactions of the principal system (system of interest) with the
environment and both these systems together form a closed system. Let the
principal quantum system be in state $\rho $\ and the environment in
standard state $\rho _{\text{env}}.$ It is further assumed that
system-environment initial state is product state of the form $\rho \otimes
\rho _{\text{env}}$ that undergoes a unitary interaction $U$. Then the final
state $\acute{\rho}$\ of the system is obtained by the relation 
\begin{equation}
\acute{\rho}=\varepsilon (\rho )=\mathrm{Tr}_{\text{env}}\left[ U\left( \rho
\otimes \rho _{\text{env}}\right) U^{\dagger }\right] ,  \label{noise}
\end{equation}%
where $\mathrm{Tr}_{\text{env}}$ represents the partial trace over the
environment. \textrm{In general the final state of the principal system }$%
\varepsilon (\rho )$\textrm{\ may not be related to the initial state by
unitary transformation}. In the most general case the quantum process $%
\varepsilon $ is a trace preserving%
\index{Trace preseving map} and completely positive linear map%
\index{Completely positive map} that maps the input density matrix to output
density matrix \cite{stinespring}. By trace preserving it means that as for
the input state $\rho $ it is always true that Tr$\left( \rho \right) =1$
similarly for output state $%
\acute{\rho}$ it must also be true that Tr$\left( \acute{\rho}\right) =1.$
The positivity condition implies that the quantum process $\varepsilon $
maps the positive density matrix $\rho $ (having non-negative eigenvalues)
to positive density matrix $\acute{\rho}.$ To explain complete positivity
another system $R$ having density matrix $\rho _{R}$ is introduced so that
the initial state of the system becomes $\rho _{R}\otimes \rho .$ Then a map 
$\varepsilon $ is complete positive if the process $I\otimes \varepsilon $ ($%
I$ is identity) maps the positive operator $\rho _{R}\otimes \rho $ to
positive operator $\rho _{R}\otimes \acute{\rho}.$

For the state of environment $\rho _{\text{env}}=\left| e_{0}\right\rangle
\left\langle e_{0}\right| $ with $\left| e_{k}\right\rangle $ as orthonormal
basis for the finite dimensional state space of the environment, Eq. (\ref%
{noise}) gives 
\begin{eqnarray}
\varepsilon (\rho ) &=&\mathrm{Tr}_{\text{env}}\left[ U\left( \rho \otimes
\left| e_{0}\right\rangle \left\langle e_{0}\right| \right) U^{\dagger }%
\right]  \notag \\
&=&\tsum \left\langle e_{k}\right| U\left( \rho \otimes \left|
e_{0}\right\rangle \left\langle e_{0}\right| \right) U^{\dagger }\left|
e_{k}\right\rangle  \notag \\
&=&\tsum E_{k}\rho E_{k}^{\dagger },  \label{operator-sum}
\end{eqnarray}%
where $E_{k}=\left\langle e_{k}\right| U\left| e_{0}\right\rangle $ are
Kraus operators acting on the principal system. These operators satisfy the
completeness relation 
\begin{equation}
\tsum E_{k}E_{k}^{\dagger }=I,
\end{equation}%
for trace preserving processes and the representation (\ref{operator-sum})
is called Kraus representation%
\index{Kraus representation} or operator sum representation%
\index{Operator sum representation} for process $\varepsilon .$ It is one of
the powerful mathematical representation for quantum operations \cite{chuang}%
.

\section{Quantum Correlated Noise}

One of the important examples of quantum noise is decoherence%
\index{Decoherence} that is a non-unitary dynamics and results due to the
coupling of principal system with the environment. Decoherence in form of
phase damping%
\index{Quantum channel!phase damping} or dephasing is much interesting. It
is uniquely quantum mechanical and describes the loss of quantum information
without loss of energy. The energy eigenstate of the system do not change\
as a function of time during this process but the system accumulates a phase
proportional to the eigenvalue. With the passage of time the relative phase%
\index{Relative phase} between the energy eingenstates may be lost. In pure
dephasing process a qubit transforms as 
\begin{equation}
a\left| 0\right\rangle +b\left| 1\right\rangle \rightarrow a\left|
0\right\rangle +be^{i\phi }\left| 1\right\rangle ,  \label{initial state}
\end{equation}%
where $\phi $\ is the phase kick%
\index{Phase kick}. If this phase kick, $\phi $ is assumed to be a random
variable with Gaussian distribution of mean zero and variance $2\lambda $
then the density matrix of system after averaging over all the values of $%
\phi $ is \cite{chuang}%
\begin{equation}
\left[ 
\begin{array}{cc}
\left| a\right| ^{2} & ab^{\ast } \\ 
a^{\ast }b & \left| b\right| ^{2}%
\end{array}%
\right] \rightarrow \left[ 
\begin{array}{cc}
\left| a\right| ^{2} & ab^{\ast }e^{-\lambda } \\ 
a^{\ast }be^{-\lambda } & \left| b\right| ^{2}%
\end{array}%
\right] .  \label{dephasing}
\end{equation}%
It is evident from the above equation that in this process the phase kicks
cause the\emph{\ }off-diagonal elements of the density matrix to decay
exponentially to zero with time. \textrm{Which means that a quantum system
initially prepared in a pure state }%
\begin{equation}
a\left| 0\right\rangle +b\left| 1\right\rangle
\end{equation}%
\textrm{decays to in an incoherent superposition of states of the form }%
\begin{equation}
\left| a\right| ^{2}\left| 0\right\rangle \left\langle 0\right| +\left|
b\right| ^{2}\left| 1\right\rangle \left\langle 1\right|
\end{equation}%
\textrm{\ } In the operator sum representation%
\index{Operator sum representation}\ the dephasing process can be expressed
as \cite{chuang,kraus} 
\begin{equation}
\rho _{f}=\overset{1}{\underset{i=0}{\tsum }}A_{i%
\text{ }}\rho _{\text{in}}A_{i}^{\dagger },  \label{kraus-1}
\end{equation}%
where\emph{\ } 
\begin{eqnarray}
A_{0} &=&\sqrt{1-\frac{p}{2}}I,  \notag \\
A_{1} &=&\sqrt{\frac{p}{2}}\sigma _{z},  \label{kraus}
\end{eqnarray}%
are the Kraus operators%
\index{Kraus operators},$\ I$ is the identity operator and $\sigma _{z}$\ is
the Pauli matrix. Recognizing\emph{\ }$1-p=e^{-\lambda },$ let $N$\ qubits
are allowed to pass through such a channel then Eq. (\ref{kraus-1})\ becomes 
\begin{equation}
\rho _{f}=\overset{N}{\underset{k_{1},%
\text{ ...,\ }k_{n}=0}{\tsum }}\left( A_{k_{n}}\otimes
......A_{k_{1}}\right) \rho _{\text{in }}\left( A_{k_{1}}^{\dagger }\otimes
......A_{k_{n}}^{\dagger }\right) .  \label{kraus-operators}
\end{equation}%
Now if noise is correlated with memory of degree $\mu ,$ the Kraus operators%
\index{Kraus operators!with memory} for two qubit system become \cite{palma}

\begin{equation}
A_{i,j}=%
\sqrt{p_{i}\left[ \left( 1-\mu \right) p_{j}+\mu \delta _{ij}\right] }%
A_{i}\otimes A_{j}.  \label{kraus-memory}
\end{equation}%
Physically, this expression means that with the\emph{\ }probability $1-\mu $
the noise is uncorrelated and can be completely specified by the Kraus
operators $A_{i,j}^{u}=\sqrt{p_{i}p_{j}}A_{i}\otimes A_{j}$ whereas with
probability $\mu $ the noise is correlated and is specified by Kraus
operators of the form $A_{ii}^{c}=\sqrt{p_{i}}A_{i}\otimes A_{i}.$

\section{Quantization of Games in Presence of Correlated Noise}

The protocol for quantum games in presence of decoherence%
\index{Decoherence} is developed by\emph{\ }Flitney and Abbott\emph{\ }\cite%
{flitney-3}. An initial entangled state is prepared by the arbiter and
passed on to the players through a dephasing quantum channel%
\index{Quantum channel!dephasing}. On receiving the quantum state players
apply their local operators (strategies) and return it back to arbiter again
through a dephasing quantum channel. Then arbiter performs the measurement
and announces their payoffs.

The initial quantum state of game is given by Eq. (\ref{state in}) and the
strategies of the players are given by Eq. (\ref{combination}) with unitary
operators $R_{i}$, $P_{i}$ defined as:

\begin{align}
R_{i}\left\vert 0\right\rangle & =e^{i\phi _{i}}\left\vert 0\right\rangle ,%
\text{ \ \ }R_{i}\left\vert 1\right\rangle =e^{-i\phi _{i}}\left\vert
1\right\rangle ,  \notag \\
P_{i}\left\vert 0\right\rangle & =e^{i\left( \frac{\pi }{2}-\psi \right)
}\left\vert 1\right\rangle ,\text{ \ \ \ \ \ }P_{i}\left\vert 1\right\rangle
=e^{i\left( \frac{\pi }{2}+\psi \right) }\left\vert 0\right\rangle ,
\label{operators correlated}
\end{align}%
where $-\pi \leq \phi _{i},\psi _{i}\leq \pi .$\emph{\ }This is extension of
generalized quantization scheme%
\index{Quantization scheme!generalized} to three strategy set of parameters
in accordance with Ref. \cite{flitney}\emph{. }The payoff operators used by
the arbiter to determine the payoff for Alice and Bob are%
\begin{equation}
P=\$_{00}P_{00}+\$_{01}P_{01}+\$_{10}P_{10}+\$_{11}P_{11},
\label{payoff operator correlated}
\end{equation}%
where for $m,n=0,1$ operators $P_{mn}=\left\vert \psi _{mn}\right\rangle
\left\langle \psi _{mn}\right\vert $ are explained in Eqs. (\ref{oper a})
with\emph{\ }$\delta \in \left[ 0,%
\frac{\pi }{2}\right] $ and $\$_{ij}$ are the elements of payoff matrix in $%
ith$ row and $jth$ column . As stated in section (\ref{GQS}) these operators
reduce to that of Eisert's scheme for $\delta $ equal to $\gamma ,$ which
represents the entanglement of the initial state \cite{eisert}. And for $%
\delta =0$ above operators transform into that of Marinatto and Weber's
scheme \cite{marinatto}. Using Eqs. (\ref{state in}), (\ref{payoff}), (\ref%
{kraus-memory}) and (\ref{payoff operator correlated}) the payoffs come out
to be 
\begin{eqnarray}
\$(\theta _{i},\phi _{i},\psi _{i}) &=&c_{1}c_{2}\left[ \eta \$_{00}+\chi
\$_{11}+\left( \$_{00}-\$_{11}\right) \mu _{p}^{(1)}\mu _{p}^{(2)}\xi \cos
2(\phi _{1}+\phi _{2})\right]  \notag \\
&&+s_{1}s_{2}\left[ \eta \$_{11}+\chi \$_{00}-\left( \$_{00}-\$_{11}\right)
\mu _{p}^{(1)}\mu _{p}^{(2)}\xi \cos 2(\psi _{1}+\psi _{2})\right]  \notag \\
&&+c_{1}s_{2}\left[ \eta \$_{01}+\chi \$_{10}+\left( \$_{01}-\$_{10}\right)
\mu _{p}^{(1)}\mu _{p}^{(2)}\xi \cos 2(\phi _{1}-\psi _{2})\right]  \notag \\
&&+c_{2}s_{1}\left[ \eta \$_{10}+\chi \$_{01}-\left( \$_{01}-\$_{10}\right)
\mu _{p}^{(1)}\mu _{p}^{(2)}\xi \cos 2(\phi _{2}-\psi _{1})\right]  \notag \\
&&+\frac{\mu _{p}^{(2)}\left( \$_{00}-\$_{11}\right) }{4}\sin \theta
_{1}\sin \theta _{2}\sin \delta \sin \left( \phi _{1}+\phi _{2}+\psi
_{1}+\psi _{2}\right)  \notag \\
&&+\frac{\mu _{p}^{(2)}\left( \NEG{\$}_{10}-\$_{01}\right) }{4}\sin \theta
_{1}\sin \theta _{2}\sin \delta \sin \left( \phi _{1}-\phi _{2}+\psi
_{1}-\psi _{2}\right)  \notag \\
&&+\frac{\mu _{p}^{(1)}\left( -\$_{00}-\$_{11}+\$_{01}+\$_{10}\right) }{4}%
\sin \theta _{1}\sin \theta _{2}\sin \gamma \sin \left( \phi _{1}+\phi
_{2}-\psi _{1}-\psi _{2}\right) ,  \notag \\
&&  \label{payoff correlated}
\end{eqnarray}%
where 
\begin{eqnarray*}
\eta &=&\cos ^{2}\left( \delta /2\right) \cos ^{2}\left( \gamma /2\right)
+\sin ^{2}\left( \delta /2\right) \sin ^{2}\left( \gamma /2\right) , \\
\chi &=&\cos ^{2}\left( \delta /2\right) \sin ^{2}\frac{\gamma }{2}+\sin
^{2}\left( \delta /2\right) \cos ^{2}\left( \gamma /2\right) , \\
\xi &=&1/2\left( \sin \delta \sin \gamma \right) , \\
c_{i} &=&\cos ^{2}\frac{\theta _{i}}{2}, \\
s_{i} &=&\sin ^{2}\frac{\theta _{i}}{2}, \\
\mu _{p}^{(i)} &=&\left( 1-\mu _{i}\right) \left( 1-p_{i}\right) ^{2}+\mu
_{i}.
\end{eqnarray*}%
The payoff for the players can be found by putting the appropriate values
for $\$_{ij}$ (elements of the payoff matrix for the corresponding game) in
Eq. (\ref{payoff correlated}). These payoffs become the classical payoffs
for $\delta =\gamma =0$ and for $\delta =\gamma =\frac{\pi }{2}$and $\mu =0$
these payoffs transform to the results of Flitney and Abbott \cite{flitney-3}%
. It is known that decoherence%
\index{Decoherence!-} has no effect on the Nash equilibrium%
\index{Nash equilibrium (NE)!-} of the game but it causes a reduction in the
payoffs \cite{chen,flitney-3}. In our case it is interesting to note that
this reduction of the payoffs depends on the degree of memory%
\index{Memory} $\mu .$ As $\mu $ increases from zero to one the effect of
noise reduces until finally for $\mu =1$ the payoffs become as that for
noiseless game irrespective of any value of $p_{i}$. It is further to be
noted that in comparison to memoryless case \cite{flitney-3} the quantum
phases $\phi _{i},\psi _{i}$\ do not vanish even for maximum value of
decoherence, i.e.,. for $p_{1}=p_{2}=1$.

To see further the effects of memory in quantum games%
\index{Quantum game!with correlated noise} we consider a situation in which
Alice is restricted to play classical strategies, i.e., $\phi _{1}=\psi
_{1}=0,$ whereas Bob is capable of playing the quantum strategies as well.
Under these circumstances following four cases for the different
combinations of $\delta $ and $\gamma $ are worth noting:

\textbf{Case (i)} When $\delta =\gamma =0$ then it is clear from Eq. (\ref%
{payoff correlated}) payoffs \ are the same as in the case of\emph{\ }%
classical game \cite{eisert1}. These payoffs, as expected, are independent
of the dephasing probabilities $p_{i}$, the quantum strategies $\phi
_{2},\psi _{2}$ and the memory.

\textbf{Case (ii)} When $\delta =0,\gamma \neq 0$ then $\eta =\cos ^{2}%
\frac{\gamma }{2},\chi =\sin ^{2}\frac{\gamma }{2},$and $\xi =0.$ Using
payoff matrix (\ref{matrix-prisoner}) for the game of Prisoners' Dilemma%
\index{Prisoners' dilemma!with correlated noise} and Eq.\emph{\ }(\ref%
{payoff correlated}) the payoffs for the two players are: 
\begin{eqnarray}
\$^{A}(\theta _{1},\theta _{2},\phi _{2},\psi _{2}) &=&c_{1}c_{2}\left(
3-2\sin ^{2}%
\frac{\gamma }{2}\right) +s_{1}s_{2}\left( 1+2\sin ^{2}\frac{\gamma }{2}%
\right)  \notag \\
&&+5c_{1}s_{2}\sin ^{2}\frac{\gamma }{2}+5c_{2}s_{1}\left( 1-\sin ^{2}\frac{%
\gamma }{2}\right)  \notag \\
&&+\frac{\mu _{p}^{(1)}}{4}\sin \theta _{1}\sin \theta _{2}\sin \gamma \sin
\left( \phi _{2}-\psi _{2}\right) ,  \notag \\
\$^{B}(\theta _{1},\theta _{2},\phi _{2},\psi _{2}) &=&c_{1}c_{2}\left(
3-2\sin ^{2}\frac{\gamma }{2}\right) +s_{1}s_{2}\left( 1+2\sin ^{2}\frac{%
\gamma }{2}\right)  \notag \\
&&+5c_{1}s_{2}\left( 1-\sin ^{2}\frac{\gamma }{2}\right) +5c_{2}s_{1}\sin
^{2}\frac{\gamma }{2}  \notag \\
&&+\frac{\mu _{p}^{(1)}}{4}\sin \theta _{1}\sin \theta _{2}\sin \gamma \sin
\left( \phi _{2}-\psi _{2}\right) .
\end{eqnarray}%
\ In this case the optimal strategy for the quantum player, Bob, is $\phi
_{2}-\psi _{2}=\frac{\pi }{2}.$ Though his choice for $\theta _{2}$ depends
on Alice's choice for\emph{\ }$\theta _{1},$but he can play $\theta _{2}=%
\frac{\pi }{2},$ without being bothered about Alice's choice as rational
reasoning leads Alice to play\emph{\ }$\theta _{1}=\frac{\pi }{2}$. Under
these choices of moves the payoffs for the two players are equal: 
\begin{eqnarray}
\$^{A}(\frac{\pi }{2},\frac{\pi }{2},\phi _{2}-\psi _{2} &=&\frac{\pi }{2}%
)=\$^{B}(\frac{\pi }{2},\frac{\pi }{2},\phi _{2}-\psi _{2}=\frac{\pi }{2}) 
\notag \\
&=&\frac{9}{4}+\frac{\mu _{p}^{(1)}}{4}\sin \gamma .
\end{eqnarray}%
It is evident that the quantum player has no advantage over the classical
player. Similarly for the Chicken game%
\index{Chicken game!with correlated noise} putting the payoffs from payoff
matrix (\ref{matrix chicken})\ we get: 
\begin{eqnarray}
\$^{A}(\theta _{1},\theta _{2},\phi _{2},\psi _{2}) &=&c_{1}c_{2}\left(
3-3\sin ^{2}%
\frac{\gamma }{2}\right) +s_{1}s_{2}\left( 3\sin ^{2}\frac{\gamma }{2}\right)
\notag \\
&&+c_{1}s_{2}\left( 3\sin ^{2}\frac{\gamma }{2}+1\right) +c_{2}s_{1}\left(
4-3\sin ^{2}\frac{\gamma }{2}\right)  \notag \\
&&+\frac{\mu _{p}^{(1)}}{2}\sin \theta _{1}\sin \theta _{2}\sin \gamma \sin
\left( \phi _{2}-\psi _{2}\right) , \\
\$^{B}(\theta _{1},\theta _{2},\phi _{2},\psi _{2}) &=&c_{1}c_{2}\left(
3-3\sin ^{2}\frac{\gamma }{2}\right) +s_{1}s_{2}\left( 3\sin ^{2}\frac{%
\gamma }{2}\right)  \notag \\
&&+c_{1}s_{2}\left( 4-3\sin ^{2}\frac{\gamma }{2}\right) +c_{2}s_{1}\left(
1+3\sin ^{2}\frac{\gamma }{2}\right)  \notag \\
&&+\frac{\mu _{p}^{(1)}}{2}\sin \theta _{1}\sin \theta _{2}\sin \gamma \sin
\left( \phi _{2}-\psi _{2}\right) ,
\end{eqnarray}%
and it can be shown using the same argument as for the game of Prisoners'
Dilemma that the quantum player does not have any advantage over classical
player in the Chicken game as well.

For the case of the quantum Battle of Sexes%
\index{Battle of sexes!with correlated noise} using values from payoff
matrix (\ref{matrix-BoS}) the payoffs become 
\begin{eqnarray}
\$^{A}(\theta _{1},\theta _{2},\phi _{2},\psi _{2}) &=&c_{1}c_{2}\left(
2-\sin ^{2}%
\frac{\gamma }{2}\right) +s_{1}s_{2}\left( 1+\sin ^{2}\frac{\gamma }{2}%
\right)  \notag \\
&&-\frac{3\mu _{p}^{(1)}}{4}\sin \theta _{1}\sin \theta _{2}\sin \gamma \sin
\left( \phi _{2}-\psi _{2}\right) ,  \notag \\
\$^{B}(\theta _{1},\theta _{2},\phi _{2},\psi _{2}) &=&c_{1}c_{2}\left(
1+\sin ^{2}\frac{\gamma }{2}\right) +s_{1}s_{2}\left( 2-\sin ^{2}\frac{%
\gamma }{2}\right)  \notag \\
&&-\frac{3\mu _{p}^{(1)}}{4}\sin \theta _{1}\sin \theta _{2}\sin \gamma \sin
\left( \phi _{2}-\psi _{2}\right) .
\end{eqnarray}%
Here the optimal strategy for Bob is $\phi _{2}-\psi _{2}=-\frac{\pi }{2}$
and $\theta _{2}=\frac{\pi }{2}$, keeping in view that the best strategy for
Alice is $\theta _{1}=\frac{\pi }{2}.$ The corresponding payoffs of the
players are again equal for these choices, i.e., 
\begin{eqnarray}
\$^{A}(\frac{\pi }{2},\frac{\pi }{2},\phi _{2}-\psi _{2} &=&-\frac{\pi }{2}%
)=\$^{B}(\frac{\pi }{2},\frac{\pi }{2},\phi _{2}-\psi _{2}=-\frac{\pi }{2}) 
\notag \\
&=&\frac{3}{4}+\frac{3}{4}\mu _{p}^{(1)}\sin \gamma .
\end{eqnarray}%
It is clear that for the case $\delta =0,\gamma \neq 0$ the quantum player
has no advantage over the classical player for three games considered above%
\emph{. }It is interesting\emph{\ }because\emph{\ }the game starts from an
entangled state and the payoffs are also the functions of the quantum
phases, $\phi _{i},\psi _{i},$ dephasing probability, $p_{1}$ and the degree
of memory%
\index{Memory}, $\mu _{1}$, of the quantum channel between Bob and arbiter.

\textbf{Case (iii)} When $\delta \neq 0,\gamma =0$\ then using Eq. (\ref%
{payoff correlated}) and the values from the payoff matrices given in
subsection (\ref{examples}) the payoffs for the two players in games of
Prisoners' Dilemma%
\index{Prisoners' dilemma!with correlated noise}, Chicken%
\index{Chicken game!with correlated noise} and Battle of Sexes%
\index{Battle of sexes!with correlated noise} are 
\begin{eqnarray}
\$^{A}(\theta _{1},\theta _{2},\phi _{2},\psi _{2}) &=&c_{1}c_{2}\left(
3-2\sin ^{2}%
\frac{\delta }{2}\right) +s_{1}s_{2}\left( 1+2\sin ^{2}\frac{\delta }{2}%
\right)  \notag \\
&&+\frac{7\mu _{p}^{(2)}}{4}\sin \theta _{1}\sin \theta _{2}\sin \delta \sin
\left( \phi _{2}+\psi _{2}\right) ,  \notag \\
\$^{B}(\theta _{1},\theta _{2},\phi _{2},\psi _{2}) &=&c_{1}c_{2}\left(
1+\sin ^{2}\frac{\delta }{2}\right) +s_{1}s_{2}\left( 2-\sin ^{2}\frac{%
\delta }{2}\right)  \notag \\
&&-\frac{3\mu _{p}^{(2)}}{4}\sin \theta _{1}\sin \theta _{2}\sin \delta \sin
\left( \phi _{2}+\psi _{2}\right) ,
\end{eqnarray}%
\begin{eqnarray}
\$^{A}(\theta _{1},\theta _{2}) &=&c_{1}c_{2}\left( 3-3\sin ^{2}\frac{\delta 
}{2}\right) +s_{1}s_{2}\left( 3\sin ^{2}\frac{\delta }{2}\right)
+c_{1}s_{2}\left( 1+3\sin ^{2}\frac{\delta }{2}\right)  \notag \\
&&+c_{2}s_{1}\left( 4-3\sin ^{2}\frac{\delta }{2}\right) ,  \notag \\
\$^{B}(\theta _{1},\theta _{2},\phi _{2},\psi _{2}) &=&c_{1}c_{2}\left(
3-3\sin ^{2}\frac{\delta }{2}\right) +s_{1}s_{2}\left( 3\sin ^{2}\frac{%
\delta }{2}\right) +c_{1}s_{2}\left( 4-3\sin ^{2}\frac{\delta }{2}\right) 
\notag \\
&&+c_{2}s_{1}\left( 1+3\sin ^{2}\frac{\delta }{2}\right) +\frac{3\mu
_{p}^{(2)}}{2}\sin \theta _{1}\sin \theta _{2}\sin \delta \sin \left( \phi
_{2}+\psi _{2}\right) ,  \notag \\
&&
\end{eqnarray}%
and\emph{\ }%
\begin{eqnarray}
\$^{A}(\theta _{1},\theta _{2},\phi _{2},\psi _{2}) &=&c_{1}c_{2}\left(
2-\sin ^{2}\frac{\delta }{2}\right) +s_{1}s_{2}\left( 1+\sin ^{2}\frac{%
\delta }{2}\right)  \notag \\
&&+\frac{3\mu _{p}^{(2)}}{4}\sin \theta _{1}\sin \theta _{2}\sin \delta \sin
\left( \phi _{2}+\psi _{2}\right) ,  \notag \\
\$^{B}(\theta _{1},\theta _{2},\phi _{2},\psi _{2}) &=&c_{1}c_{2}\left(
1+\sin ^{2}\frac{\delta }{2}\right) +s_{1}s_{2}\left( 2-\sin ^{2}\frac{%
\delta }{2}\right)  \notag \\
&&-\frac{3\mu _{p}^{(2)}}{4}\sin \theta _{1}\sin \theta _{2}\sin \delta \sin
\left( \phi _{2}+\psi _{2}\right) ,
\end{eqnarray}%
respectively.\emph{\ }It is evident from the above expressions for the
payoffs that the optimal strategy for Bob, the quantum player, is $\phi
_{2}+\psi _{2}=-\frac{\pi }{2}$ with,\ $\theta _{2}=\frac{\pi }{2},$for
Prisoners' Dilemma%
\index{Prisoners' dilemma!with correlated noise} and Battle of Sexes%
\index{Battle of sexes!with correlated noise}. But corresponding payoff for
Alice is less. However, she can overcome this by playing $\theta _{1}=0$ or $%
\pi ,$\ so that the payoffs for both the players become independent of the
quantum phases $\phi _{2},\psi _{2}$. So there remain no option for the
quantum player to enhance his payoff by exploiting the quantum move. However
in the case of Chicken game%
\index{Chicken game!with correlated noise} the quantum player can enhance
his payoff without effecting the payoff of classical player. But again the
classical player has the ability to prevent quantum strategies by playing $%
\theta _{1}=0$ or $\pi .$ So there remains no advantage for playing quantum
strategies. It is also interesting to note that though by playing this move
Alice could force the payoffs of the two players to be independent of
dephasing factor $p_{2}$\ and the degree of memory%
\index{Memory} $\mu _{2}$, however, the game remains different from its
classical counterpart.

\textbf{Case (iv)} When $\delta =\gamma =%
\frac{\pi }{2},$then Eq. (\ref{payoff correlated}) with $\mu _{1}=\mu
_{2}=0, $ gives the results of Flitney and Abbott \cite{flitney-3} and the
quantum player is better off for $p<1$. However, when decoherence%
\index{Decoherence!-} increases this advantage diminishes and vanishes for
maximum decoherence,\emph{\ }i.e.,\emph{\ }$p=1$. But in our case when $\mu
\neq 0$, the quantum player is always better off even for maximum noise,
i.e., $p=1,$ which was not possible in memoryless case$.$ Furthermore it is
worth noting that as the degree of memory increases from $0$ to $1$ the
effect of noise on the payoffs starts decreasing and for $\mu =1$ it behaves
like a noiseless game.

In the case of Prisoners' Dilemma%
\index{Prisoners' dilemma!with correlated noise}, the optimal strategy for
Bob is to play $\phi _{2}=%
\frac{\pi }{2}$ and $\psi _{2}=0.$ His choice for, $\theta _{2},$ is $\frac{%
\pi }{2}$, independent of Alice's move. The payoffs for Alice and Bob as a
function of decoherence probability $p_{1}=p_{2}=p$ at $\mu =\frac{1}{2},$ is%
\begin{eqnarray}
\$^{A}(\theta _{1},\theta _{2},\phi _{2},\psi _{2}) &=&c_{1}c_{2}\left[
2+\mu _{p}^{2}\cos 2\phi _{2}\right] +s_{1}s_{2}\left[ 2-\mu _{p}^{2}\cos
2\psi _{2}\right]  \notag \\
&&+\frac{5}{2}c_{1}s_{2}\left[ 1-\mu _{p}^{2}\cos 2\psi _{2})\right] +\frac{5%
}{2}c_{2}s_{1}\left[ 1+\mu _{p}^{2}\cos 2\phi _{2}\right]  \notag \\
&&+\frac{\mu _{p}}{4}\sin \theta _{1}\sin \theta _{2}\sin \left( \phi
_{2}-\psi _{2}\right) -\frac{3\mu _{p}}{4}\sin \theta _{1}\sin \theta
_{2}\sin \left( \phi _{2}+\psi _{2}\right) ,  \notag \\
&&
\end{eqnarray}%
\begin{eqnarray}
\$^{B}(\theta _{1},\theta _{2},\phi _{2},\psi _{2}) &=&c_{1}c_{2}\left[
2+\mu _{p}^{2}\cos 2\phi _{2}\right] +s_{1}s_{2}\left[ 2-\mu _{p}^{2}\cos
2\psi _{2}\right]  \notag \\
&&+\frac{5}{2}c_{1}s_{2}\left[ 1+\mu _{p}^{2}\cos 2\psi _{2}\right] +\frac{5%
}{2}c_{2}s_{1}\left[ 1-\mu _{p}^{2}\cos 2\phi _{2}\right]  \notag \\
&&+\frac{7\mu _{p}}{4}\sin \theta _{1}\sin \theta _{2}\sin \left( \phi
_{2}+\psi _{2}\right) +\frac{\mu _{p}}{4}\sin \theta _{1}\sin \theta
_{2}\sin \left( \phi _{2}-\psi _{2}\right) ,  \notag \\
&&
\end{eqnarray}%
where%
\begin{equation*}
\mu _{p}=\frac{1+\left( 1-p\right) ^{2}}{2}.
\end{equation*}%
It is obvious from above payoffs\ that quantum player Bob can always out
perform Alice, for all values of $p$. Similarly for the case of Chicken%
\index{Chicken game!with correlated noise} and Battle of Sexes%
\index{Battle of sexes!with correlated noise} game, it can be proved that
the classical player can be out performed by Bob, at $\phi _{2}=%
\frac{\pi }{2},\psi _{2}=0$\ and $\theta _{2}=\frac{\pi }{2}$ and $\phi
_{2}=-\frac{\pi }{2},\psi =0$ and $\theta _{2}=\frac{\pi }{2},$ respectively$%
.$

\section{Summary}

Quantum games with correlated noise are studied under the generalized
quantization scheme with three parameter set of strategies. Three games,
Prisoners' Dilemma, Battle of Sexes and Chicken game are studied with one
player restricted to classical strategy while other allowed to play quantum
strategies. It is shown that the effects of the memory and decoherence
become effective for the case, $\gamma =\delta =\frac{\pi }{2},$ for which
quantum player out perform classical player. It is also shown that memory
controls payoffs reduction due to decoherence and for the limit of maximum
memory decoherence becomes ineffective.

\chapter{\label{crypto}Quantum Key Distribution}

Cryptography is the science of secret communication. Over the centuries it
developed from the protocols of simple transposition and substitution to
modern cryptographic schemes like one-time pads%
\index{One time pads} and public key cryptosystems%
\index{Public key cryptosystem} \cite{dirk,kahn,rivest}. All of these
schemes rely on a secret key that is shared between the sender and intended
receiver prior to any secure communication between them. However, the
security of the key can never be ensured and if it becomes known then any
one can decrypt the message. In\emph{\ }one-time pads protocol, for example,
the sender and receiver physically exchange the key and store it at some
secure location. In such an exchange the key can be copied either during the
exchange or from the secure location. In public key cryptosystems, such as,
RSA%
\index{RSA}, the receiver generates a pair of keys: a \textit{public key}
and a \textit{private key} \cite{rivest}. The security of the communication
relies on determining the prime factors%
\index{Prime factors} of a large integer. It is generally believed that the
number of steps a classical computer would need to factorize an N decimal
digit, grows exponentially with N. With recent advances in quantum
computing, it is now possible to factorize very large numbers much faster.
As a result the security of RSA will be at risk. This problem can easily be
fixed by quantum cryptography.

Quantum cryptography%
\index{Quantum cryptography!-} offers an entirely new technique for secure
key distribution%
\index{Quantum key distribution} where security relies upon the laws of
quantum physics instead of computational complexity. There are two different
protocols for quantum cryptography: one developed by Bennett \textit{et al}. 
\cite{bennett-0,bennett-01,bennett} which is based on the no-cloning theorem%
\index{No cloning theorem} and the uncertainty principle%
\index{Uncertainty principle}; while the other was presented by Ekert 
\index{Quantum cryptography!Ekert protocol}\cite{ekert} which involves
quantum entanglement%
\index{Entanglement} and the violation of Bell's theorem%
\index{Bell's theorem} \cite{bell}. In this protocol the Bell's inequalities%
\index{Bell's inequality} are used to detect the presence of Eve to ensure
secure key distribution. Quantum cryptography can also be thought of as a
game between the sender and receiver, who want to communicate, and the
eavesdroppers \cite{ekert}.\emph{\ }In this chapter we present a new
protocol for quantum key distribution%
\index{Quantum key distribution!by quantum games} based on quantum game
theory. Here the disturbance in predefined values of the elements of a
decoding matrix (payoff matrix) detects the presence of Eve.

\section{Classical Cryptography}

The important and widely used protocols of classical cryptography are
one-time pads%
\index{One time pads} and RSA public cryptography%
\index{RSA}. Next we introduce them one by one.

\subsection{One-Time Pads}

Despite of being secure it is one of the simplest cryptosystems. It is
rumored that it remained in use for communicating diplomatic information
between Washington and Moscow \cite{welsh}. In this cryptosystem, prior to
any communication, Alice and Bob who are interested in secret communication
meet in a safe place and share a large number of secret keys printed in form
of booklets or pads. The keys are random numbers picked uniformly in the
range $0$ to $l-1$, where $l$ is the number of symbols in the alphabet. Then
they return home with the pad of keys in their possession.

When Alice wants to convey a secret message to Bob she follows the following
steps.

\begin{enumerate}
\item The message $M_{text}$ consisting of $N$\ symbols is converted to a
sequence of $N$\ integers $M=\left\{ m_{1},m_{2},....,m_{N}\right\} $.

\item A key $K=\left\{ k_{1},k_{2},.....,k_{N}\right\} $\ is selected from a
page $P$ of her secret key pad shared with Bob.

\item The message $M$ is encrypted to $E=\left\{
e_{1},e_{2},....,e_{N}\right\} $ using formula 
\begin{equation*}
e_{i}=m_{i}+k_{i}%
\func{mod}l,
\end{equation*}%
where $l$ is the number of symbols in the alphabet.

\item The encrypted message $E$ along with keys pad page $P\ $i.e.\ $\left(
E,P\right) $ is sent to Bob.
\end{enumerate}

Bob on receiving the message performs the following steps.

\begin{enumerate}
\item The message $M=\left\{ m_{1},m_{2},....,m_{N}\right\} $ is decrypted
by taking the key $K=\left\{ k_{1},k_{2},.....,k_{N}\right\} $ used by Alice
from page $P$ of the shared key pads and the relation 
\begin{equation*}
m_{i}=e_{i}-k_{i}+l\func{mod}l.
\end{equation*}

\item The message is converted back to $M_{text}$ from this sequence of
integers.
\end{enumerate}

\subsection{RSA Public Cryptography}

This cryptosystem was invented by Ronald Rivest, Adi Shamir and Leonard
Adleman hence it bears the name RSA cryptosystem \cite{rivest}. In this
protocol a person wishing to receive secret messages creates a pair of keys
known as public key and private key. The public key is publicized but the
private key is kept secret. If somebody is interested in sending secret
message he takes the public key of the intended recipient to encrypt the
message. Upon receiving the scrambled message the receiver decrypts it with
the help of his private key. In the following we give the mathematical
details required to know how the keys are generated and how the messages are
encrypted and decrypted.

In order to generate the public and private keys the receiver takes two
large prime numbers $p,q$ and finds their product $n=pq.$ Next he finds an
integer $d$ that is coprime to $(1-p)(1-q),$ and computes $e$ with the help
of the relation 
\begin{equation}
ed\equiv 1\func{mod}(1-p)(1-q).  \label{public key}
\end{equation}%
The public key to be broadcasted is a pair of numbers $(e,n)$ and the
private key is the pair of numbers $(d,n).$ The interested party in sending
secret messages converts the text $M_{text}$ to a sequence of integers, $%
M_{i}$ and then encrypts it by the use of the following formula 
\begin{equation}
E_{i}=\left( M_{i}\right) ^{e}\func{mod}n.
\end{equation}%
On reception the receiver decrypts this message using 
\begin{equation}
M_{i}=\left( E_{i}\right) ^{d}\func{mod}n,
\end{equation}%
and then converts it back to original text.

To break the code in RSA\ cryptosystem one requires the private key, $(d,n)$
with the help of public key, $(e,n).$ That can be accomplished very easily
with the help of equation (\ref{public key}) subject to the condition if one
can find the prime factors of\ $n.$i.e. the prime numbers $p$ and $q$. But
it is believed that performing prime factorization of a very large number is
difficult by any classical computer. Therefore, RSA is secure. However with
the advent of quantum computer it will not remain difficult to find prime
factors of large numbers hence RSA cryptosystem will not remain secure. At
that time we will need quantum cryptography.

\section{Quantum Cryptography}

In the next we explain two simple protocols that utilize two different
quantum phenomenon to protect the secret information from being tampered.

\subsection{BB84 Protocol}

This protocol%
\index{Quantum cryptography!BB84 protocol} was introduced by Charles H.
Bennett and Gilles Brassard in 1984 hence it bears the name BB84 \cite%
{bennett-0}. Security in this protocol relies on the inability to measure
non-orthogonal quantum states perfectly. The task is accomplished by coding
logical bit 0 into two different non-orthogonal quantum states and similarly
1 into two other non-orthogonal states such that the encoding states for 0
and 1 are pairwise orthogonal.

Let the information be encoded into polarization states of individual
photons such that

\begin{eqnarray}
0 &\longrightarrow &\left\{ 
\begin{array}{c}
\left| H\right\rangle \\ 
\left| A\right\rangle =%
\frac{\left| H\right\rangle +\left| V\right\rangle }{\sqrt{2}}%
\end{array}%
\right. ,  \notag \\
1 &\longrightarrow &\left\{ 
\begin{array}{c}
\left| V\right\rangle \\ 
\left| D\right\rangle =\frac{\left| H\right\rangle -\left| V\right\rangle }{%
\sqrt{2}}%
\end{array}%
\right. ,  \label{coding}
\end{eqnarray}%
where $\left| H\right\rangle $ and $\left| V\right\rangle $ represent the
horizontal and vertical polarizations of a photon respectively. The states $%
\left| H\right\rangle $ and $\left| V\right\rangle $ are also termed as
rectilinear bases where as the states $\left| A\right\rangle $ and $\left|
D\right\rangle $ are called diagonal bases.

It can be seen from Eq. (\ref{coding}) that the four polarization states are
pairwise orthogonal i.e. 
\begin{equation}
\left\langle V\right. \left| H\right\rangle =\left\langle A\right. \left|
D\right\rangle =0.
\end{equation}%
Furthermore if the measurement is performed in bases identical to the bases
in which a photon is prepared it gives deterministic results otherwise
random outcomes are achieved i.e. 
\begin{eqnarray*}
\left\langle H\right. \left| H\right\rangle &=&\left\langle V\right. \left|
V\right\rangle =\left\langle A\right. \left| A\right\rangle =\left\langle
D\right. \left| D\right\rangle =1, \\
\left| \left\langle H\right. \left| A\right\rangle \right| ^{2} &=&\left|
\left\langle H\right. \left| D\right\rangle \right| ^{2}=\left| \left\langle
V\right. \left| A\right\rangle \right| ^{2}=\left| \left\langle V\right.
\left| D\right\rangle \right| ^{2}=\frac{1}{2}.
\end{eqnarray*}%
Prior to any key distribution Alice and Bob agree that $\left|
H\right\rangle $ and $\left| A\right\rangle $ stand for bit value $0$
whereas $\left| V\right\rangle $ and $\left| D\right\rangle $ stand for $1$.
Then the sender, Alice generates a sequence of random numbers and encodes
them using the predefined four polarization states. The polarization states
for coding are used randomly and independently. Upon receiving the photons
Bob performs the measurements using the rectilinear or diagonal bases
randomly and independently of Alice. Statistically their bases match in
about 50\% cases which gives Bob deterministic results. Then they contact on
a public channel and tell each other which bases they have used. Whenever
their bases coincide they record the results otherwise they discard it. In
case there is no eavesdropper in the channel then Bob receives the same bit
that Alice has transmitted.

If there is an eavesdropping, Eve in the way from Alice to Bob who performs
measurements in the bases similar to Bob to see what bits are being sent.
She intercepts the photon chooses the bases randomly and performs the
measurement to decode the bit. In order to remain hidden from Alice and Bob
sight Eve will transmit the photon in same polarization state in which she
received it. In this situation she will make an error for quarter of the
time. But in this scenario for the cases where the bases of Alice and Bob
match the results of Alice and Bob will not be correlated which uncovers the
presence of Eve so they abort communication.

\subsection{Ekert Protocol}

This protocol was presented by Artur Ekert \cite{ekert-1} in 1991%
\index{Quantum cryptography!Ekert protocol}. It works as follows: Alice and
Bob share a large number of two qubit entangled states of the form 
\begin{equation}
\left| \psi \right\rangle =%
\frac{\left| 01\right\rangle -\left| 10\right\rangle }{\sqrt{2}}.
\end{equation}%
In order to share secret key Alice and Bob perform measurements on their
respective qubits. Let they, for example, perform measurement on their
qubits at angles of $\left\{ 0^{\circ },45^{\circ },90^{\circ }\right\} $
and $\left\{ 45^{\circ },90^{\circ },135^{\circ }\right\} $ respectively in
a plane perpendicular to axis connecting them. The sequence of these
measurements are performed in complete random way. However for the choice of
same orientations of their detector and in case there is no eavesdropping in
the way then their results are always totally anticorrelated. On the other
hand when the orientations of their detectors do not match then quantum
mechanics tells the way for the calculation of correlation coefficient. In
this case if $P_{\pm \mp }\left( \theta _{i}^{a},\theta _{j}^{b}\right) $ is
the probability of getting spin up (+1) and spin down (-1) then the
correlation coefficient $E\left( \theta _{i}^{a},\theta _{j}^{b}\right) $\
is found as 
\begin{equation}
E\left( \theta _{i}^{a},\theta _{j}^{b}\right) =P_{++}\left( \theta
_{i}^{a},\theta _{j}^{b}\right) +P_{--}\left( \theta _{i}^{a},\theta
_{j}^{b}\right) -P_{+-}\left( \theta _{i}^{a},\theta _{j}^{b}\right)
-P_{-+}\left( \theta _{i}^{a},\theta _{j}^{b}\right)
\label{correlation function}
\end{equation}%
With the help of Eq. (\ref{correlation function}) we can find a function 
\begin{equation}
S=E\left( \theta _{1}^{a},\theta _{3}^{b}\right) +E\left( \theta
_{1}^{a},\theta _{2}^{b}\right) +E\left( \theta _{2}^{a},\theta
_{3}^{b}\right) -E\left( \theta _{2}^{a},\theta _{2}^{b}\right)
\end{equation}%
This function $S$ was proposed by Clauser, Horne, Shimony and Holt for the
generalized Bell theorem, known as CHSH\ inequality \cite{chsh}. Quantum
mechanics demands that 
\begin{equation}
S=-2\sqrt{2}
\end{equation}%
After the measurements have been performed then Alice and Bob contact on a
public channel to know what orientations of their detectors they have used.
They divide these results into two groups. The first group corresponds to
the results where they used different orientations of their detectors and
the second group belongs the results where they performed the measurement in
matching orientations of their detectors. For the first group they announce
publicly what results they obtained. Since the orientations of detectors was
selected randomly and independently therefore, the correlation between their
results according to the principles of quantum mechanics and in the absence
of eavesdropping should come out to be $-2\sqrt{2}$ \cite{chsh}. This
assures the legitimate users that the results they obtained in the second
group, where they selected the same orientations of the detectors, are
totally anticorrelated and can be converted into a secret key. However if
Alice and Bob find a significant departure from the expected correlations,
that quantum mechanics demands, it indicates the presence of eavesdropper.
So they will have to abort the communication. Experimental demonstration of
this protocol has been accomplished by Rariry \textit{et al}.\ in 1994 \cite%
{rarity}.

In the next section we present a new scheme for Quantum Key Distribution
(QKD) based on the mathematical frame work of our Generalized Quantization
Scheme for Games.

\section{A New Scheme for Quantum Key Distribution}

This new scheme for Quantum Key Distribution (QKD) is derived from the
quantum game theoretic setup. In fact it is not a quantum game anymore as
there is no strategic competition between the two players, Alice and Bob.
Alice and Bob shares multiple copies of maximally entangled states. As a
first step, Alice and Bob identify a set of unitary operators to code
various symbols that needs to be communicated. Then Bob simulates all
mutually agreed upon operators of Alice against his choices, at his end and
evaluates the expectation values of all decoding operators and constructs a
decoding bi-matrix. Bob chooses his operators with the consideration to
avoid any overlap among the various expectation values of bi-matrix
elements. The expectation values depends on the operators used by Alice and
Bob. This decoding bi-matrix will later be used to identify the operator
used by Alice.

In the second set of the scheme. Alice applies a local unitary operator on
her part of the shared entangled qubit and pass it on to Bob. Bob applies
his local unitary operator on his part of the qubit. On receiving the
Alice's part of the qubit, Bob evaluates the expectation values of the two
decoding operators and compares the pair of expectation values with the
already simulated bi-matrix elements. The comparison would reveal the
unitary operator used by Alice. Repeating this process Alice and Bob would
be able to share a string of bits which could act as key for secret
communication.

In this scheme, it is possible for Eve to perform a measurement while
transmission of the Alice's part of the qubit to Bob. However, any such
attempt even on some copies of the maximally entangled qubits shared between
Alice and Bob would result in the change in the expectation values. A
careful choice of the two decoding operators would enable Bob to detect the
presence of Eve. Presence of Eve could be communicated back to Alice via any
classical channel to ignore that particular attempt.

One peculiar feature of our protocol is the evaluation of expectation
values. This requires multiple copies of the qubit to communicate a required
symbol. This in principle could effect the efficiency and security of our
protocol.\textrm{\ }Here we show the number of copies required to
communicate a symbol with sure detection of presence of Eve.

Consider a situation where Alice intends to send a symbol $m_{1}$\ to Bob
with whom she shares a maximally entangled state of the form (\ref{Bell
states-a}). She applies one of the mutually agreed unitary operator, $%
U_{A}\left( \theta _{A},\alpha _{A},\beta _{A}\right) =I$\ on her part of
qubit and sends her part to Bob.\ Lets assume there is Eve who tries to read
the transmitted symbol by performing a measurement on Alice's part of the
qubit, in the computational basis\ $\left| 0\right\rangle $, $\left|
1\right\rangle $.\ Eve knows the unitary operators used by Alice as these
operators were mutually agreed by Alice and Bob prior to any key generation.
But she is unaware of the operators used by Bob and the decoding bi-matrix
used by Bob to determine the unitary operator applied by Alice. Upon
measurement she would get either $\left| 0\right\rangle $\ or $\left|
1\right\rangle $\ with equal probability. In order to remain hidden from the
scene Eve, depending on her measurement results, would send either $\left|
0\right\rangle $\ or $\left| 1\right\rangle $\ to Bob.\ If, for example, on
receiving the qubits, Bob decides to apply the identity operator $I$\ on his
part of the entangled qubit, then the final state in his possession would be
either $\left| 00\right\rangle $\ or $\left| 11\right\rangle $\ with equal
probability. Let the expectation value or bi-matrix elements associated with
the state $\left| 00\right\rangle $\ and $\left| 11\right\rangle $\ be $%
\left( a,b\right) $\ and $\left( c,d\right) ,$\ respectively.\ According to
the requirement of protocol Alice would have to send $n$\ copies of qubits
to transmit the symbol $m_{1}$. If Eve succeeds in intercepting $i$\ copies
out of total $n$\ copies then the bi-matrix element $\left( a,b\right) $%
\thinspace associated with the state $\left| 00\right\rangle $\ becomes%
\begin{equation}
\left( \alpha _{i},\beta _{i}\right) =\left( \frac{a\left( n-i\right) +ci}{n}%
,\frac{b\left( n-i\right) +di}{n}\right)  \label{averages}
\end{equation}%
and the expectation value with including error introduced by Eve would
become:\textrm{\ }%
\begin{eqnarray}
f\left( n\right) &=&\left( \underset{i=0}{\overset{n}{\sum }}P_{i}\alpha
_{i},\underset{i=0}{\overset{n}{\sum }}P_{i}\beta _{i}\right)  \notag \\
&=&\left( \frac{a+c}{2},\frac{b+d}{2}\right)  \label{average or errors}
\end{eqnarray}%
\textrm{\ }where 
\begin{equation}
P_{i}=\frac{1}{2^{n}}\left( 
\begin{array}{c}
n \\ 
i%
\end{array}%
\right)  \label{binomial}
\end{equation}%
is the probability for binomial distribution.\ It is interesting to note
that the expression (\ref{average or errors}) is independent of $n,$\ the
number of copies\textrm{. }Comparing the values given in Eq. (\ref{averages}%
) and those already obtained through simulation, Bob can detect the presence
of Eve.\textrm{\ }Now to estimate of the number of copies required to
reliably detect the presence of Eve we take the help of standard deviation.
By the use of Eq. (\ref{average or errors}) the standard deviation comes out
to be 
\begin{eqnarray}
\left( \sigma _{1},\sigma _{2}\right) &=&\left( \sqrt{\underset{i=0}{\overset%
{n}{\sum }}P_{i}\alpha _{i}^{2}-\left( \underset{i=0}{\overset{n}{\sum }}%
P_{i}\alpha _{i}\right) ^{2}},\sqrt{\underset{i=0}{\overset{n}{\sum }}%
P_{i}\beta _{i}^{2}-\left( \underset{i=0}{\overset{n}{\sum }}P_{i}\beta
_{i}\right) ^{2}}\right)  \notag \\
&=&\left( \frac{\left( a-c\right) }{2\sqrt{n}},\frac{\left( b-d\right) }{2%
\sqrt{n}}\right) ,  \label{std deviation}
\end{eqnarray}%
which is inversely proportional to $\sqrt{n}.$ Standard deviation can be
reduced by increasing the number of copies.

\subsection{The Description of Protocol}

Let Alice and Bob share the initial quantum states of the form of Eq. (\ref%
{state in}). The local unitary operators of Alice and Bob derived from our
generalized quantization scheme 
\index{Quantization scheme!generalized}\emph{\ }are represented by Eq. (\ref%
{combination}) with $R$ and $P$ defined as:

\begin{align}
R_{A}\left| 0\right\rangle & =e^{i\alpha _{A}}\left| 0\right\rangle ,%
\text{ \ }\ \ \ \ \ \ \ \ \ \ \ \text{\ }R_{A}\left| 1\right\rangle
=e^{-i\alpha _{A}}\left| 1\right\rangle ,  \notag \\
P_{A}\left| 0\right\rangle & =e^{i\left( \frac{\pi }{2}-\beta _{A}\right)
}\left| 1\right\rangle ,\text{ \ \ \ \ \ \ }P_{A}\left| 1\right\rangle
=e^{i\left( \frac{\pi }{2}+\beta _{A}\right) }\left| 0\right\rangle ,  \notag
\\
R_{B}\left| 0\right\rangle & =\left| 0\right\rangle ,\text{ \ }\ \ \ \ \ \ \
\ \ \ \ \ \ \ \ \ \text{\ }R_{B}\left| 1\right\rangle =\left| 1\right\rangle
,  \notag \\
P_{B}\left| 0\right\rangle & =\left| 1\right\rangle ,\text{ \ \ \ \ }\ \ \ \
\ \ \ \ \ \ \ \ \text{\ }P_{B}\left| 1\right\rangle =-\left| 0\right\rangle ,
\label{operators crypto}
\end{align}%
where $-\pi \leq \alpha _{A},\beta _{A}\leq \pi .$\emph{\ }The operators
used by Bob for the measurement are\emph{\ }%
\begin{equation}
P^{k}=\$_{00}^{k}P_{00}+\$_{01}^{k}P_{01}+\$_{10}^{k}P_{10}+%
\$_{11}^{k}P_{11},  \label{payoff operator crypto}
\end{equation}%
where $k=A,B$ and for $m,n=0,1$ the operators $P_{mn}=\left| \psi
_{mn}\right\rangle \left\langle \psi _{mn}\right| $ are given in Eqs. (\ref%
{oper a}) with\ $\delta \in \left[ 0,\frac{\pi }{2}\right] $\ and $%
\$_{ij}^{k}$\ are the elements of coding matrix in $ith$ row and $jth$
column.\textrm{\ }It is also important to note that Bob chooses the coding
matrix on his own will without being known to Alice. Therefore, it will be
difficult for Eve to construct the decoding operators of the form of Eq. (%
\ref{payoff operator crypto}).The rationale in choosing this coding matrix
is to avoid or to reduce the overlap of the expectation values of the
decoding operators. The results of measurements performed by Bob are
recorded as\emph{\ }%
\begin{equation}
\$_{k}(\theta _{i},\alpha _{A},\beta _{A})=\text{Tr}(P^{k}\rho _{f})\text{,}
\label{payoff formula crypto}
\end{equation}%
\emph{\ } where Tr represents the trace of a\emph{\ }matrix .

The presence of Eve can be modeled as a phase damping channel%
\index{Quantum channel!phase damping} \cite{chuang,flitney-3}.\ The quantum
state after the Eve's measurement transforms to\emph{\ }%
\begin{equation}
\rho =\overset{2}{\underset{i=0}{\tsum }}A_{i%
\text{ }}\rho _{in}\text{ }A_{i}^{\dagger },  \label{kraus1}
\end{equation}%
where $A_{0}=\sqrt{p}\left\vert 0\right\rangle \left\langle 0\right\vert ,$ $%
A_{1}=\sqrt{p}\left\vert 1\right\rangle \left\langle 1\right\vert $ and $%
A_{2}=\sqrt{1-p}\hat{I}$ are the Kraus operators%
\index{Kraus operators}. Using Eqs. (\ref{state in}), (\ref{payoff operator
crypto}), (\ref{payoff formula crypto}) and (\ref{kraus}) we get 
\begin{align}
\$_{k}(\theta _{i},\alpha _{A},\beta _{A})& =c_{1}c_{2}\left[ \eta
\$_{00}^{k}+\chi \$_{11}^{k}+\left( \$_{00}^{k}-\$_{11}^{k}\right) \mu
_{p}\xi \cos 2\alpha _{A}\right]  \notag \\
& +s_{1}s_{2}\left[ \eta \$_{11}^{k}+\chi \$_{00}^{k}-\left(
\$_{00}^{k}-\$_{11}^{k}\right) \mu _{p}\xi \cos 2\beta _{A}\right]  \notag \\
& +c_{1}s_{2}\left[ \eta \$_{01}^{k}+\chi \$_{10}^{k}+\left(
\$_{01}^{k}-\$_{10}^{k}\right) \mu _{p}\xi \cos 2\alpha _{A}\right]  \notag
\\
& +c_{2}s_{1}\left[ \eta \$_{10}^{k}+\chi \$_{01}^{k}-\left(
\$_{01}^{k}-\$_{10}^{k}\right) \mu _{p}\xi \cos 2\beta _{A}\right]  \notag \\
& +\left( 
\frac{\$_{00}^{k}-\$_{11}^{k}+\$_{10}^{k}-\$_{01}^{k}}{4}\right) \mu
_{p}\sin \theta _{1}\sin \theta _{2}\sin \delta \sin \left( \alpha
_{A}+\beta _{A}\right)  \notag \\
& +\left( \frac{-\$_{00}^{k}-\$_{11}^{k}+\$_{01}^{k}+\$_{10}^{k}}{4}\right)
\sin \theta _{1}\sin \theta _{2}\sin \gamma \sin \left( \alpha _{A}-\beta
_{A}\right) ,  \notag \\
&  \label{payoff crypto}
\end{align}%
where we have defined: 
\begin{subequations}
\label{abrs}
\begin{eqnarray*}
\eta &=&\cos ^{2}\left( \delta /2\right) \cos ^{2}\left( \gamma /2\right)
+\sin ^{2}\left( \delta /2\right) \sin ^{2}\left( \gamma /2\right) , \\
\chi &=&\cos ^{2}\left( \delta /2\right) \sin ^{2}\frac{\gamma }{2}+\sin
^{2}\left( \delta /2\right) \cos ^{2}\left( \gamma /2\right) , \\
\xi &=&1/2\left( \sin \delta \sin \gamma \right) ,\text{ }c_{i}=\cos ^{2}%
\frac{\theta _{i}}{2}, \\
s_{i} &=&\sin ^{2}\frac{\theta _{i}}{2},\text{ }\mu _{p}=1-p.
\end{eqnarray*}%
The elements of the bi-matrix can be found by putting the appropriate values
for $\$_{ij}^{k}$ (elements of the coding matrix) in Eq. (\ref{payoff crypto}%
).

Alice, the sender, in our protocol, applies unitary operators $U_{A}\left(
\theta _{A},\alpha _{A},\beta _{A}\right) $, whereas Bob, the intended
receiver applies the\emph{\ }unitary operator $U_{B}\left( \theta
_{B}\right) ,$\ see Eq. (\ref{combination}). Prior to any key distribution
Alice and Bob agree on exact form of the unitary to be used by Alice by
fixing values of the set $\left( \theta _{A},\alpha _{A},\beta _{A}\right) $%
\ which may stands for four symbols $m_{1},m_{2},m_{3},m_{4}$. On the other
hand Bob applies his unitary for $\theta _{B}=0$ or $\pi .$ These choices
help Bob in forming a well defined decoding bi-matrix. Bob has the option to
apply two or more unitary local operators according to his own will.

Let Alice and Bob agree on the following four unitaries for the four symbols
to be sent by Alice 
\end{subequations}
\begin{eqnarray*}
U_{A}\left( 0,0,0\right) &\Rightarrow &m_{1}, \\
U_{A}\left( \frac{\pi }{3},\frac{\pi }{2},\frac{\pi }{2}\right) &\Rightarrow
&m_{2}, \\
U_{A}\left( \frac{\pi }{2},\frac{\pi }{2},\frac{\pi }{2}\right) &\Rightarrow
&m_{3}, \\
U_{A}\left( \pi ,\pi ,\pi \right) &\Rightarrow &m_{4}.
\end{eqnarray*}%
Now if Bob chooses $\$_{00}^{A}=\$_{00}^{B}=3,$ $\$_{01}^{A}=\$_{10}^{B}=0,$ 
$\$_{10}^{A}=\$_{01}^{B}=5,$ $\$_{11}^{A}=\$_{11}^{B}=1$\emph{\ }in Eq. (\ref%
{payoff crypto}) and apply $U_{B}\left( 0\right) $ and $U_{B}\left( \pi
\right) $ on his part of the qubit and then simulate all unitaries allowed
for Alice, at his end, he would get the following decoding bi-matrx: 
\begin{equation}
\overset{}{%
\begin{tabular}{l}
$m_{1}$ \\ 
$m_{2}$ \\ 
$m_{3}$ \\ 
$m_{4}$%
\end{tabular}%
\overset{%
\begin{tabular}{ll}
$U_{B}\left( 0\right) $ & $U_{B}\left( \pi \right) $%
\end{tabular}%
}{\overset{}{\left[ 
\begin{tabular}{ll}
$\left( 3,3\right) $ & $\left( 0,5\right) $ \\ 
$\left( \frac{3}{4},2\right) $ & $\left( \frac{9}{2},\frac{3}{4}\right) $ \\ 
$\left( \frac{1}{2},3\right) $ & $\left( 4,\frac{3}{2}\right) $ \\ 
$\left( 5,0\right) $ & $\left( 1,1\right) $%
\end{tabular}%
\right] }}}  \label{no-eavesdropping}
\end{equation}%
Now in the second part after applying local operators on her qubit, Alice
sends her qubit to Bob, who applies one of his local unitary operator,
according to his own will, and then calculates the expectation value of
coding operators (\ref{payoff operator crypto}). These measured values are
compared with the elements of the decoding bi-matrix (\ref{no-eavesdropping}%
). Since Bob is well aware of his own action therefore he will have to
compare only one column of the matrix (\ref{no-eavesdropping}). By doing
this he can very easily find the unitary operator applied by Alice and hence
he find the corresponding secret key element that Alice wants to transmit%
\textrm{.}\emph{\ }Repeating this process a secret key is transferred to
Bob. It is interesting to note that Alice has the option of transmitting
four different symbols $m_{1},m_{2},m_{3},m_{4}$\ for key formation while
using only a two dimensional quantum system.

Whenever there is an eavesdropper in the way and performs measurement on the
qubits then the decoding matrix from Eq. (\ref{payoff crypto}) takes the
form 
\begin{equation}
\overset{%
\begin{tabular}{llllll}
$U_{B}\left( 0\right) $ &  &  &  &  & $U_{B}\left( \pi \right) $%
\end{tabular}%
}{%
\begin{tabular}{l}
$m_{1}$ \\ 
$m_{2}$ \\ 
$m_{3}$ \\ 
$m_{4}$%
\end{tabular}%
\overset{}{\left[ 
\begin{tabular}{ll}
$\left( 3-p,3-p\right) $ & $\left( \frac{5}{2}p,5-\frac{5}{2}p\right) $ \\ 
$\left( \frac{3}{4}+\frac{11}{8}p,2+\frac{1}{8}p\right) $ & $\left( \frac{9}{%
2}-\frac{17}{8}p,\frac{3}{4}+\frac{13}{8}p\right) $ \\ 
$\left( \frac{1}{2}+\frac{7}{4}p,3-\frac{3}{4}p\right) $ & $\left( 4-\frac{7%
}{4}p,\frac{3}{2}+\frac{3}{4}p\right) $ \\ 
$\left( 5-\frac{5}{2}p,\frac{5}{2}p\right) $ & $\left( 1+p,1+p\right) $%
\end{tabular}%
\right] }}  \label{eavesdropper}
\end{equation}%
It is clear that the decoding\ matrix (\ref{eavesdropper}) is different from
decoding \ matrix (\ref{no-eavesdropping}) for all values of $p>0$. When for
any action (unitary operator) of Alice, Bob performs the measurement using
operators (\ref{payoff operator crypto})\ he finds that the measured
elements are different than the elements of the reference matrix (\ref%
{no-eavesdropping}) he already has in his library. It informs him about the
activity of eavesdropping and they abort communication. It is important to
note that no term in any column of matrix (\ref{eavesdropper}) for a given
unitary operator of Alice can create a value matching some other term in
matrix (\ref{no-eavesdropping}). Thus a little chance exists for the
activity of eavesdropping to give ambiguous results and hence to remain
secret from Bob. One of the most common eavesdropping strategy is catch
resend attack%
\index{Catch resend attack}. In this attack if Eves succeeds in finding the
bit she resends a similar bit to Bob. But in our case if it so happens then
the correlation between Alice and Bob will break and the elements of
decoding matrix (payoff matrix) will change that reveals eavesdropping. Let,
for example, Alice applies unitary operator $U_{A}\left( 0,0,0\right) $\ on
her qubit and sends it to Bob. During transmission Eve performs a
measurement on the qubit and gets either $0$\ or $1$. On the basis of her
measurement results Eve sends either $\left| 0\right\rangle $\ or $\left|
1\right\rangle $\ to Bob. If Bob applies $U_{B}\left( 0\right) $\ before
measurement then the final state received by him is $\left| 00\right\rangle $%
\ or $\left| 11\right\rangle $\ with $50\%$\ probability. If Alice sends $n$%
\ copies in order to transmit $U_{A}\left( 0,0,0\right) $\ and Eve
interrupts $i$\ of them. Then with the help of Eq. (\ref{averages}) matrix (%
\ref{no-eavesdropping}) become 
\begin{equation}
\left( \alpha _{i},\beta _{i}\right) =\left( 
\frac{3\left( n-i\right) +5i}{n},\frac{3\left( n-i\right) }{n}\right)
\end{equation}%
and by Eq. (\ref{average or errors})\ we get 
\begin{equation}
f\left( n\right) =\left( 4,\frac{3}{2}\right) ,  \label{element}
\end{equation}%
which is independent of $n,$\ the number of copies. Furthermore it to be
noted that the matrix element given in Eq. (\ref{element}) is not an element
of the bi-matrix (\ref{no-eavesdropping}) for $U_{B}\left( 0\right) $. Hence
presence of Eve no more remains hidden. Now the question that how much
resources i.e. number of copies of input state, Bob will require for Eve's
detection. Using Eq. (\ref{std deviation}) and matrix (\ref{no-eavesdropping}%
) we get%
\begin{equation}
\left( \sigma _{1},\sigma _{2}\right) =\left( \frac{1}{\sqrt{n}},\frac{1.5}{%
\sqrt{n}}\right)
\end{equation}%
Therefore for this case nine to ten copies will be sufficient for Eve's
detection.

\section{Summary}

In summary we devised a quantum key distribution protocol that based on the
mathematical set up of quantum game theory. It is also interesting to note
that we can send four symbols while using only a two dimensional system that
is not possible in other quantum key distribution protocols. It is to be
added that quantum games have experimental realization \cite{du-2,zhou}
therefore, this technique is not beyond the reach of today's technology.

\chapter{\label{QST}Quantum State Tomography}

All information about a quantum system is contained in the state of a
system. However the state is not an observable in quantum mechanics \cite%
{peres} therefore, it is not possible to perform all measurements on the
single state to extract the whole information about the system. No-cloning
theorem%
\index{No cloning theorem} does not allow to create a perfect copy of the
system without prior knowledge about its state \cite{wootters}. Hence, there
remains no way, even in principle, to infer the unknown quantum state of a
single system \cite{ariano}. However it is possible to estimate the unknown
quantum state of a system when many identical copies of the system are
available. This procedure of reconstructing an unknown quantum state through
a series of measurements on a number of identical copies of the system is
called quantum state tomography%
\index{Quantum state tomography}. In this process each measurement gives a
new dimension of the system and therefore, infinite number of copies are
required to reconstruct the exact state of a quantum system.

Some of the main tasks of quantum information theory are to study the
effects of decoherence \cite{white}, to optimize the performance of quantum
gates \cite{brien}, to quantify the amount of information that various
parties can obtain by quantum communication protocols \cite{langford} and
utilization of quantum error correction protocols in real world situations
effectively \cite{altepeter}. In all these cases a complete characterization
of the quantum state is required \cite{adamson}. For which quantum state
tomography is one of the best tools. The problem of quantum state tomography
was first addressed by Fano \cite{fano} who recognized the need to measure
two non commuting observables. However it remained mere speculation until
original proposal for quantum tomography and its experimental verification 
\cite{ariano,vegal,raymer}. Since than it being applied successfully\ to the
measurement of photon statistics of a semiconductor laser \cite{munroe},
reconstruction of density matrix of squeezed vacuum \cite{schiller} and
probing the entangled states of light and ions \cite{paris}\textrm{.}

In this chapter by making use of the mathematical framework of generalized
quantization scheme a technique for quantum state tomography%
\index{Quantum state tomography!by quantum games} is developed.\textrm{\
Strictly speaking this arrangement is not a game but the mathematical setup
of quantum games is used as a tool} \ \textrm{It works as follows: Alice
sends an unknown pure quantum state }$\rho $\textrm{\ to Bob who appends it
with }$\left| 0\right\rangle \left\langle 0\right| $\textrm{\ that results
the initial state }$\rho _{in}=\left| 0\right\rangle \left\langle 0\right|
\otimes \rho .$\textrm{\ Bob applies unitary operator }$U=U_{A}\otimes U_{B}$%
\textrm{\ on the appended quantum state and finds the expectation values of
payoff operators }$P^{A}$\textrm{\ and }$P^{B}$\textrm{.\ The results are
recorded in the form of a bi matrix (payoff matrix) having elements }$\left(
\$_{A},\$_{B}\right) $\textrm{.\ It is observed that for a particular set of
unitary operators (strategies) and for a certain game these elements become
equal to to Stokes parameters (see subsection \ref{stokes}) of the given
quantum state }$\rho $\textrm{. In this way an unknown quantum state can be
measured and reconstructed. It means that finding the payoffs is measurement
of input quantum state.}

\section{Quantization Scheme for Games and Quantum State Tomography}

Let Alice forwards an unknown quantum state of the form of Eq. (\ref{initial
state}) to Bob who appends it to 
\begin{equation}
\rho _{in}=\left| 0\right\rangle \left\langle 0\right| \otimes \rho
\end{equation}%
where $\rho =\left| \psi \right\rangle \left\langle \psi \right| $ and then
applies the unitary operators 
\begin{equation}
U_{k}=\cos 
\frac{\theta _{k}}{2}R_{k}+\sin \frac{\theta _{k}}{2}P_{k}
\end{equation}%
where $R_{k}$, $P_{k}$\emph{\ }are defined as:

\begin{align}
R_{k}\left| 0\right\rangle & =e^{i\alpha _{k}}\left| 0\right\rangle ,\text{
\ \ }R_{k}\left| 1\right\rangle =e^{-i\alpha _{k}}\left| 1\right\rangle , 
\notag \\
P_{k}\left| 0\right\rangle & =-\left| 1\right\rangle ,\text{ \ \ \ \ \ }%
P_{k}\left| 1\right\rangle =\left| 0\right\rangle .  \label{operators}
\end{align}%
with $0\leq \theta \leq \pi $\emph{\ }and$\ k=A$, $B.$ \textrm{After the
application of the operator }$U=\left( U_{A}\otimes U_{B}\right) $\textrm{\
the final state becomes} 
\begin{equation}
\rho _{f}=\left( U_{A}\otimes U_{B}\right) \rho _{in}\left( U_{A}\otimes
U_{B}\right) ^{\dagger }.  \label{final-state}
\end{equation}%
The operators used \textrm{by Bob to} perform the measurement are 
\begin{eqnarray}
P_{00} &=&\left| 00\right\rangle \left\langle 00\right| ,P_{01}=\left|
01\right\rangle \left\langle 01\right| ,  \notag \\
P_{10} &=&\left| 10\right\rangle \left\langle 10\right| ,P_{11}=\left|
11\right\rangle \left\langle 11\right|
\end{eqnarray}%
so that the payoff operators for Alice and Bob become

\begin{equation}
P^{k}=\$_{00}^{k}P_{00}+\$_{01}^{k}P_{01}+\$_{10}^{k}P_{10}+%
\$_{11}^{k}P_{11},  \label{payoff operator}
\end{equation}%
where $\$_{ij}^{k}$ are the entries of payoff matrix in $ith$ row and $jth$
column for player $k$. In our generalized quantization scheme (chap. \ref%
{general}), payoffs for the players are calculated as 
\begin{equation}
\$^{k}(\theta _{i},\alpha _{i},\theta ,\phi )=\text{Tr}(P^{k}\rho _{f})\text{%
,}  \label{payoff formula}
\end{equation}%
where Tr represents the trace of a\emph{\ }matrix, $k=A,B$ and $i=A,B$.
Using Eqs. (\ref{initial-state}), (\ref{payoff operator}) and (\ref{payoff
formula}) the payoffs come out to be 
\begin{align}
\$^{k}(\theta _{i},\alpha _{i},\theta ,\phi )& =\left( \$_{00}^{k}\chi
+\$_{11}^{k}\Omega +\$_{01}^{k}\xi +\$_{10}^{k}\eta \right) \cos ^{2}\frac{%
\theta }{2}+\left( \$_{00}^{k}\xi +\$_{11}^{k}\eta +\$_{01}^{k}\chi \right. +
\notag \\
& \left. \$_{10}^{k}\Omega \right) \sin ^{2}\frac{\theta }{2}+\left[ \left\{
\left( \$_{00}^{k}-\$_{01}^{k}\right) \beta +\left(
\$_{10}^{k}-\$_{11}^{k}\right) \Theta \right\} \cos \alpha _{2}\right] \sin
\theta \cos \phi +  \notag \\
& \left[ \left\{ \left( \$_{00}^{k}-\$_{01}^{k}\right) \beta +\left(
\$_{10}^{k}-\$_{11}^{k}\right) \Theta \right\} \sin \alpha _{2}\right] \sin
\theta \sin \phi ,  \label{payoff}
\end{align}%
where 
\begin{eqnarray}
\chi &=&\cos ^{2}\frac{\theta _{A}}{2}\cos ^{2}\frac{\theta _{B}}{2},\text{
\ }\xi =\cos ^{2}\frac{\theta _{A}}{2}\sin ^{2}\frac{\theta _{B}}{2},  \notag
\\
\Omega &=&\sin ^{2}\frac{\theta _{A}}{2}\sin ^{2}\frac{\theta _{B}}{2},\text{
\ }\eta =\sin ^{2}\frac{\theta _{A}}{2}\cos ^{2}\frac{\theta _{B}}{2}, 
\notag \\
\beta &=&\frac{1}{2}\cos ^{2}\frac{\theta _{A}}{2}\sin \theta _{2},\text{ \ }%
\Theta =\frac{1}{2}\sin ^{2}\frac{\theta _{A}}{2}\sin \theta _{2}.
\end{eqnarray}%
For $\$_{00}^{A}=\$_{10}^{A}=\$_{01}^{B}=\$_{11}^{B}=1,\$_{11}^{A}=%
\$_{01}^{A}=\$_{00}^{B}=\$_{10}^{B}=-1$ with the help of Eq. (\ref{payoff})
we have the following cases:

Step (1) When $\theta _{A}=\theta _{B}=\alpha _{B}=\frac{\pi }{2}$ we get 
\begin{eqnarray}
\$^{A} &=&\sin \theta \sin \phi ,  \notag \\
\$^{B} &=&-\sin \theta \sin \phi .  \label{case-1}
\end{eqnarray}%
Comparing the result (\ref{case-1}) with Eq. (\ref{stokes parameters}) we
see that the payoff of Alice is one of the Stokes parameters.

Step (2) When $\theta _{A}=\theta _{B}=\frac{\pi }{2}$ and $\alpha _{2}=0$
then Eq. (\ref{payoff}) reduces to 
\begin{eqnarray}
\$^{A} &=&\sin \theta \cos \phi ,  \notag \\
\$^{B} &=&-\sin \theta \cos \phi .  \label{case-2}
\end{eqnarray}%
Comparing Eqs. (\ref{case-2})\ and (\ref{stokes parameters}) it is evident
that it is also one the Stokes parameters.

Step (3) When $\theta _{A}=\theta _{B}=0$ then Eq. (\ref{payoff}) gives 
\begin{eqnarray}
\$^{A} &=&\cos \theta ,  \notag \\
\$^{B} &=&-\cos \theta .  \label{case-3}
\end{eqnarray}%
Comparison of the result (\ref{case-3}) with Eq. (\ref{stokes parameters})
shows the payoff of Alice is third Stokes parameter.

It is clear from Eqs. (\ref{case-1}) (\ref{case-2}) and (\ref{case-3}) that
the payoffs are equal to the Stokes parameters of quantum state. In this way
finding the expectation value of a single observable helps us to reconstruct
the quantum state. Furthermore the standard deviation for all of the above
cases is bounded above by 1\textrm{.} It shows that quantum game theory can
be helpful in quantum state tomography. Furthermore this technique is simple
and not beyond the reach of recent technology \cite{du-2,zhou}.

\section{Summary}

The state of the quantum system contains all the information about the
system. In classical mechanics it is possible in principle, to devise a set
of measurements that can fully recover the state of the system. In quantum
mechanics two fundamental theorems, Heisenberg uncertainty principle and no
cloning theorem forbid to recover the state of a quantum system without
having some prior knowledge. This problem, however, can be solved with the
help of quantum state tomography. Where an unknown quantum state is
estimated through a series of measurements on a number of identical copies
of a system. Here we showed that how an unknown quantum state can be
reconstructed by making use of mathematical framework of generalized
quantization scheme of games. In our technique Alice sends an unknown pure
quantum state to Bob who appends it with $\left| 0\right\rangle \left\langle
0\right| $ and then applies the unitary operators on the appended quantum
state and finds the payoffs for Alice and Bob.\ It is shown that for a
particular set of unitary operators these elements become equal to Stokes
parameters of the unknown quantum state. In this way an unknown quantum
state can be measured and reconstructed.

\chapter{Conclusion}

There have been two well known quantization schemes for two person non zero
sum games. The first was introduced by Eisert \textit{et al.} \cite{eisert}\
and the second by Marinatto and Weber \cite{marinatto}. In this thesis we
presented a generalized quantization scheme for two person non zero sum
games that can be reduced to both these schemes for a separate set of
parameters. Furthermore we identified some other situations that were not
apparent in existing quantization schemes. Then we studied different aspects
of quantum games using this quantization scheme. Furthermore using the
mathematical framework of generalized quantization scheme we proposed a
protocol for quantum key distribution and quantum state tomography.

Summary of the main results is as follows.

\begin{enumerate}
\item In an interesting comment on Marinatto and Weber quantization scheme
Benjamin \cite{benjamin} pointed out that in quantum Battle of Sexes%
\index{Battle of sexes!-} the dilemma still exists as the payoffs at both
the Nash equilibria are same and hence both the Nash equilibria are equally
acceptable to the players. The players still have the chance of playing
mismatched strategies and falling into the worst case payoff scenario%
\index{Worst case payoff}. We showed that this worst case payoffs scenario
is not due to the quantization scheme itself but it is due to the
restriction on the initial state parameters of the game. If the game is
allowed to start from a more general initial entangle state then a condition
on the initial state parameters can be set in a manner that the payoffs for
the mismatched or the worst case situation are different for different
players which results as a unique solution of the game.

\item We developed a generalized quantization scheme%
\index{Quantization scheme!generalized} for two person non zero sum games.
It gives a relationship between two apparently different quantization
schemes introduced by Eisert \textit{et al}. \cite{eisert} and Marinatto and
Weber \cite{marinatto}. To introduce this quantization scheme the game of
Battle of Sexes%
\index{Battle of sexes!generalized quantization scheme!-} has been used as
an example but the scheme is applicable to all other games as well. Separate
set of parameters are identified for which this scheme reduces to that of
Marinatto and Weber \cite{marinatto} and Eisert \textit{et al.} \cite{eisert}
schemes. Furthermore there have\ been identified some other interesting
situations which are not apparent within the exiting two quantizations
schemes.

\item We analyzed the effects of measurement on quantum games%
\index{Quantum game!role of measurement} under the generalized quantization
scheme. It was observed that as in the case of quantum channels there were
four types of classical channel capacities \cite{king} similarly quantum
games could have four types of payoffs for the different combinations of
entangled / product input state and entangled / product measurement basis.
Furthermore we also established a relation among these payoffs.

\item We studied quantum games%
\index{Quantum game!with correlated noise} in presence of quantum correlated
noise in the context of our generalized quantization scheme. It was observed
that in the limit of maximum correlation the effect of decoherence vanished
and the quantum game behaved as a noiseless game.

\item Quantum key distribution%
\index{Quantum key distribution} is the technique that allows two parties to
share a random bit sequence with a high level of confidence. Later on this
random bit of sequence works as a key for secure communication between them.
Making use of the mathematical set up of generalized quantization scheme a
protocol for quantum key distribution had been proposed that can transmit
four symbols while using two dimensional quantum system.

\item The state of the quantum system contains all the information about the
system. In classical mechanics it is possible in principle, to devise a set
of measurements that can fully recover the state of the system. In quantum
mechanics two fundamental theorems, Heisenberg uncertainty principle and no
cloning theorem forbid to recover the state of a quantum system without
having some prior knowledge. This problem, however, can be solved with the
help of quantum state tomography. It is a procedure to reconstruct an
unknown quantum state through a series of measurements on a number of
identical copies of the system. Each measurement gives a new dimension of
system. We showed that how an unknown quantum state can be reconstructed by
making use of the framework of quantum game theory.\ It is shown that for
particular set of unitary operators (strategies) and payoff matrix the
payoffs of the players become equal to Stokes parameters of the unknown
quantum state. In this way an unknown quantum state can be measured and
reconstructed.
\end{enumerate}

\end{document}